\documentclass[12pt]{article}
\usepackage{jheppub} 
\usepackage{xcolor}
\usepackage{subcaption}

\definecolor{ForestGreen}{RGB}{34,139,34}
\usepackage[T1]{fontenc}
\usepackage{esint}
\usepackage{physics}
\usepackage{amsmath}
\usepackage{tikz}
\usepackage{mathrsfs}
\usetikzlibrary{arrows.meta, decorations.markings}
\newtheorem{theorem}{Theorem}

\tikzset{
    pil/.style={->, thick, shorten <=2pt, shorten >=2pt,},
    mid arrow/.style={postaction={decorate,decoration={
        markings,
        mark=at position .5 with {\arrow[#1]{stealth'}}
    }}},
}

\newcommand{\eg}{{\it e.g.,}\ }
\newcommand{\ie}{{\it i.e.,}\ }

\usepackage{cancel}
\usepackage{mdframed,framed}

\newcommand{\m}[1]{\mathcal{#1}}
\newcommand{\mh}[1]{\hat{\mathcal{#1}}}
\newcommand{\beqn}{\begin{eqnarray}}
\newcommand{\eeqn}{\end{eqnarray}}
\newcommand{\bee}{\begin{equation} \begin{aligned}}
\newcommand{\eee}{ \end{aligned} \end{equation}}

\usepackage{tikz-cd}
\usepackage{pict2e}
\usepackage{mathtools}
\newcommand{\defeq}{\vcentcolon=}

\usepackage{tocloft}
\newcommand{\rd}{\text{d}}

\newcommand{\chkM}{{\color{red} \,\checkmark\kern-5pt{}_{M}}}

\newcommand{\be}{\begin{equation}}
\newcommand{\ee}{\end{equation}}
\newcommand{\bea}{\begin{eqnarray}}
\newcommand{\eea}{\end{eqnarray}}

\usepackage{ragged2e}
\usepackage{blindtext}

\newcommand{\mt}[1]{\textrm{\tiny #1}}
\newcommand{\reef}[1]{(\ref{#1})}

\newcommand{\beqa}{\begin{eqnarray}}
\newcommand{\eeqa}{\end{eqnarray}}

\renewcommand{\(}{\left(}
\renewcommand{\)}{\right)}
\renewcommand{\[}{\left[}
\renewcommand{\]}{\right]}

\newcommand{\tlt}{{\tilde t}}
\newcommand{\ts}{{\tilde s}}
\newcommand{\J}{\mathcal{J}}
\newcommand{\Jh}{{\mh J}}
\newcommand{\tlm}{{\tilde m}}

\title{The Geodesics Less Traveled: Nonminimal RT Surfaces and Holographic Scattering}

\author[a, b]{Jacqueline Caminiti,}
\author[a, b, c]{Caroline Lima}
\author[a]{and Robert C. Myers}

\affiliation[a]{Perimeter Institute for Theoretical Physics, \\
	Waterloo, Ontario, N2L 2Y5, Canada}
\affiliation[b]{Department of Physics \& Astronomy, University of Waterloo, \\
	Waterloo, Ontario, N2L 3G1, Canada}
\affiliation[c]{Institute for Quantum Computing, University of Waterloo, \\ Waterloo, Ontario, N2L 3G1, Canada}

\emailAdd{jcaminiti@perimeterinstitute.ca,
clima@perimeterinstitute.ca, rmyers.perimeter@gmail.com}

\abstract{The connected wedge theorem \cite{May:2019odp,May:2022clu} states that in order to have a scattering process in the bulk, it is necessary to have $O(1/G_N)$ mutual information between certain ``decision” regions in the boundary theory.
While this large mutual information is not generally sufficient to imply scattering, \cite{Caminiti:2024ctd} showed that for a certain class of geometries, bulk scattering is implied by a certain relation between two (possibly non-minimal) Ryu–Takayanagi surfaces. 
Here, we show that the 2-to-2 version of the theorem becomes an equivalence in pure AdS$_3$: large mutual information between appropriate boundary subregions is both necessary and sufficient for bulk scattering. 
This result allows us to extend the findings of \cite{Caminiti:2024ctd} to a broader class of asymptotically AdS$_3$ spacetimes, which we illustrate with the spinning conical defect geometry. In contrast, we find that matter sources can disrupt this converse relation, and that the $n$-to-$n$ version of the theorem with $n>2$ lacks a converse even in the AdS$_3$ vacuum.}

\begin{document}
\maketitle
\flushbottom

\section{Introduction}
\label{sec:intro}

Over the past decade, quantum information has emerged as a powerful tool in exploring holography and quantum gravity.
Extensive investigations have revealed that quantum entanglement of the microscopic degrees of freedom in the boundary theory plays an essential role in the emergence of a smooth bulk spacetime geometry in holography. 

The key result underlying these investigations is the Ryu-Takayanagi (RT) formula \cite{Ryu:2006bv,Ryu:2006ef} (see also \cite{Rangamani:2016dms}), which relates the entanglement entropy of a boundary subregion to the area of a spacelike extremal surface in the bulk spacetime.
Detailed calculations have shown that at least in certain settings, the entanglement structure of a boundary timeslice determines the spacetime geometry of the corresponding Cauchy surface in the bulk \cite{Balasubramanian:2013lsa,Myers:2014jia,Czech:2014wka,Balasubramanian:2014sra,Headrick:2014eia}. Further, making use of a `first law' of entanglement entropy \cite{Blanco:2013joa}, it can be shown \cite{Lashkari:2013koa,Faulkner:2013ica,Swingle:2014uza,Faulkner:2017tkh} that the Ryu-Takayanagi prescription requires that spacetimes dual to small excitations of the CFT vacuum state must satisfy Einstein’s equations perturbatively expanded around pure AdS spacetime. Hence holographic entanglement entropy plays a key role in understanding both the bulk geometry and its dynamics.

The important role of entanglement in the dynamics and causal structure of the bulk theory was highlighted in recent work on holographic scattering \cite{May:2019yxi, May:2019odp, May:2021nrl, May:2022clu}. The essential result \cite{May:2019odp,May:2022clu} is summarized as the connected wedge theorem (CWT), which observes that strong entanglement of two or more boundary subregions 
is a necessary precondition for certain bulk processes involving timelike-separated input and output regions.
The CWT is proven for general asymptotically AdS$_3$ geometries where the HRRT formula \cite{Hubeny:2007xt} evaluates the entanglement entropy in the boundary theory and the bulk obeys the null energy condition.\footnote{The CWT also extends to semiclassical spacetimes with quantum matter \cite{May:2019odp,May:2022clu} using generalized entropy and quantum extremal surfaces \cite{Faulkner:2013ana,Engelhardt:2014gca}, and assuming the quantum focusing conjecture \cite{Bousso:2015mna} and quantum maximin formula \cite{Akers:2019lzs}.}

\begin{figure}
    \centering
    \qquad\qquad
    \subfloat[\label{fig:connected-scatteringintro}]{
    \includegraphics[scale=0.3]{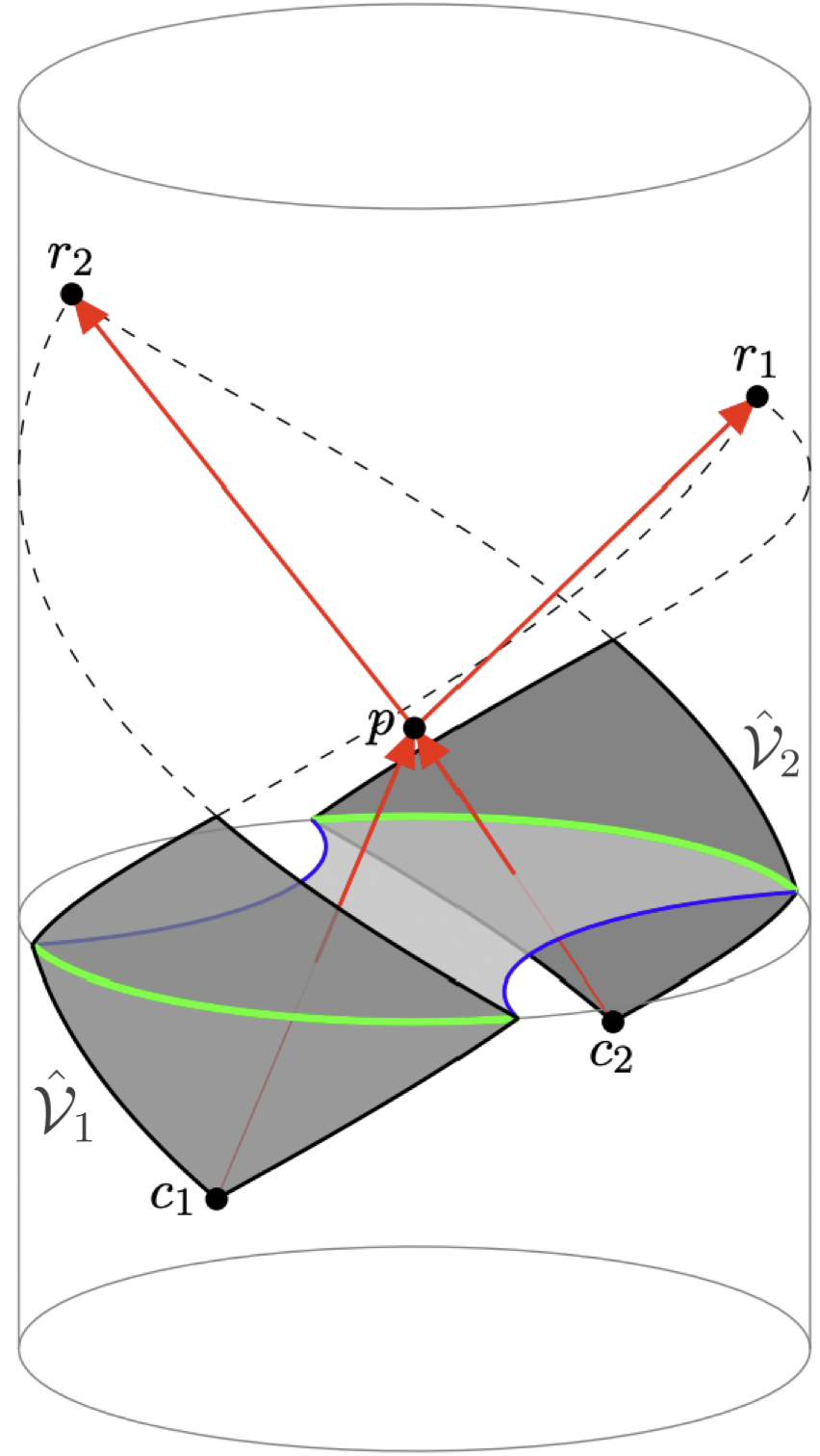}
    }
    \hfill
    \subfloat[\label{fig:disconnected-scatteringintro}]{
    \includegraphics[scale=0.3]{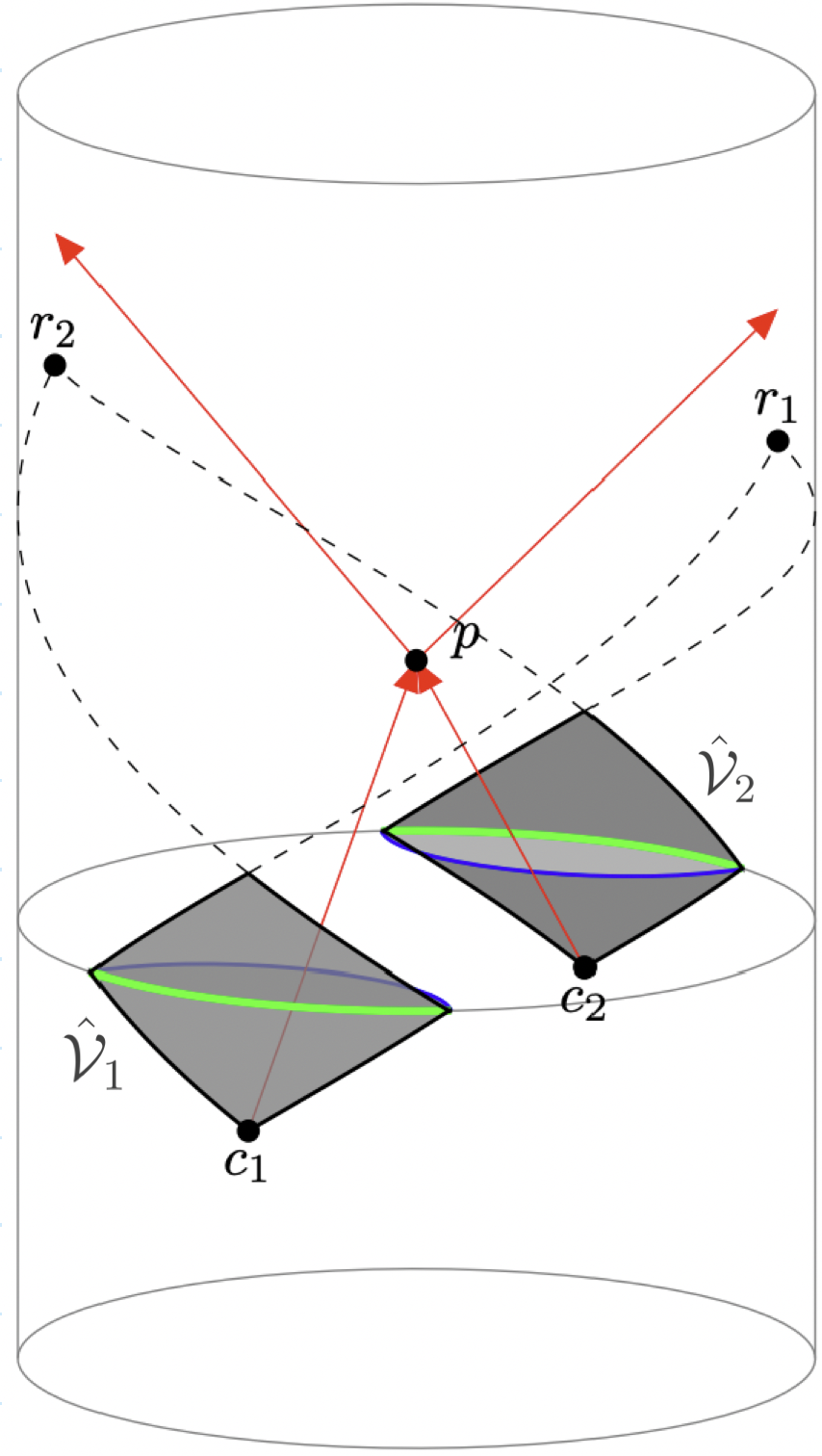}
    }
    \qquad\qquad
    \caption{(a) By the CWT, a bulk-only scattering process from $c_1, c_2$ to $r_1, r_2$ requires associated boundary regions (green) to be strongly correlated, as captured by the connectivity of the entanglement wedge. 
    (b) When there is no bulk-only scattering, $\mh{V}_1$ and $\mh{V}_2$ need not be strongly correlated. 
    In particular, they can have  disconnected minimal-area extremal surfaces. This figure is reproduced from \cite{May:2019yxi}.}
    \label{fig:phasetransition3dintro}
\end{figure}

The simplest example of the CWT involves two input points $c_1$, $c_2$ and two output points $r_1$, $r_2$ specified at the conformal boundary. The CWT is then concerned with processes where probe inputs enter at $c_1$ and $c_2$ and interact deep in the bulk, and the resulting outputs are retrieved at $r_1$ and $r_2$, but the probes cannot come into causal contact if they are constrained to travel within the conformal boundary --
see figure \ref{fig:phasetransition3dintro}.
In this case, the CWT dictates that particular regions associated with the inputs (\ie the ``decision regions'') denoted $\mh{V}_1$ and $\mh{V}_2$ are strongly entangled, as characterized by the corresponding entanglement wedge ${\cal E}({\mh{V}_1\cup \mh{V}_2})$ being connected. 

Hence, from the boundary perspective, the process is enabled by the strong entanglement of the degrees of freedom in the two decision regions. The intuition behind this result is the following: 
The process which proceeds by a local interaction of the probes in in the bulk must somehow be encoded in the boundary CFT by the basic tenets of holography.
However, since the probes cannot actually meet in the boundary theory, this process must be encoded in a  nonlocal manner.
In the context of quantum information, it has been understood that nonlocal protocols to achieve certain quantum computation tasks require specific entanglement resources -- see \eg \cite{Buhrman:2014aou,Kent:2011jim}.
Hence, one is led to expect that bulk scattering processes can be encoded as a nonlocal quantum protocol in the boundary, requiring some minimum amount of entanglement among particular boundary subregions as a resource. The connected entanglement wedge emerging from the CWT implies the required amount of entanglement is characterized by a mutual information $I(\mh{V}_1:\mh{V}_2)$ of order ${\cal O}(1/G_N)$.

We emphasize that the CWT does not provide an if-and-only-if statement.
That is, while large mutual information (\ie a connected wedge) is necessary for these bulk-only scattering processes, it is not sufficient. Counterexamples are known \cite{May:2019odp, Caminiti:2024ctd}, yielding the intuition that not all correlations contributing to $I(\mh{V}_1:\mh{V}_2)$  are useful in facilitating the nonlocal quantum protocols. Further,
counterexamples to the converse also exist for a stronger version of the CWT known as the regions theorem \cite{May:2021nrl}, which will be introduced in section \ref{sec:CWT_PureAdS} and is our main focus in this work. The absence of a converse raises the question: what boundary quantity, if any, serves as a diagnostic for bulk scattering? One hopes that answering this question will improve our understanding of the structure of entanglement needed for the corresponding nonlocal quantum protocols.

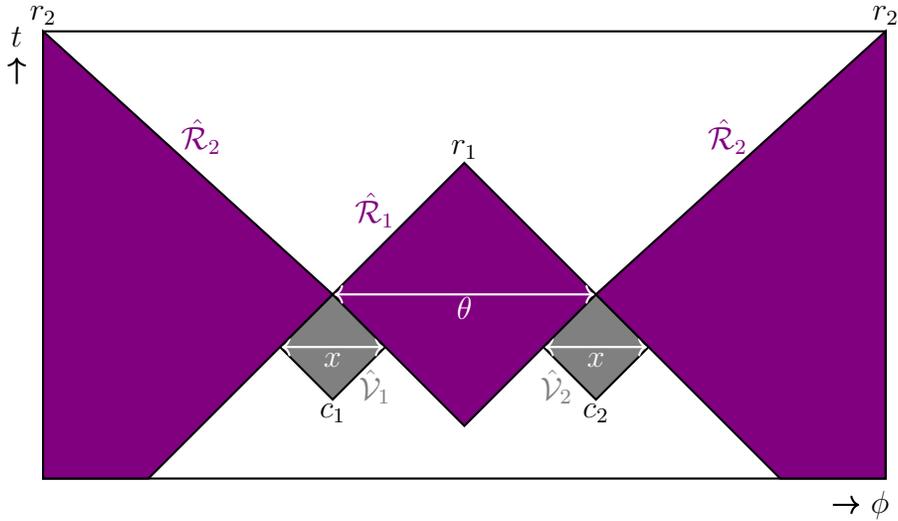
\begin{figure}[t]
    \centering
    \begin{tikzpicture}[scale=0.7]
\draw[thick] (-8, 5) rectangle (8, -3.5);

\draw[->, thick] (-8.5, 4) -- (-8.5, 4.5) node[above] {$t $};
\draw[->, thick] (7,-4) -- (7.5,-4) node[right] {$\phi$};

\draw[fill=gray, thick]
  (-1.5,-1) -- (-2.5, 0) -- (-3.5, -1) -- (-2.5, -2) -- cycle;
  \node[white] at (-2.5,-1.25) {$x$};
  \draw[<->, thick, white] (-1.53,-1) -- (-3.47, -1);
  \node[gray] at (-1.7, -1.75) {$\mh{V}_1$};
  \node at (-2.5, -2.25) {$c_1$};
  
\draw[fill=gray, thick]
  (1.5,-1) -- (2.5, 0) -- (3.5, -1) -- (2.5, -2) -- cycle;
  \node[white] at (2.5,-1.25) {$x$};
  \draw[<->, thick, white] (1.53,-1) -- (3.47,-1);
  \node[gray] at (1.75, -1.75) {$\mh{V}_2$};
  \node at (2.5, -2.25) {$c_2$};

\draw[fill=violet, thick]
  (-2.5,0) -- (0,2.5) -- (2.5,0) -- (0,-2.5) -- cycle;
\node[violet] at (-1.7, 1.6) {$\mh{R}_1$};
\draw[<->, white, thick] (-2.47,0) -- (2.47,0);
\node[white] at (0,-0.25) {$\theta$};
\node at (0, 2.75) {$r_1$};

\draw[fill=violet, thick]
  (-2.5,0) -- (-6,-3.5) -- (-8, -3.5) -- (-8, 5) -- cycle;
\node[violet] at (-5, 3) {$\hat{\m{R}}_2$};
\node at (-8, 5.3) {$r_2$};

\draw[fill=violet, thick]
  (2.5,0) -- (6,-3.5) -- (8, -3.5) -- (8, 5) -- cycle;
\node[violet] at (5, 3) {$\mh{R}_2$};
\node at (8, 5.3) {$r_2$};
  
\end{tikzpicture}
\caption{The 2-to-2 scattering setup on the boundary cylinder, which was considered in \cite{Caminiti:2024ctd}. Both input regions, $\mh{V}_1$ and $\mh{V}_2$, are on a fixed time slice, have the same size $x$, and their midpoints are separated by an angle $\theta$. The output regions, $\mh{R}_1$ and $\mh{R}_2$, are maximized to cover an entire Cauchy slice. Note that the vertical edges on the left and and right are identified.}
\label{fig:optimal}
\end{figure}
\begin{figure}[htbp]
\centering
\begin{subfigure}[b]{0.3\textwidth}
    \centering
    \includegraphics[width=\textwidth]{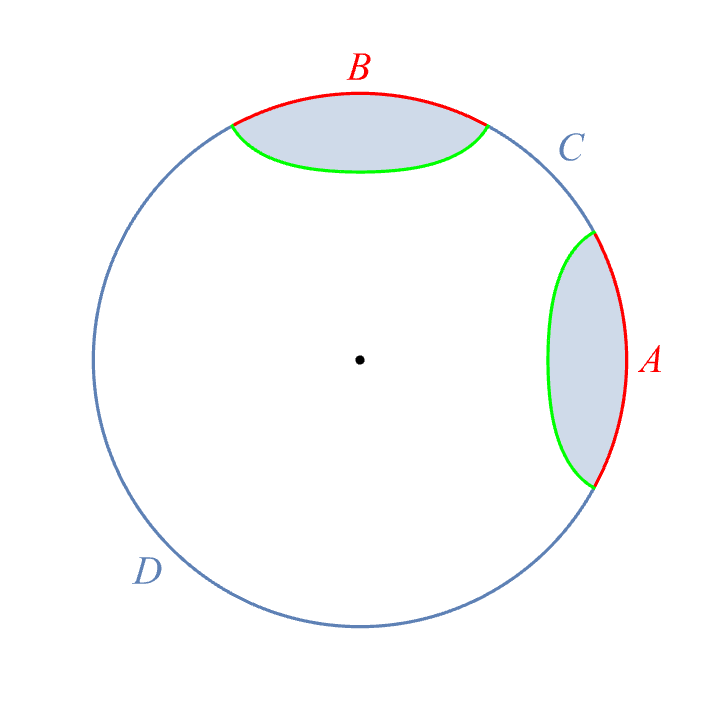}
    \caption{}
    \label{fig:dcircs_d}
\end{subfigure}
\hfill
\begin{subfigure}[b]{0.3\textwidth}
    \centering
    \includegraphics[width=\textwidth]{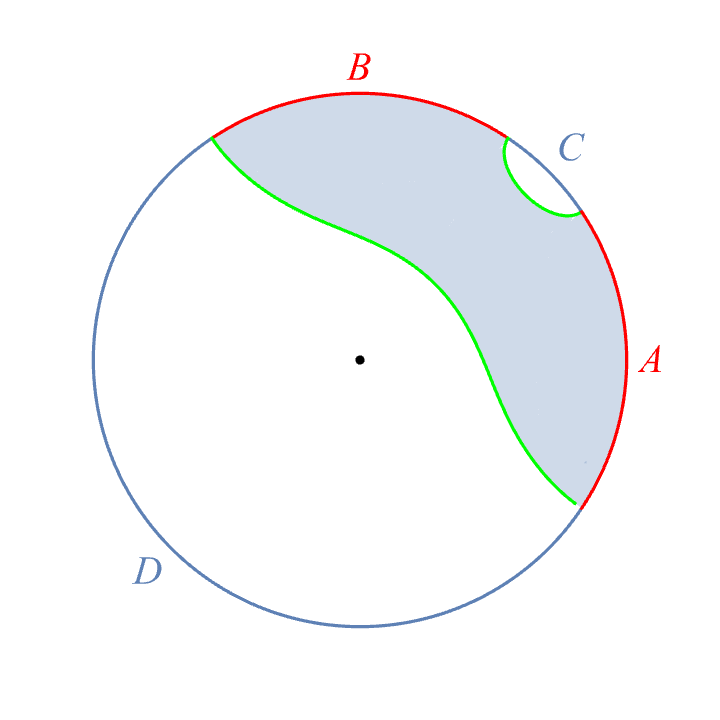}
    \caption{}
    \label{fig:dcircs_u}
\end{subfigure}
\hfill
\begin{subfigure}[b]{0.3\textwidth}
    \centering
    \includegraphics[width=\textwidth]{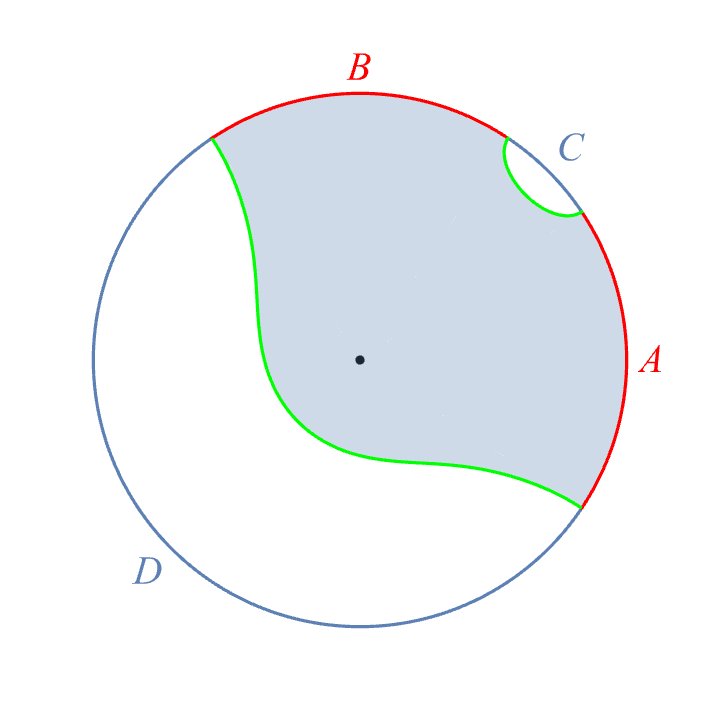}
    \caption{}
    \label{fig:dcircs_o}
\end{subfigure}
\caption{A constant-$t$ slice of the conical defect geometry showing, from left to right, the $d$, $u$, and $o$ candidates for $\m{E}(A\cup B)$. Figure copied from \cite{Caminiti:2024ctd}.}
\label{fig:dcircs}
\end{figure}

Towards this goal, \cite{Caminiti:2024ctd} examined the regions version of the CWT in asymptotically AdS$_3$ spacetimes containing a (static) massive object, namely a conical defect or a BTZ black hole. Figure \ref{fig:optimal} illustrates the 2-to-2 scattering setup on the boundary cylinder considered in \cite{Caminiti:2024ctd}. The two input regions $\mh{V}_1$ and $\mh{V}_2$ lie on a fixed time slice and have the same size $x$, and their midpoints are separated by an angle $\theta$. The output regions, $\mh{R}_1$ and $\mh{R}_2$, are maximized to cover an entire Cauchy slice, as illustrated.
In this situation, there are three possible topologies for the entanglement wedge ${\cal E}({\mh{V}_1\cup \mh{V}_2})$: a disconnected topology (denoted $d$) and two connected topologies (denoted $u$ and $o$).
The $u$ and $o$ candidates differ by whether or not the entanglement wedge does ($o$) or does not ($u$) include the massive object at the center of the spacetime --
see figure \ref{fig:dcircs}. 

The analysis in \cite{Caminiti:2024ctd} revealed that a curious relationship\footnote{The inequality $u<d$ is a convenient shorthand to indicate that the area of the extremal surface associated with the $u$ configuration is smaller than that of the corresponding $d$ configuration.} $u<d$ was sufficient to imply holographic scattering in these nontrivial AdS$_3$ backgrounds. Of particular note was that the $u<d$ condition still implied scattering even when $u$ did not describe the actual entanglement wedge, \ie the corresponding extremal surface was not the minimal RT surface. 
In the present work, we test the robustness of the $u<d$ converse to the CWT with a view towards understanding better what kind of entanglement allows for holographic scattering.

To summarize our results, we begin by observing that in pure AdS$_3$, the 2-to-2 CWT has a converse, as anticipated in \cite{May:2021nrl}. 
We then explore the consequences of this result for holographic scattering in a general asymptotically AdS$_3$ vacuum solution, $\mathcal{M}$.
Our key insight is that any open, simply connected subregion $\mathcal{N}\subset \mathcal{M}$ which does not contain defects or singularities is equivalent to a portion of pure AdS$_3$, and so the CWT converse in pure AdS$_3$ has implications for this broader class of geometries.
Building on this insight, we generalize the $u<d$ result of \cite{Caminiti:2024ctd} to all asymptotically AdS$_3$ vacuum solutions.
Specifically, we show that a $u<d$-type relation among nonminimal surfaces contained in $\mathcal{N}$ implies holographic scattering in pure states.
Notably, this extension applies to, \eg  spinning defect or spinning BTZ backgrounds, with arbitrary choices of spacelike
separated input regions (not only those shown in figure \ref{fig:optimal}).

Next, we test the robustness of the $u<d$ result beyond the case of locally AdS$_3$ solutions by adding matter which deforms the local geometry.
In particular, we analyze holographic scattering in a static asymptotically AdS$_3$ background containing a spherical shell of matter.
We find that the correspondence between holographic scattering and nonminimal surfaces persists in certain regimes, specifically when the relevant RT surfaces are contained in a subregion $\mathcal{N}$  equivalent to a portion of pure AdS$_3$.
However, the correspondence breaks down when the relevant RT surfaces probe the matter shell, and we interpret this failure in section \ref{sec:discuss}.

Lastly, we extend our analysis to the $n$-to-$n$ CWT \cite{May:2022clu}.
We find that for $n>2$, the CWT does not admit a converse even in the AdS$_3$ vacuum. This indicates that $u<d$-type inequalities should not be expected to govern  higher-$n$ scattering.

The remainder of the paper is organized as follows:
In section \ref{sec:CWT_PureAdS},
we review the connected wedge theorem and demonstrate that the 2-to-2 CWT naturally has a converse in pure AdS$_3$.
Building on this observation, in section \ref{sec:genulessd}, we generalize the $u<d$ result of \cite{Caminiti:2024ctd} to general asymptotically AdS$_3$ vacuum solutions.
In section \ref{sec:starb}, we show that the $u<d$ result fails to generalize for asymptotically AdS$_3$ solutions with matter by analyzing holographic scattering in a simple background containing a spherical shell. 
In section \ref{sec:3to3}, we extend our analysis to consider the $n$-to-$n$ CWT 
and demonstrate that there is no converse with $n>2$, even in pure AdS$_3$.
Finally, we discuss our results and possible future directions in section \ref{sec:discuss}, in particular drawing connections to future work \cite{ap,ap2}. 
Additional technical details are provided in the appendices:
Appendix \ref{app:basics} provides useful expressions for spacelike and null geodesics in global AdS$_3$ coordinates.
Appendix \ref{sec:spinning} examines how the results of section \ref{sec:genulessd} apply for the spinning conical defect. Appendix \ref{sec:deetail} contains further details on the shell geometry and the calculations discussed in section \ref{sec:starb}.

\paragraph{Summary of notation}

\begin{itemize}
\itemsep0.25em 
\item We use curly capital letters $\m{A}$, $\m{B}$, $\m{C}$, \ldots for codimension-0 bulk spacetime regions.
\item We use curly, hatted capitals $\mh{A}$, $\mh{B}$, $\mh{C}$, \ldots for codimension-0 boundary regions.
\item We use plain capital letters $A$, $B$, $C$, \ldots for codimension-1 boundary regions. 
\item We use $\J^{\pm}(\m{S})$ to denote the causal future or past of the bulk region $\m{S}$ and $\mh{J}^{\pm}(\mh{S})$ to denote the causal future or past of $\mh{S}$ taken in the boundary. Bulk sets like $\J^+(\m{S})$ are taken in the conformally completed spacetime that includes the asymptotic boundary, so they can include boundary points.
\item The RT surface associated to region $A$ with causal development $\mh{A}$ is denoted $\gamma_A$, or equivalently $\gamma_{\mh{A}}$.
\item The entanglement wedge of a boundary region $\mh{S}$ is denoted by $\m{E}(\mh{S})$.
\item The causal wedge $\J^+(\mh{S})\cap \J^-(\mh{S})$ of a boundary region $\mh{S}$ is denoted by $\m{C}(\mh{S})$.
\item The domain of dependence of a bulk or boundary region is denoted by $\m{D}(\cdot)$.
\end{itemize}

\section{\texorpdfstring{The connected wedge theorem in pure AdS$_3$}{The connected wedge theorem in pure AdS3}}
\label{sec:CWT_PureAdS}

In this section, we introduce the regions-based connected wedge theorem \cite{May:2021nrl} and show that for 2-to-2 scattering in pure AdS$_3$, the converse of the connected wedge theorem always holds: namely, having a connected entanglement wedge is enough to imply holographic scattering is possible.

\subsection{The connected wedge theorem}\label{subsec:CWT}

We state the connected wedge theorem (CWT) \cite{May:2021nrl}:

\begin{theorem}
Pick four regions ${\mh C}_1$, ${\mh C}_2$, ${\mh R}_1$, ${\mh R}_2$ on the boundary of an asymptotically AdS$_3$ spacetime.
From these, define the decision regions
\begin{equation}
\begin{aligned}
    \mh{V}_1 &= \mh{J}^+(\mh{C}_1)\cap\mh{J}^-(\mh{R}_1)\cap \mh{J}^-(\mh{R}_2)\,,\\
    \mh{V}_2 &= \mh{J}^+(\mh{C}_2)\cap\mh{J}^-(\mh{R}_1)\cap \mh{J}^-(\mh{R}_2)\,.
\end{aligned}
\label{eq:bdyregions}
\end{equation}
Assume that ${\mh C}_i \subseteq \mh{V}_i$.
Assume also that the bulk geometry obeys the null energy condition and the HRRT surface can be found using the maximin formula \cite{Wall:2012uf}.
Define the bulk scattering region
\begin{equation}
    J_{1,2\to1,2}
    =
    \m{J}^+(\m{C}_1) \cap \m{J}^+(\m{C}_2) \cap \m{J}^-(\m{R}_1) \cap \m{J}^-(\m{R}_2) \,,
    \label{eq:scatregion}
\end{equation}
where $\m{C}_i = \m{E}(\mh{C}_i)$ and $\m{R}_i = \m{E}(\mh{R}_i)$.
Then $J_{1,2\to1,2} \ne \emptyset$ implies $\m{E}(\mh{V}_1\cup \mh{V}_2)$ is connected.
\label{thm:cwt}
\end{theorem}

The CWT is primarily of interest in the case where the boundary scattering region
\beqn
\hat{J}_{1,2\to 1,2}=\mh{J}^+(\mh{C}_1)\cap\mh{J}^+(\mh{C}_2)\cap\mh{J}^-(\mh{R}_1)\cap \mh{J}^-(\mh{R}_2)=\mh{V}_1\cap\mh{V}_2
\label{eq:bscatt}
\eeqn
vanishes. 
That is, probes from ${\mh C}_1$ and ${\mh C}_2$ cannot directly be brought together, made to interact, and then sent to ${\mh R}_1$ and ${\mh R}_2$ within the boundary. However, it may be that scattering is still possible in the bulk, and the CWT indicates that such bulk-only scattering processes require strong correlations among the $\mh{V}_i$ subregions.

Note that the example discussed in the introduction and depicted in figure \ref{fig:phasetransition3dintro} involves a special case of the CWT where the boundary regions consist of single points: $\mh{C}_1=\{c_1\}$, 
$\mh{R}_2=\{r_2\}$,
etc.
This points version of the theorem was developed first \cite{May:2019odp}, but we focus on the regions theorem throughout the main text because it is more general.
We henceforth reserve the labels $c_i$ to refer to the past-most points of the $\mh{C}_i$ regions, and $r_i$ to refer to the future-most points of the $\mh{R}_i$ regions.
We emphasize that the two versions of the CWT are the same in pure AdS$_3$, the case of interest in this section, because the equivalence of causal wedges and entanglement wedges in pure AdS$_3$ implies
\begin{equation}
   \m{J}^+(\m{E}(\mh{V}_i))
   = \m{J}^+(c_i)
   \qquad 
   \m{J}^-(\m{E}(\mh{R}_i))
   = \m{J}^-(r_i)\,.
   \label{eq:equivalence}
\end{equation}

To recap, the CWT stipulates that bulk scattering implies strong correlations between the corresponding boundary regions.
As indicated in the introduction, we are interested to ask if there is a converse statement, \ie to what extent are there boundary diagnostics which imply bulk scattering processes is possible. To pursue this question, one should maximize the output regions $\mh{R}_i$, as this makes it more likely to obtain a nontrivial bulk scattering region \eqref{eq:scatregion}.
Specifically, going forward, we set $\mh{C}_i= \mh{V}_i$.
After fixing $\mh{V}_1$ and $\mh{V}_2$, then $r_1$ and $r_2$ are fixed, 
and the maximal choice of the $\mh{R}_i$  is given by
\begin{equation}
\begin{aligned}
    \mh{R}_i &= \hat{J}^-(r_i) \backslash \left[ \hat{J}^-(\mh{V}_1)
    \cup \hat{J}^-(\mh{V}_2)\right]
    \,,\qquad i = 1,2\,.
    \label{eq:RRR}
\end{aligned}
\end{equation}
We will assume these choices going forward.
For typical configurations of this type, see figures \ref{fig:optimal} and \ref{fig:iffsetup}.

\subsection{The converse in pure AdS: 2-to-2 case}\label{subsec:converseAdS}

We now demonstrate that in pure AdS$_3$, holographic scattering is not only necessary but also sufficient for a connected entanglement wedge.
This was demonstrated in appendix B of \cite{May:2021nrl} for the case where $\mh{V}_1$ and $\mh{V}_2$ are the causal developments of equally-sized intervals on a constant time-slice; our approach is very similar to the approach in this reference.\footnote{We thank Alex May for conversations on this topic.}

To begin, we have two input regions, $\mh{V}_1$ and $\mh{V}_2$, on the boundary of AdS$_3$, \eg see figure \ref{fig:phasetransition3dintro}. We take these two regions to be the causal development of two general but spacelike-separated intervals on the boundary.\footnote{We assume that the intervals are separated as we are primarily interested in the situation where $\hat{J}_{1,2\to1,2}=\emptyset$, but our analysis can also accommodate the situation where the input regions overlap.} Following the previous discussion, we have implicitly assumed that the input and decision regions coincide (\ie $\mh{C}_i=\mh{V}_i$) and we also maximize the the output regions $\mh{R}_i$, choosing them to be complementary causal diamonds on the boundary cylinder as in eq.~\reef{eq:RRR}.

Now, our analysis is simplified by working in Poincar\'e coordinates in the bulk. 
Hence we note that while our general $\mh{V}_1$ and $\mh{V}_2$ were implicitly chosen on the full Lorentzian cylinder ($S^1 \times \mathbb{R}$) on the conformal boundary of AdS$_3$, they can always be mapped to two-dimensional Minkowski space ($\mathbb{R}^{1,1}$). We make a corresponding coordinate transformation from global to Poincar\'e coordinates in the bulk. 
We may choose this mapping such that $\mh{R}_1$ is conformally mapped to a causal diamond inside the Minkowski space. In contrast, with this choice, $\mh{R}_2$ extends beyond the region on the cylinder that is mapped to Minkowski space. Hence (a portion of) the second output region is conformally mapped to two semi-infinite causal diamonds straddling spatial infinity (the image of a single point on the cylinder). Now crucially, as we will see, the objects involved in bulk scattering, \eg the RT surface for $\mh{R}_1$, all sit within the Poincar\'e patch, and so to prove the CWT converse in pure AdS$_3$ it suffices to work in the Poincar\'e patch.

Recall that in Poincar\'e coordinates, the AdS$_3$ metric becomes\footnote{Here and below, we adopt units in which the AdS radius of curvature equals one.}
\begin{equation}
    ds^2 = \frac{1}{z^2}(-dT^2 + dX^2 + dz^2)\,.
    \label{eq:Poincare}
\end{equation}
With the null coordinates,
\begin{align}
    U&= T- X\,, \label{eq:NULL}\\
    V&= T+X\,, \nonumber
\end{align}
the boundary metric for two-dimensional Minkowski space becomes simply
\begin{align}
    \dd s^2=- \dd U\,\dd V\,.
\end{align}

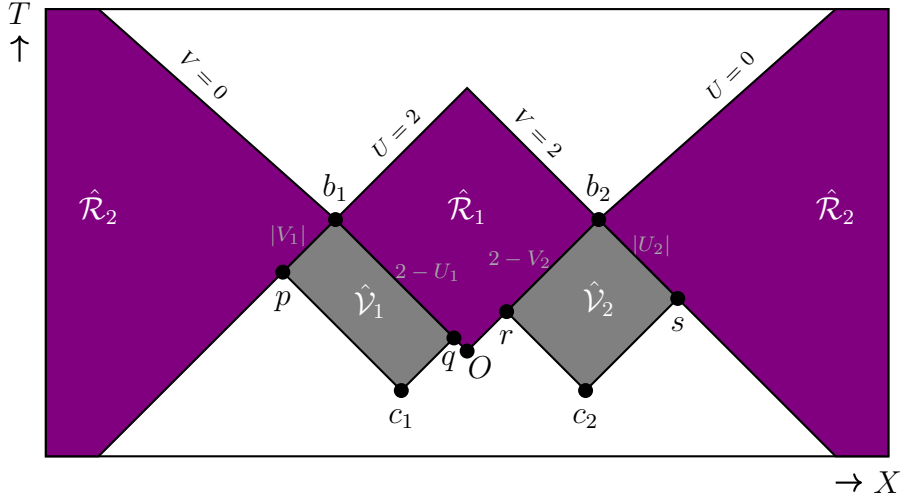
\begin{figure}[t]
    \centering
    \begin{tikzpicture}[scale=0.7]
\draw[thick] (-8, 5) rectangle (8, -3.5);

\draw[->, thick] (-8.5, 4) -- (-8.5, 4.5) node[above] {$T$};
\draw[->, thick] (7,-4) -- (7.5,-4) node[right] {$X$};

\draw[fill=violet, thick]
  (-2.5,1) -- (0,3.5) -- (2.5,1) -- (0,-1.5) -- cycle;
\node[white] at (0, 1.25) {$\mh{R}_1$};
\node[rotate=45] at (-1.35,2.65) {\scriptsize $U=2$};
\node[rotate=-45] at (1.35,2.65) {\scriptsize $V=2$};
\node[rotate=-45] at (-5,3.75) {\scriptsize $V=0$};
\node[rotate=45] at (5,3.75) {\scriptsize $U=0$};

\draw[fill=violet, thick]
  (-2.5,1) -- (-7,-3.5) -- (-8, -3.5) -- (-8, 5)--(-7,5) -- cycle;
\node[white] at (-7, 1.25) {$\hat{\m{R}}_2$};

\draw[fill=violet, thick]
  (2.5,1) -- (7,-3.5) -- (8, -3.5) --(8, 5)-- (7, 5) -- cycle;
\node[white] at (7, 1.25) {$\mh{R}_2$};
\draw[fill=gray, thick]
  (-0.25,-1.25) -- (-2.5, 1) -- (-3.5, 0) -- (-1.25, -2.25) -- cycle;
  \node[white] at (-1.85, -.55) {$\mh{V}_1$};
  \node[circle, fill=black, inner sep=2pt, label=above:$b_1$] at (-2.5, 1) {};
  \node[circle, fill=black, inner sep=2pt, label=below:$c_1$] at (-1.25, -2.25) {};
  \node[circle, fill=black, inner sep=2pt, label=below:$p$] at (-3.5, 0) {};
  \node[circle, fill=black, inner sep=2pt] at (-0.25,-1.25) {};
  \node at (-0.35,-1.7)  {$q$};
  \node[circle, fill=black, inner sep=2pt] at (0,-1.5) {};
  \node at (.25,-1.8)  {$O$};
  \node[gray!70] at (-3.4, 0.7)  {\scriptsize $|V_1|$};
  \node[gray!70] at (-.75, 0)  {\scriptsize $2-U_1$};
  \node[gray!70] at (1, .25)  {\scriptsize $2-V_2$};
  \node[gray!70] at (3.5, .5)  {\scriptsize $|U_2|$};
  
\draw[fill=gray, thick]
  (0.75,-.75) -- (2.5, 1) -- (4, -.5) -- (2.25, -2.25) -- cycle;
  \node[white] at (2.5, -.5) {$\mh{V}_2$};
  \node[circle, fill=black, inner sep=2pt, label=above:$b_2$] at (2.5, 1) {};
  \node[circle, fill=black, inner sep=2pt, label=below:$r$] at  (0.75,-.75) {};
  \node[circle, fill=black, inner sep=2pt, label=below:$c_2$] at  (2.25, -2.25) {};
  \node[circle, fill=black, inner sep=2pt, label=below:$s$] at  (4, -.5) {};
\end{tikzpicture}
\caption{Scattering regions in two-dimensional Minkowski spacetime. While $\mh{R}_1$ is a finite causal diamond, $\mh{R}_2$ consists of two semi-infinite causal diamonds.}
\label{fig:iffsetup}
\end{figure}

Now we can further arrange the configuration of scattering regions as shown in figure \ref{fig:iffsetup}, using the remaining boundary symmetries. First, we use space and time translations to position the past-most point $O$ of the first output region, $\mh{R}_1$, at the origin. Next, using a boost, we can position the spatial corners, $b_1$ and $b_2$, of this region (\ie $\partial\mh{R}_1\cap\partial\mh{R}_2$)  to be on the same time slice. Finally with a scale transformation,  we can conveniently choose this slice to be $T=1$. In null coordinates $(U,V)$, we then have
\begin{equation}
    b_1=(2,0)
    \qquad \textrm{and}\qquad
    b_2=(0,2)\,.
\end{equation}
As discussed above, ${\mh R}_2$ is the causal complement of $\mh{R}_1$ and is not fully contained within the Poincar\'e patch.

Now ${\mh R}_1$ and  ${\mh R}_2$ are circumscribed by the four null lines $U=2$, $V=2$, $U=0$, and $V=0$.
Further, ${\mh V}_1$ must have its future-most boundaries contained in the lines $U=2$ and $V=0$, leaving two free parameters $V_1$ and $U_1$, which parameterize the past boundaries of this region.
Similarly, ${\mh V}_2$ must have its future-most boundaries at $U=0$ and $V=2$, leaving the two free parameters $V_2$ and $U_2$ to describe its past boundaries.
Note, in our conventions, $V_1,U_2<0$  while $0<U_1,V_2<2$ -- see figure \ref{fig:iffsetup}. 

Now using these null coordinates, the spatial corners of $\mh{V}_1$ are given by
\begin{equation}
    p
    = (2, V_1)
    \qquad \textrm{and}\qquad
    q
    = (U_1, 0)\,,
\end{equation}
and for ${\mh V}_2$, we have
\begin{equation}
    r
    = (0, V_2)
    \qquad \textrm{and}\qquad
    s
    = (U_2, 2)\,.
\end{equation}
In both cases, we can regard the input regions 
to be the domain of dependence of the interval running between the corresponding points above.

We now turn to the entanglement wedge of ${\mh V}_1\cup {\mh V}_2$.
For a single boundary interval $A$ with endpoints $(U_i, V_i)$ and $(U_f, V_f)$ in Poincar\'e AdS$_3$, the length of the corresponding RT surface is
\begin{align}
    S_A = \frac{c}{3} \log (\frac{\sqrt{|U_f - U_i| |V_f - V_i|}}{\epsilon}),
    \label{eq:SA}
\end{align}
where $c$ is the CFT central charge and $\epsilon$ is the short distance UV cutoff. (In the bulk, we have a cutoff surface at $z=\epsilon$.)

For the combined boundary subregion ${\mh V}_1\cup {\mh V}_2$, there are two possible RT candidates.
The first RT candidate is the disconnected candidate $d$; it consists simply of the RT surface for ${\mh V}_1$ and the RT surface for ${\mh V}_2$.
The corresponding entanglement entropy is
\begin{equation}
    S_d = \frac{c}{3}\log(\frac{\sqrt{| V_1 U_2 (2-U_1)(2-V_2)|}}{\epsilon^2})\,.
    \label{eq:sd}
\end{equation}
The second candidate is the connected candidate $u$; it consists of the RT surface extending from $p$ to $s$ and the RT surface extending from $q$ to $r$.
The corresponding entanglement entropy is
\begin{align}
    S_u = \frac{c}{3}\log(\frac{\sqrt{|U_1 V_2 (2-V_1)(2-U_2)|}}{\epsilon^2})\,.
    \label{eq:su}
\end{align}
Comparing eqs.~\eqref{eq:sd} and \eqref{eq:su}, the connected phase has the minimal area (\ie $u<d$) precisely when
\begin{align}\label{eq:dequalsu}
      V_1 \,U_2\, (2-U_1 - V_2)>U_1 \,V_2 \,(2-V_1 - U_2)\,.
\end{align}

We would like to compare eq.~\eqref{eq:dequalsu} to the condition for a non-empty bulk scattering region, \ie $J_{1,2\to 1,2}\ne\emptyset$.
We can simplify the expression \eqref{eq:scatregion} for the scattering region by observing that $\mh{R}_1$ and $\mh{R}_2$ share the same RT surface: $\gamma_{\mh{R}_1}=\gamma_{\mh{R}_2}$.
Hence,\footnote{For a more detailed justification of eq.~\eqref{eq:keystone}, see \cite{Caminiti:2024ctd}.}
\begin{equation}
    J^-({\cal R}_1)\cap J^-({\cal R}_2) = J^-(\gamma_{\mh{R}_1})\,,
    \label{eq:keystone}
\end{equation}
and using eq.~\eqref{eq:equivalence}, the scattering region can be written as
\begin{equation}
    J_{1,2\to1,2}
    =
    \m{J}^+(c_1) \cap \m{J}^+(c_2) \cap \m{J}^-(\gamma_{\mh{R}_1}) \,.
    \label{eq:ccr}
\end{equation}
Here $c_1$ and $c_2$ correspond to the past-most points of $\mh{V}_1$ and $\mh{V}_2$, respectively -- see figure \ref{fig:iffsetup}.  
Now, eq.~\eqref{eq:ccr} says that holographic scattering is possible if and only if two probe systems leaving from $c_1$ and $c_2$ respectively can meet at $\gamma_{\mh{R}_1}$.
The threshold for scattering thus occurs when \textit{null} rays emanating from $c_1$ and $c_2$ intersect precisely at $\gamma_{\mh{R}_1}$.
Note that $\gamma_{\mh{R}_1}$ is contained in the future lightsheet emanating from the past-most point of $\mh{R}_1$, which we have positioned at the origin $(U,V,z)=(0,0,0)$.
Hence, we can determine the scattering threshold by considering three lightcones emanating from the conformal boundary at $z=0$, with vertices $(0,0)$, $c_1= (U_1, V_1)$, and $c_2=(U_2, V_2)$ respectively. These lightcones are given by
\begin{align}
    0&=z^2 - UV \label{eq:lc0}\\
    0&=z^2-(U-U_1)(V-V_1)\\
    0&=z^2-(U-U_2)(V-V_2)\,,
\end{align}
and all three lightcones intersect at a single point with $U$ and $V$ coordinates given by
\begin{align} \label{eq:intersectionlightcones}
    (U_I, V_I)=\left(\frac{U_1 U_2 (V_2 -V_1)}{U_1 V_2 - U_2 V_1},\frac{V_1 V_2 (U_1 -U_2)}{U_1 V_2 - U_2 V_1}\right).
\end{align}
The threshold for holographic scattering occurs when $(U_I, V_I)$ occurs precisely at the RT surface for ${\mh R}_1$, given by
\begin{align}
    &T= \frac{U+V}{2} = 1\\
    &X^2+z^2 = \frac{(V-U)^2}{2} + z^2 =1.
\end{align}
If $(U_I, V_I)$ satisfies the first equation, then it  automatically satisfies the second equation, because by definition $(U_I, V_I)$ lies within the lightcone emanating from $(0,0)$, \ie we can re-express eq.~\reef{eq:lc0} as $z^2 + X^2 = T^2$. The first equation, ${U_I + V_I} = 2$, yields the $u=d$ condition from eq.~\eqref{eq:dequalsu}.
More generally, one can verify the scattering condition ${U_I + V_I} < 2$ matches the $u<d$ inequality given there.

Therefore, as advertised, the CWT converse holds for the most general configuration of 2-to-2 scattering regions in the pure AdS$_3$ spacetime.

\section{\texorpdfstring{The generalized $u<d$ converse}{The generalized u<d converse}}
\label{sec:genulessd}

In the previous section, we have shown that the 2-to-2 CWT has a converse in pure AdS$_3$.
In this section, we demonstrate the consequences of this result for a CWT converse in general AdS$_3$ vacuum solutions, yielding an extension of the $u<d$ result of \cite{Caminiti:2024ctd} to a broader class of backgrounds.

In three spacetime dimensions, vacuum solutions of Einstein's equations with negative cosmological constant, such as conical defect and BTZ black hole solutions, are locally AdS$_3$ \cite{Brown:1988am}, but may have nontrivial topology.
Consider one such vacuum solution, $\m{M}$, and consider an open, simply connected region $\m{N}\subset \m{M}$ which does not contain defects or singularities.
Because the manifold is locally AdS$_3$, $\m{N}$ can be endowed with coordinates under which the line element is that of AdS$_3$ in \eg standard global coordinates; in other words, $\m{N}$ is a portion of pure AdS$_3$.

Now consider a holographic scattering setup in $\m{M}$ obeying the following three assumptions:
\begin{enumerate}
    \item $\m{N}$ contains a connected entanglement wedge candidate $\m{E}^u_{{\mh V}_1\cup{\mh V}_2}$ for $\mh{V}_1\cup\mh{V}_2$,
    \item the top ridge of $\m{E}^u_{{\mh V}_1\cup{\mh V}_2}$ (see figure \ref{fig:inN} below) is the RT surface $\gamma_{{\mh R}_1}$ for ${\mh R}_1$,
    \item the boundary state is pure.\footnote{Note that we allow for the possibility that $\m{M}$ contains multiple asymptotically AdS$_3$ boundaries. An example of holographic scattering in the multi-boundary case is the two-sided BTZ setup in \cite{Caminiti:2024ctd}.}
\end{enumerate}
We denote the RT candidate associated with $\m{E}^u_{{\mh V}_1\cup{\mh V}_2}$ as $u$ because much like the $u$ candidate in figure \ref{fig:dcircs_u}, it corresponds to a connected phase, and by virtue of being contained in $\m{N}$, it avoids any interesting structures, such as defects or horizons, in the global solution. 
We note that $\m{E}^u_{{\mh V}_1\cup{\mh V}_2}$ may not represent the true entanglement wedge, denoted $\m{E}_{{\mh V}_1\cup{\mh V}_2}$, since there may exist RT candidates for ${\mh V}_1\cup{\mh V}_2$ which lie outside of $\m{N}$ and have less area than $u$.
Similarly, in writing assumption 2, we have used that the top ridge of $\m{E}^u_{{\mh V}_1\cup{\mh V}_2}$ automatically furnishes an RT candidate for ${\mh R}_1$;\footnote{To see this, observe that the top ridge of $\m{E}^u_{{\mh V}_1\cup{\mh V}_2}$ is the intersection of two lightsheets emitted orthogonally from $u$. 
These lightsheets have zero expansion because $u$ is an extremal surface and there are no focusing effects from the Raychaudhuri equation in pure AdS$_3$.} however, it need not correspond to the true RT surface $\gamma_{\mh{R}_1}$, since other RT candidates for ${\mh R}_1$, with less area, may exist outside of $\m{N}$.
Hence, our assumption that $\gamma_{\mh{R}_1}$ is the top ridge of $\m{E}^u_{{\mh V}_1\cup{\mh V}_2}$ is not redundant with the previous assumption, though see further discussion on this point in section \ref{sec:discuss}.

The above assumptions come with two consequences which will be important for our analysis.
Firstly, assumption 1 implies  $\m{N}$ contains a disconnected entanglement wedge candidate 
for $\mh{V}_1\cup\mh{V}_2$, with corresponding RT surface denoted $d$. 
Here, $u$ and $d$ are uniquely defined because in pure AdS$_3$, each spacelike boundary interval has a unique homologous extremal surface (and $\m{N}$ is an open portion of pure AdS$_3$).
In particular, the $d$ entanglement wedge is simply the union of the $\mh{V}_1$ causal wedge and the $\mh{V}_2$ causal wedge.
Secondly, the assumption that the boundary state is pure implies $\gamma_{{\mh R}_1}=\gamma_{{\mh R}_2}$ by virtue of ${\mh R}_1$ and ${\mh R}_2$ being causal complements \cite{Hubeny:2012wa}.
We remind the reader that purity is not intrinsic to $\m{N}$; rather, purity requires constraints on the global nature of the bulk geometry, such as the existence or nonexistence of horizons and singularities.

We now argue that $u<d$ implies holographic scattering, as follows:
As in eq.~\eqref{eq:keystone} above, purity of the state implies
\begin{equation}
    J^-({\cal R}_1)\cap J^-({\cal R}_2) = J^-(\gamma_{{\mh R}_1})\,.
    \label{eq:keystone2}
\end{equation}
Hence, the bulk scattering region \eqref{eq:scatregion} is
\begin{equation}
    J_{12\to12}
    =
    J^+(\m{V}_1) \cap J^+(\m{V}_2) \cap J^-(\gamma_{\mh{R}_1})\supset
    J^+(c_1) \cap J^+(c_2) \cap J^-(\gamma_{\mh{R}_1})
    \,,
    \label{eq:coness}
\end{equation}
where in the second step we are using that the causal wedge of $\mh{V}_i$ is contained in the entanglement wedge of $\mh{V}_i$, for $i=1,2$.
The scattering region in the final expression does not actually depend on large-scale features of ${\cal R}_1$ and ${\cal R}_2$ -- it depends only on the relation of $c_1$ and $c_2$ to $\gamma_{\mh{R}_1}$.
Indeed, it lies entirely within the $u$ entanglement wedge, and hence by assumption is contained in $\m{N}$ --
see figure \ref{fig:inN} and caption.

\begin{figure}
    \centering
    \includegraphics[width=0.6\linewidth]{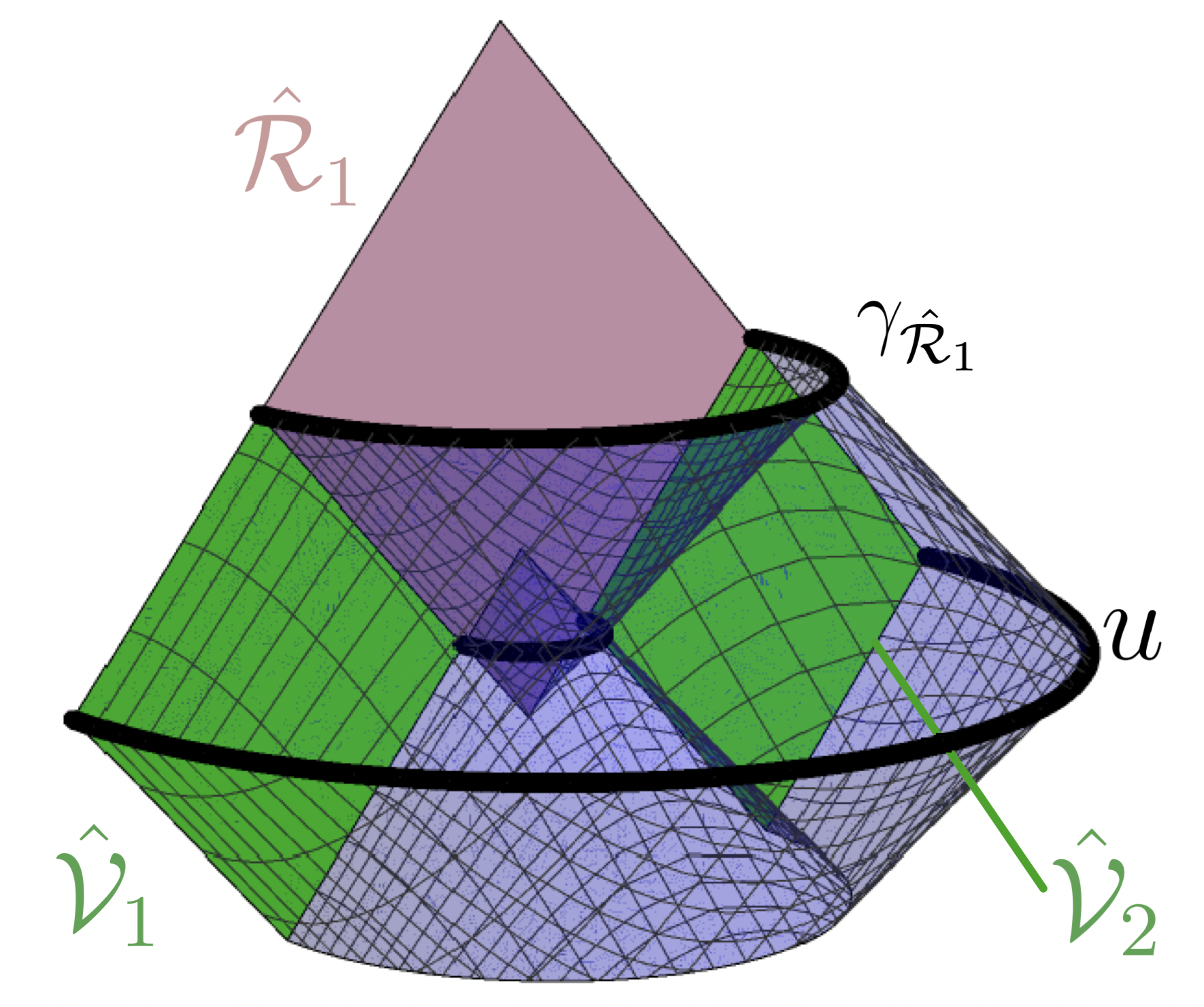}
    \caption{The $u$ entanglement wedge is the codimension-0 bulk subregion bounded by the purple lightsheets. 
    Its topmost ridge is an RT candidate for $\mh{R}_1$, and by assumption, the minimal RT candidate.
    The $u$ entanglement wedge contains the scattering region described by the final expression in eq.~\eqref{eq:coness}.
    To see why, first note that the two future-most purple lightsheets coincide with $\partial J^-(\gamma_{\mh{R}_1})$. 
    Then, note the scattering region does not extend below the past-most purple lightsheets because the scattering region is cut off by $J^+(c_1) \cap J^+(c_2)$.
    }
    \label{fig:inN}
\end{figure}

Since the scattering problem just described, together with the $u$ and $d$ surfaces, lies entirely within $\m{N}$, and since we have already shown the CWT has a converse in pure AdS$_3$, we conclude the $u<d$ condition implies holographic scattering here.

We emphasize that even when scattering occurs, the $u$ RT candidate may not be the true RT surface for $\mh{V}_1\cup \mh{V}_2$.
Namely, it is possible that there exists another RT candidate which is not contained within $\m{N}$ and yet has smaller area than $u$.
Hence, the converse of the CWT in vacuum geometries sometimes may consist of a constraint on non-minimal extremal surfaces.

This result generalizes and clarifies the conical defect and BTZ examples considered in \cite{Caminiti:2024ctd}.
For instance, \cite{Caminiti:2024ctd} found that $u<d$ implies scattering in heavy conical defect geometries when ${\mh V}_1$ and ${\mh V}_2$ lie in a constant time-slice and are taken to be of equal size (\eg see figure \ref{fig:optimal} below).
Our result reveals $u<d$ implies scattering for any choice of spacelike separated ${\mh V}_1$ and ${\mh V}_2$, as a simple consequence of the local AdS$_3$ structure of the geometry.

We emphasize that the general arguments in this section apply well beyond the static BTZ and conical defect solutions studied in \cite{Caminiti:2024ctd}; other nontrivial vacuum solutions where our results would apply include, \eg spinning defects (\eg \cite{Miskovic:2009uz, Compere:2018aar, Edelstein:2011vu, Briceno:2021dpi, Li:2024rma}), spinning BTZ black holes \cite{Banados:1992wn,Banados:1992gq}, and orbiting defects and black holes \cite{Balasubramanian:1999zv}. 
Again, in these geometries, the scattering setup extends beyond that considered in \cite{Caminiti:2024ctd}; we can choose the input regions, $\mh{V}_1$ and $\mh{V}_2$, to be any spacelike separated causal diamonds on the boundary cylinder while the output regions, $\mh{R}_1$ and $\mh{R}_2$ would be the causal diamonds extending between the future tips of these input regions. 
In appendix \ref{sec:spinning}, we examine the example of spinning defects in detail.

What are the limitations of this result?
Recall that from assumptions 1-3, the scattering region described by the final expression in eq.~\eqref{eq:coness} is necessarily contained in the $u$ entanglement wedge, and hence also in $\m{N}$.
As a consequence, the scattering region does not contain matter sources (which deform the local geometry).\footnote{Interestingly, since all of our three assumptions are always satisfied in the heavy defect geometry (see \cite{Caminiti:2024ctd}), we can conclude that scattering regions in the defect geometry never contain the defect itself.
Physically, this arises because probe systems approaching the defect at $r=0$ experience significant time delays, and large time delays enable boundary scattering to occur.} 
This is a clue that the presence of matter sources will disrupt the relationship between extremal surfaces and bulk scattering; we will elaborate on this theme in the following section.

\section{Converse with matter?}
\label{sec:starb}

The close relationship between holographic scattering and RT candidates discussed in section \ref{sec:genulessd}
depended on the background geometries being locally AdS$_3$.
Hence, we expect this relationship will be disrupted by the presence of matter, which deforms the local geometry.

To investigate this question, we construct a simple asymptotically AdS$_3$ geometry with nontrivial matter content: a background containing a spherically symmetric shell of matter. After briefly describing the shell geometry in section \ref{sec:stargeom}, we identify the RT surfaces in this background in section \ref{sec:rtstar}, and in section \ref{sec:simpleEx}, we identify a simple example where the matter of the shell indeed disrupts the relationship between $u<d$ and holographic scattering. Further details are discussed in appendix \ref{sec:deetail}.

\subsection{Shell geometry}
\label{sec:stargeom}

The metric of the shell geometry reads
\begin{equation}
ds^2=\left\{\begin{array}{ll}
       -f_{i}(r)\,d\tlt^2+\frac{dr^2}{f_{i}(r)}+r^2d\phi^2& \qquad\text{for}\ \  0\le r\le R\,,\\
        -f_{e}(r)\,dt^2+\frac{dr^2}{f_{e}(r)}+r^2d\phi^2& \qquad\text{for}\ \  r\ge R\,,
        \end{array}\right.
\label{eq:metrics}
\end{equation}
where
\begin{equation}
f_{i}(r)=r^2+1 \qquad{\rm and}\qquad
 f_{e}(r)=r^2-M \,,
\label{eq:blacken}
\end{equation}
with  $M \in(-1, 0)$ and $\phi\sim\phi+2\pi$. Implicitly, we have introduced an infinitely thin shell of matter at $r=R$. In the interior region ($0\le r\le R$), this metric corresponds to global AdS$_3$. 
In the exterior ($r\ge R$), the metric corresponds to a conical defect background with mass parameter $M$.\footnote{Note that we are only choosing $M<0$ for simplicity. In principle one could set $M>0$, as long as one takes care to ensure $R$ is outside of the BTZ horizon:  $R>\sqrt{M}$. One might also consider geometries where both the interior and exterior have independent mass parameters, $M_i$ and $M_e$. In this case, the null energy condition would simply require $M_i<M_e$.}

The induced metric on the shell is
\begin{equation}
d\ts^2 = -f_{i}(R)\,d\tlt^2+R^2d\phi^2=-f_{e}(R)\,dt^2+R^2d\phi^2\,.
\label{eq:shellm}
\end{equation}
Therefore, to match the two geometries, we relate the exterior and interior times as
\begin{equation}
\tlt = \sqrt{\frac{f_{e}(R)}{f_{i}(R)}}\,t\,.
\label{relate}
\end{equation}
Note that
\begin{equation}
\frac{f_{e}(R)}{f_{i}(R)}=\frac{R^2+|M|}{R^2+1}<1\,,
\label{ratio}
\end{equation}
and so clocks on the interior are running more slowly than on the exterior because of the redshift introduced by the massive shell.
In appendix \ref{sec:deetail}, we demonstrate that with this constraint, the matter sourcing the geometry satisfies the null energy condition, a requirement of theorem \ref{thm:cwt}.
We also note that this geometry does not contain horizons, defects, or singularities, so it obeys the pure state condition of section \ref{sec:genulessd}.

One may ask why the resulting geometry is static if it contains a shell of matter which seemingly would want to collapse in on itself.
Indeed, in appendix \ref{sec:deetail}, we show that the shell has positive tension which makes it want to contract. The resolution lies in recognizing that the shell functions as a domain wall separating a bubble of true vacuum with $M=-1$ from an exterior region with higher energy, $M>-1$. Since the interior has lower energy, the bubble wishes to expand as in \cite{Coleman:1980aw}. Ultimately, the geometry remains static because the positive tension of the shell precisely balances the outward pressure exerted by the bubble.

We now study entanglement wedges and holographic scattering in this shell geometry to understand how the CWT is realized in this background. To simplify the discussion, we set $M=-\mu^2$ so the exterior blackening factor reads $f_{e}(r)=r^2+\mu^2$,  with $\mu \in(0,1)$.

\subsection{RT surfaces}
\label{sec:rtstar}

In a constant-$t$ slice of the shell geometry, spacelike geodesics of angular momentum $\ell$ are characterized by 
\begin{equation}
    r(\phi) = 
    \begin{cases}
        \[\frac{1}{\ell^2}\cos^2({\mu}(\phi-\phi_0))-\frac{1}{\mu^2}\sin^2({\mu}(\phi-\phi_0))\]^{-1/2}\,, \qquad &r > R\,, \\
       \[\frac{1}{\ell^2}\cos^2 \phi-\sin^2\phi\]^{-1/2} \,, \qquad &r < R\,,
    \end{cases}
    \label{eq:shellsimple}
\end{equation}
where $\phi_0$ is a particular function of $\mu$, $\ell$, and $R$:
\begin{equation}
    \phi_0 = \tan^{-1}\left(\frac{1}{\ell}\sqrt{\frac{R^2-\ell^2}{R^2+1}}\right)
    - \frac{1}{{\mu}}\tan^{-1}\left(\frac{{\mu}}{\ell}\sqrt{\frac{R^2-\ell^2}{R^2+\mu^2}}\right)\,,
\end{equation}
as derived in appendix \ref{sec:deetail}.
Here and throughout, we take $\ell>0$, without loss of generality.

From eq.~\eqref{eq:shellsimple}, we find a general relation $r_{\min}=\ell$ between the minimum radius attained by the geodesic and the geodesic's angular momentum. 
Note, if $r_{\min}>R$, then the geodesic is entirely described by the first line of eq.~\eqref{eq:shellsimple}, and one may freely set $\phi_0=0$ by a shift of the angular coordinate. 

\begin{figure}[htbp]
    \centering
    \includegraphics[width=0.24\linewidth]{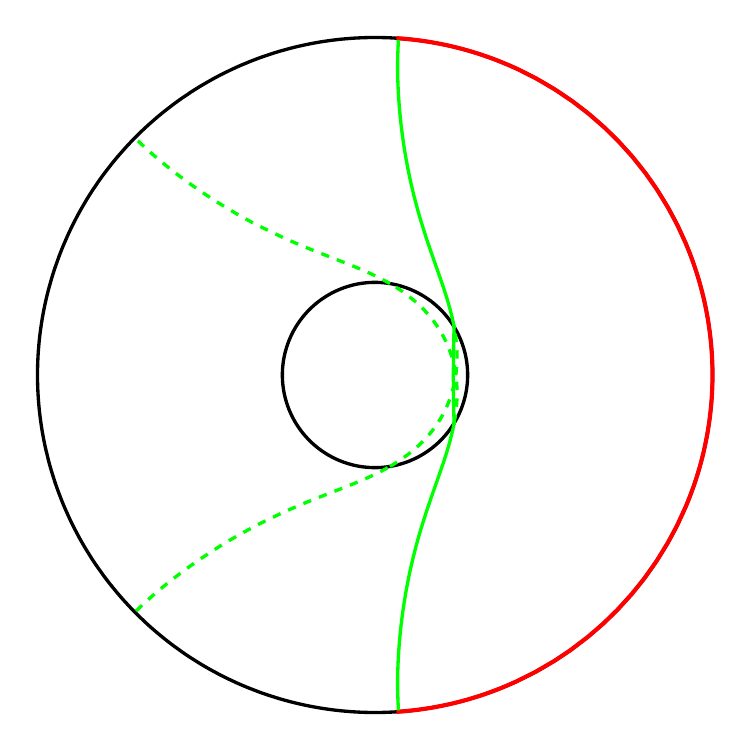}
    \hfill \includegraphics[width=0.24\linewidth]{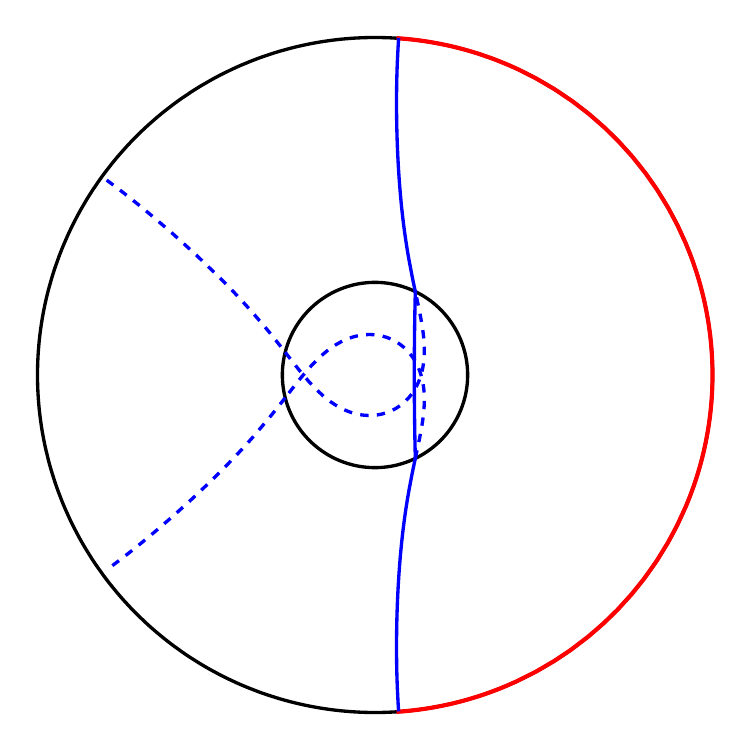}
    \hfill \includegraphics[width=0.24\linewidth]{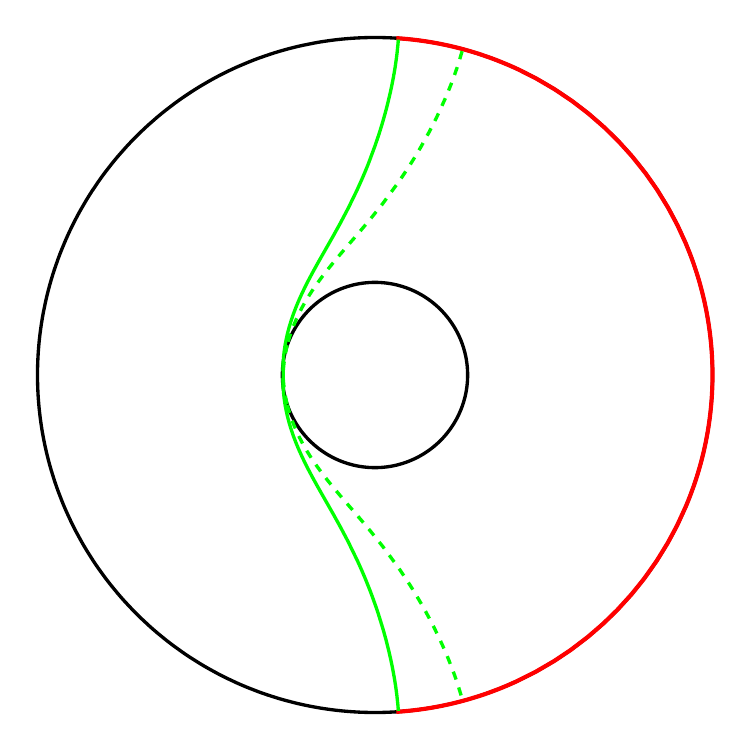}
    \includegraphics[width=0.24\linewidth]{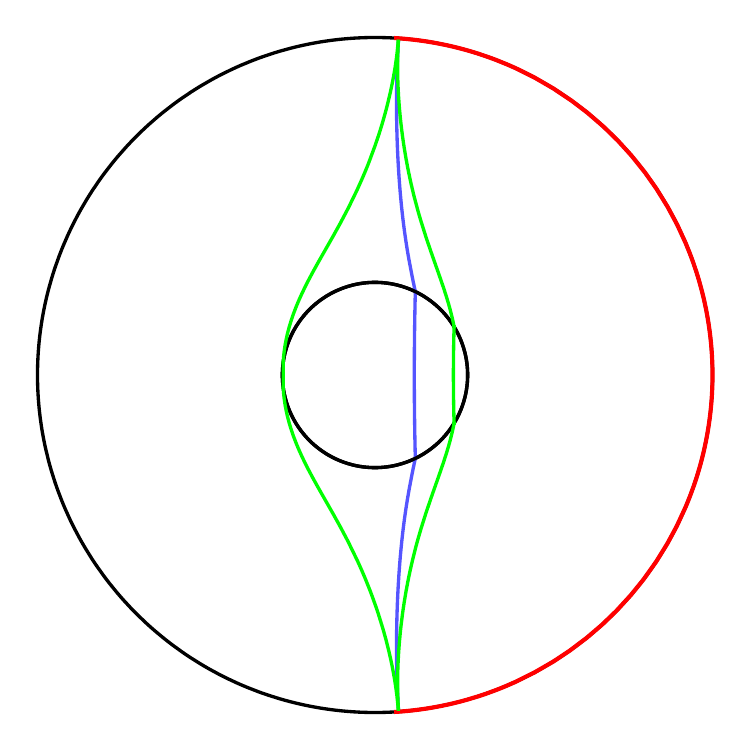}
    \caption{Spacelike geodesics in the shell geometry with $R=0.46$, $\mu^2 = 0.2$.
    The solid lines represent actual shell-crossing geodesics, while the dashed lines represent the continuation of the exterior solution; that is, how the exterior solution would behave if the shell was replaced by a defect of equal mass.
    In the final figure, we have shown all three shell-crossing geodesics in the same plot.
    They have the same opening angle $\Delta \phi\approx 0.96 \pi$.}
    \label{fig:cool}
\end{figure}

We plot some of these geodesics in figure \ref{fig:cool}.
In these plots, the radial coordinate, call it $\rho$, is related to the true radial coordinate of eq.~\eqref{eq:metrics} by $\rho=\arctan(r)$, resulting in a compactification of the geometry.
We see that while the geodesics are continuous when entering the shell, their slope is not; the curve refracts at the shell, much like a light ray passing from air to water.

To use this result for the CWT, we must be able to relate $r_{\min}=\ell$ to the angular extent of the boundary interval, $\Delta \phi$.
For $r_{\min}>R$, one straightforwardly obtains
\begin{align}
    \Delta\phi(r_{\min})=\frac{2}{{\mu}} \tan^{-1}\frac{{\mu }}{r_{\min}}
    \qquad\iff\qquad
    r_{\min}(\Delta\phi)={\mu }\cot \frac{{\mu}\Delta\phi}{2}\,,
    \label{eq:arccot}
\end{align}
while for $r_{\min}<R$, we obtain (see appendix \ref{sec:deetail})
\begin{align}
    \Delta\phi(r_{\min})= 
    \frac{2}{{\mu }}\tan^{-1}\!\left( \frac{\sqrt{R^2+\mu^2}-\sqrt{R^2-r_{\min}^2}}{\frac{r_{\min}}{\mu}\sqrt{R^2+\mu^2}+\frac{\mu}{r_{\min}}\sqrt{R^2-r_{\min}^2}  }\right)
    +
    2\tan^{-1}\!\left(\frac{1}{r_{\min}} \sqrt{\frac{R^2-r_{\min}^2}{R^2+1}}\right)
    \label{eq:cantinvert}
\end{align}
which of course in the light-shell regime ($\mu\approx 1$) and/or the large-shell regime ($R\to \infty$) is well-approximated by eq.~\eqref{eq:arccot} with $\mu\equiv1$ (namely, $\Delta\phi=2 \tan^{-1}(1/r_{\min})$).\footnote{The $R\to\infty$ limit of $\Delta \phi(r_{\min})$ is straightforward because the entire first term in eq.~\eqref{eq:cantinvert} drops out. The $\mu\to 1$ limit is a bit more complicated; instead of studying $\Delta \phi(r_{\min})$, we recommend expanding the expression $\tan(\Delta \phi(r_{\min})/2)$ and using trigonometric identities to obtain $1/r_{\min}$.}

While the $\Delta \phi$ of eq.~\eqref{eq:arccot} is monotonic in $r_{\min}$, eq.~\eqref{eq:cantinvert} is not; in some cases, there are two choices of $r_{\min}$ for the same $\Delta \phi$.
This multi-valuedness corresponds to the possibility that the same boundary interval has multiple shell-crossing RT candidates.
In general, the freedom to relabel the boundary interval via $\Delta \phi\mapsto 2\pi-\Delta \phi$ introduces additional multi-valuedness into $r_{\min}$.
That is, a single boundary angle $\Delta \phi$ can be associated not only to geodesics with $r_{\min}=r_{\min}(\Delta \phi)$, but also geodesics with $r_{\min}=r_{\min}(2\pi-\Delta \phi)$.
To capture this, we formally extend $r_{\min}$ to negative values and define
\begin{equation}
    \Delta \phi= \begin{cases}
        \Delta \phi(r_{\min})\,,\qquad &r_{\min}>0\,,\\
        2\pi-\Delta \phi(|r_{\min}|)\,,\qquad &r_{\min}<0\,.\\
    \end{cases}
    \label{eq:formal}
\end{equation}
For example, in figure \ref{fig:cool}, the geodesics shown in the first two plots have $r_{\min}>0$, while the geodesic shown in the third plot has $r_{\min}<0$.

\begin{figure}
    \centering
    \includegraphics[width=.328\linewidth]{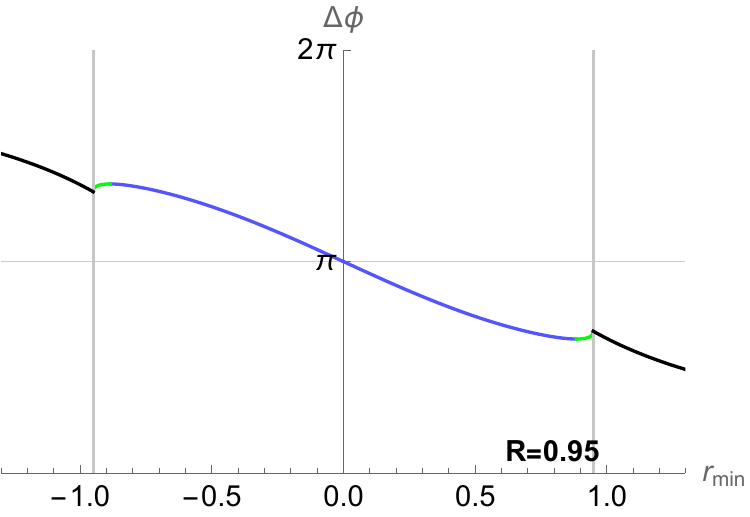}
    \hfill
    \includegraphics[width=.328\linewidth]{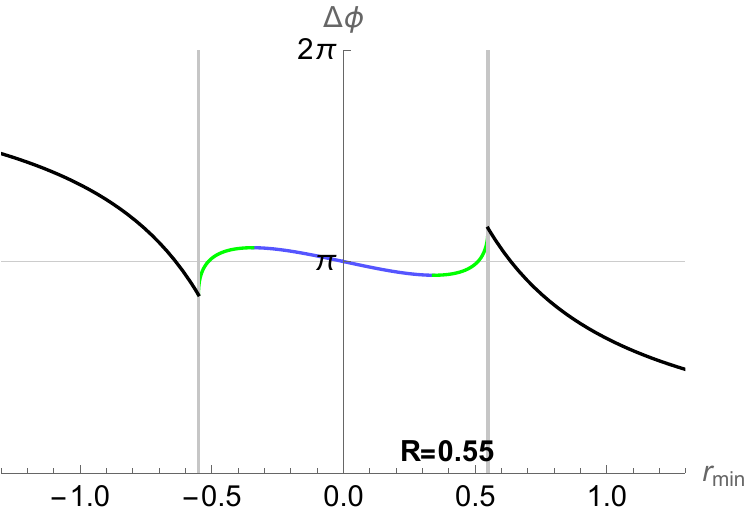}
    \hfill
    \includegraphics[width=.328\linewidth]{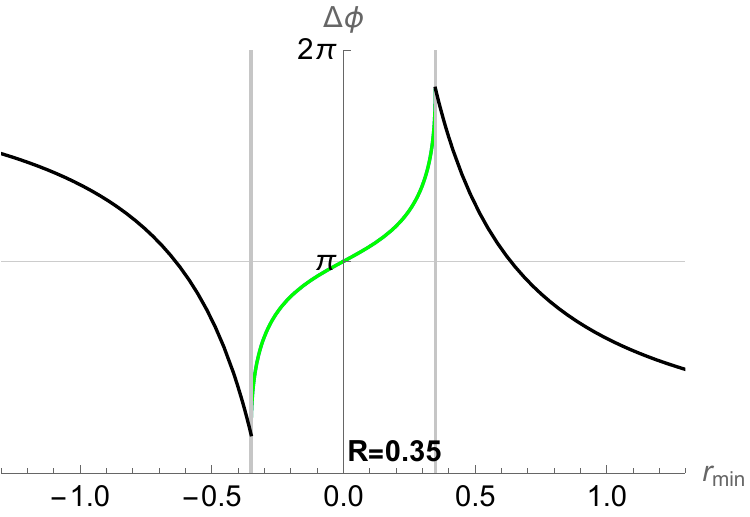}
    \caption{Geodesic opening angle $\Delta \phi$ as a function of $r_{\min}$.
    Each of the three plots is associated with a fixed choice of $R$, and in all cases the total mass of the shell is fixed to be very large ($\mu \approx 0$).
    Blue curves correspond to shell-crossing geodesics for which $\Delta \phi(r_{\min})$ is decreasing, while green curves correspond to shell-crossing geodesics for which $\Delta \phi(r_{\min})$ is increasing. Black curves correspond to geodesics which are entirely outside of the shell.
    Negative values of $r_{\min}$ are introduced in order to capture the $\Delta \phi\mapsto 2\pi-\Delta \phi$ symmetry, as in eq.~\eqref{eq:formal}.
    }
    \label{fig:nike}
\end{figure}

In figure \ref{fig:nike}, we plot the geodesic opening angle as in eq.~\eqref{eq:formal} for heavy shell geometries ($\mu \approx 0$) with various choices of $R$.
From these plots, we see that there exist either one or two shell-avoiding geodesics (black) for the same boundary angle; to see how many candidates exist for a given $\Delta \phi$, one draws a horizontal line in figure \ref{fig:nike} and counts how many times it is crossed by the curve of interest.
Additionally, there can exist either one, two, or three shell-crossing geodesics for the same boundary angle.
The shell-crossing solutions may be distinguished by whether $\Delta\phi(r_{\min})$ is a locally decreasing function (blue solution) or locally increasing (green solutions). 
Compare to figure \ref{fig:cool}, which shares the same coloring scheme.

To study RT surfaces, we must compute the lengths of these various solutions as a function of $\Delta \phi$.
For a shell-avoiding geodesic, we have (see appendix \ref{sec:deetail})
\begin{align}
    \text{Length}_{ext}(r_{\min}) = 2 \log \frac{2}{\epsilon  \sqrt{r_{\min}^2+\mu^2 }}\,,
    \label{eq:exteriorLen}
\end{align}
where we introduced a UV radial cutoff at $r=\frac{1}{\epsilon}$. 
For a shell-avoiding geodesic, $r_{\min}(\Delta \phi)$ is provided in eq.~\eqref{eq:arccot},
though there can be a second solution with $r_{\min}=r_{\min}(2\pi-\Delta \phi)$.
Minimizing over the two candidates gives the general formula
\begin{equation}
    \text{Length}_{ext}(\Delta \phi)=2 \log \left(\frac{2}{\epsilon \,{\mu}}\sin\left( \frac{{\mu}}{2}\min(\Delta \phi,2\pi-\Delta \phi)\right)\right)\,.
\end{equation}

On the other hand, the length of a shell-crossing geodesic is
\begin{align}
    \text{Length}_{in}(r_{\min}) = 2 \log \left(\frac{2}{\epsilon \sqrt{r_{\min}^2+1} }\,\frac{ \sqrt{R^2+1}+\sqrt{R^2-r_{\min}^2}}{\sqrt{R^2+\mu^2}+\sqrt{R^2-r_{\min}^2}}\right)\,,
    \label{eq:interiorLen}
\end{align}
and as discussed above, there can be as many as three\footnote{Here and throughout, we do not consider geodesics which loop many times around the defect.} shell-crossing solutions (\ie three choices of $r_{\min}$) for the same $\Delta \phi$, which can be found by numerically inverting eq.~\eqref{eq:cantinvert}.\footnote{We comment that in the large-shell ($R\to\infty$) and/or light-shell ($\mu\to0$) regime, the comment below eq.~\eqref{eq:cantinvert} gives  an approximation to $\text{Length}_{in}(\Delta \phi)$ via the relation $r_{\min} \approx \cot\frac{\Delta \phi}{2}$.\label{foot:regime}
}
In appendix \ref{sec:deetail}, we find that the only shell-crossing solution which has a chance of being the minimal RT surface is the candidate for which $\Delta \phi(r_{\min})$ is a locally decreasing function (\eg the blue curves in figures \ref{fig:cool} and \ref{fig:nike}).
Going forward, we will only be interested in this solution among all the shell-crossing solutions.

We now introduce two pieces of notation which will be useful in the following subsection.
First, we define $\Delta \phi^*=\Delta \phi^*(\mu, R)$ as the boundary angle for which
\begin{align}
    \text{Length}_{in}(\Delta \phi^*)=\text{Length}_{ext}(\Delta \phi^*)\,.
    \label{eq:LinLout}
\end{align}
By the symmetry of the problem, whenever $\Delta \phi^*$ exists, we will have $\Delta \phi^*<\pi$, and the RT surface will be shell-crossing for $\Delta\phi\in(\Delta\phi^*,2\pi-\Delta\phi^*)$, and shell-avoiding only outside this range.
We remark that, as a consequence of the analysis in appendix \ref{sec:deetail}, $r_{\min}$ behaves discontinuously as $\Delta \phi$ crosses $\Delta \phi^*$, meaning the RT surface undergoes a ``jump'' transition such that the RT surface never becomes tangent to the shell.
By abuse of notation, when $\Delta \phi^*$ does not exist we sometimes say $\Delta \phi^*>\pi$.

We close this section on RT surfaces by mentioning an important transition which can occur as $\Delta \phi$ continues to increase beyond $\Delta \phi^*$.
Observe that the shell-avoiding geodesic described by eq.~\eqref{eq:arccot} reaches deeper and deeper into the geometry as $\Delta \phi$ increases.
If $\Delta \phi$ reaches
\begin{equation}
    \Delta\phi_s:=\frac{2}{{\mu}}\, \tan^{-1}\frac{{\mu }}{R}\,,
\end{equation}
then the geodesic reaches all the way to $r_{\min}=R$. 
So, beyond $\Delta \phi=\Delta \phi_s$, the shell-avoiding geodesic ceases to exist.\footnote{Note, numerical analyses are consistent with the general relation $\Delta \phi_s > \Delta \phi^{*}$.}
These transitions will be useful in discussing the CWT in the following subsection.
To summarize: the shell-avoiding geodesic ceases to be minimal when $\Delta \phi$ crosses $\Delta \phi^*$, and ceases to exist altogether when $\Delta \phi$ crosses $\Delta \phi_s$.

\subsection{\texorpdfstring{A simple counterexample to $u<d$}{A simple counterexample to u<d}}
\label{sec:simpleEx}

Our understanding of the $u<d$ converse of the CWT in section \ref{sec:genulessd} relied on the $u$ surface belonging, by definition, to an open portion $\m{N}$ of pure AdS$_3$.
We would now like to ask if the converse holds beyond that setting, taking advantage of our new understanding of RT surfaces in the shell geometry.

In the shell geometry, we can define several RT candidates for $\mh{V}_1 \cup \mh{V}_2$.
First, we have candidates which only probe the exterior, conical defect portion of the geometry, which can be classified into $u$, $o$, and $d$ surfaces as in figure \ref{fig:dcircs}.
Then, we have the $i$ candidates, which we define as the connected candidates  with precisely one shell-crossing geodesic.
We will focus on the $u$, $o$, $d$, and $i$ candidates for now, but note that we can also have a disconnected candidate $d_2$ where both geodesics cross the shell, as well as a connected candidate $i_2$ where both geodesics cross the shell (see appendix \ref{sec:deetail} for details).
Crucially, unlike the $u$ candidate, the $i$ candidate is not contained within any open region $\m{N}$ of the shell geometry with $\m{N}$ equivalent to a portion of pure AdS$_3$.
We now find that the condition $i<d$ does not suffice to imply holographic scattering, even when the $u$ surface ceases to exist, indicating that the CWT converse in terms of nonminimal surfaces does not hold beyond the setting of locally AdS$_3$ geometries.

\begin{figure}
    \centering
    \includegraphics[width=0.5\linewidth]{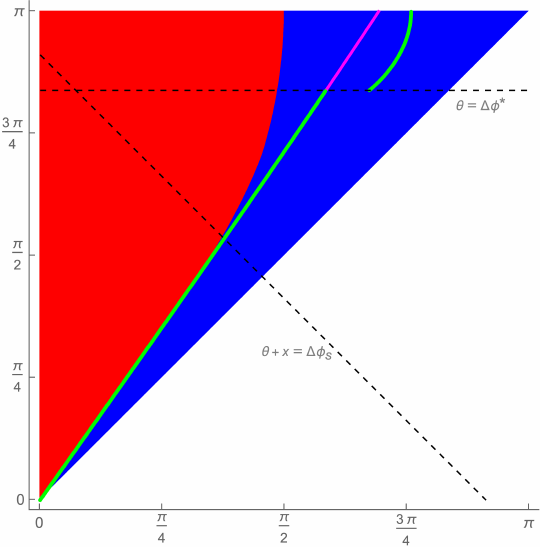}
    \caption{For $\mu^2 = 0.1$ and $R=0.65$, we have $\Delta \phi_s\approx2.86$ and $\Delta \phi^*\approx 2.63$, so one can find pairs $(\theta,x)$ satisfying both eq.~\eqref{eq:scatout} and \eqref{eq:black2} in the main text.
    To see this from the plot, note eq.~\eqref{eq:black2} is obeyed above the dashed line $\theta+x = \Delta \phi_s$, meaning that the $u$ surface does not exist.
    Meanwhile, eq.~\eqref{eq:scatout} is obeyed below the dashed line $\theta=\Delta\phi^*$, meaning that scattering happens outside of the shell.
    Now, outside of the shell, the scattering threshold (green) coincides with the naive extension (magenta) of the $u=d$ equation, eq. \eqref{eq:ulessdconic}.
    So, in the region between the two dashed lines, despite the $u$ curve not existing, the scattering threshold is described by the naive extension of the $u=d$ curve.
    For reference, we have also indicated the disconnected phase of $\m{E}(\mh{V}_1\cup\mh{V}_2)$ in red and the connected phase in blue.}
    \label{fig:counterex}
\end{figure}

To see this, imagine fixing $\mu<1/2$,\footnote{From \cite{Caminiti:2024ctd}, this condition ensures that without the shell, assumptions 1-3 of section \ref{sec:genulessd} are obeyed.} and slowly increasing the size $R$ of the shell from $R=0$.
When $R$ is small, our three assumptions in section \ref{sec:genulessd} are obeyed, and therefore section \ref{sec:genulessd} tells us the scattering region lies somewhere inside the $u$ entanglement wedge.
Now, as the shell increases in size, it may disrupt the $u$ surface without affecting the scattering region.
In particular, letting $\mh{V}_1$ and $\mh{V}_2$ be associated with intervals of equal size $x$ and separation $\theta$ on a constant-time slice of the cylinder (see figure \ref{fig:optimal}), we may end up in a scenario where scattering is at the conical defect threshold \cite{Caminiti:2024ctd}, \ie
\begin{equation}
   \cos^2 \left({\mu}\,\theta/2\right)=\cos \left({\mu}\,x\right)\,,
    \label{eq:ulessdconic}
\end{equation}
which is a naive extension of the $u=d$ formula, and yet the shell is so big that the $u$ surface simply does not exist.
To put this into equations, recall from the previous section that
\begin{equation}
    \theta < \Delta \phi^*
    \label{eq:scatout}
\end{equation}
ensures scattering happens outside the shell, meaning the threshold for scattering is the defect result eq.~\eqref{eq:ulessdconic}.\footnote{To see this, recall that the scattering region lies very close to $\gamma_{\mh{R}_1}$ near the scattering threshold.
Eq.~\eqref{eq:scatout} guarantees that $\gamma_{\mh{R}_1}$ lies outside of the shell, and so it also implies that near the scattering threshold, the holographic scattering process occurs entirely outside of the shell.} 
Meanwhile, the condition
\begin{equation}
    \theta + x > \Delta \phi_s
    \label{eq:black2}
\end{equation}
ensures the $u$ surface does not exist.
The curves corresponding to eqs.~\eqref{eq:scatout} and \eqref{eq:black2} are shown in black in figure \ref{fig:counterex}, while the scattering threshold \eqref{eq:ulessdconic} is shown in purple.

From figure \ref{fig:counterex}, we see that there is a nontrivial regime of parameter space 
where scattering is governed by eq.~\eqref{eq:ulessdconic}, and yet there is no competition among nonminimal surfaces which 
matches this equation, because the $u$ surface no longer exists.
One might try to compare eq.~\eqref{eq:ulessdconic} with the $i=d$ curve instead, but numerical analyses\footnote{See, for example, figure \ref{fig:flight} in appendix \ref{sec:deetail}.
There, the black curves represent shell-avoiding geodesics; in the present context, we focus on the black curve with positive slope and think of it as representing the portion of the $u$ geodesic with opening angle $\Delta \phi=\theta+x$.
(The black curve with negative slope represents the $o$ candidate, while the blue curve represents the $i$ candidate.)
One can verify that the naive extension of this black curve beyond $\Delta \phi=\Delta \phi_s$ has strictly larger length than all the other curves.\label{foot:physics}} 
indicate that with $\theta +x >\Delta \phi_s$, then if $i$ exists we always have $i$ strictly less than (the naive extension of) $u$.
In this regime, then, we at most have the statement that
\begin{equation}
    \textrm{holographic scattering} \implies i<d\,.
    \label{eq:ilessd}
\end{equation}
and, as claimed, $i<d$ does not suffice to imply scattering, even when the $u$ surface does not exist.
We conclude that there is no equivalence between holographic scattering and a relation among nonminimal RT surfaces in the shell geometry.

To close with a brief aside, we comment that eq.~\eqref{eq:ilessd} is nevertheless a nontrivial result.\footnote{We thank the authors of \cite{ap,ap2} for bringing this point to our attention.}
It is nontrivial because sometimes $o$ is the minimal candidate, and so the CWT need only imply $o<d$ (not $i<d$).
If we increase $\theta+x$, we can arrive at a situation where neither $i$ nor $u$ exists, in which case the reasoning in footnote \ref{foot:physics} ensures
\begin{equation}
    \textrm{holographic scattering} \implies o<d\,.
\end{equation}
Inspired by \cite{ap,ap2}, one can summarize these results by writing $u_{simple}<d$, where we define $u_{simple}$ as the unique candidate in $\{o,u,i\}$
which is contained in the simple wedge \cite{Engelhardt:2014gca,Engelhardt:2021mue} of the boundary interval with opening angle $\theta+x$.
Notably, while until now we have focused on the case $\theta<\Delta \phi^*$, we analyze phase diagrams and holographic scattering in the shell geometry in appendix \ref{sec:deetail}, and we find that for $\theta>\Delta \phi^*$, the $u_{simple}<d$ result appears to persist on a case-by-case basis.
See further discussions on this point in section \ref{sec:discuss}.

\section{\texorpdfstring{No converse for $n$-to-$n$ scattering}{No converse for n-to-n scattering}}
\label{sec:3to3}

In the previous sections, a key result was that the 2-to-2 connected wedge theorem has a converse in pure AdS$_3$.
In general, one can ask whether this is true more generally, \ie for the $n$-to-$n$ connected wedge theorem, with $n>2$.
In this section, we first examine the specific example of the 3-to-3 CWT in pure AdS$_3$ and then generalize this analysis to $n$-to-$n$ scattering.
Finally, we comment on an extension of our result to
the conical defect geometry. In all of these setups, we find counterexamples to the CWT converse.

\subsection{\texorpdfstring{The $n$-to-$n$ CWT}{The n-to-n CWT}}

Let us begin by stating the $n$-to-$n$ connected wedge theorem (adapted from \cite{May:2022clu}).

\begin{theorem}[$n$-to-$n$ CWT] \label{thm:n-to-n-first-statement}
Let $\m{M}$ be an asymptotically AdS$_{3},$ hyperbolic spacetime satisfying the null energy condition, with at least one AdS$_{3}$ boundary on which $c_1, \dots, c_n$, $r_1, \dots, r_n$ are specified,  such that the sets 
    \begin{align}
    \mh{V}_j & \equiv \Jh^+(c_j)\cap \hat\J^-(r_1) \cap \cdots \cap \hat\J^-(r_n) \,,
    \qquad
    \mh{W}_j \equiv \Jh^+(c_1)\cap \cdots \hat\J^+(c_n) \cap  \hat\J^-(r_j)\,,
    \label{eq:allforone}
\end{align}
are nonempty. At the same time, the following boundary scattering regions vanish
    \begin{eqnarray}
        \hat{J}_{j,k \to 1, \cdots,n} &=& \Jh^+(c_j)\cap \Jh^+(c_k)\cap \hat\J^-(r_1) \cap \cdots \cap \hat\J^-(r_n)=\emptyset\,,
        \nonumber\\
        \hat{J}_{1, \cdots, n \rightarrow j, k} &=& \Jh^+(c_1)\cap \cdots \cap\hat\J^+(c_n) \cap  \hat\J^-(r_j)\cap\Jh^-(r_k)=\emptyset\,.
        \label{eq:no2toall}
    \end{eqnarray}

Defining the analogous, bulk scattering regions ${J}_{j,k \to 1, \cdots,n}$, one can construct the $2$-to-all causal graph as
\begin{equation}
    \Gamma_{2\to \textrm{all}}
    =
    \{(j,k): J_{j,k \to 1,...,n} \ne\emptyset\}\,,
    \label{eq:causalgraf}
\end{equation}
That is, $\Gamma_{2\to \textrm{all}}$
consists of vertices $\{1,...,n\}$ where vertices $j$ and $k$ are connected if bulk scattering from $\m{V}_j,\m{V}_k$ to $\m{W}_1,..., \m{W}_n$ is possible.
    
    Now if the $2$-to-all causal graph $\Gamma_{2\rightarrow\text{all}}$ is connected, and if the HRRT surface of $\mh{V}_1 \cup \cdots \cup \mh{V}_n$ can be found using the maximin formula, then the entanglement wedge  $\m{E}(\mh{V}_1 \cup \dots \cup \mh{V}_n)$ is connected.
\end{theorem}

\noindent
For example, specializing to $n=3$,
the sufficient condition for $\m{E}_{\mh{V}_1 \cup \mh{V}_2 \cup \mh{V}_3}$ to be fully connected is that at least two of the following bulk scattering regions
\begin{equation}
\begin{aligned}
(a) \; J_{12\to123}\,,\qquad
(b) \; J_{23\to123}\,,\qquad
(c) \; J_{13\to123}\,,
\end{aligned}
\label{eq:abc}
\end{equation}
are nonempty; this ensures the $\Gamma_{2\to \textrm{all}}$ graph is connected.

Physically, the $n$-to-$n$ CWT says that a sufficiently large network of bulk causal connections requires strong correlations (\ie  mutual information of $\m{O}(1/G_N)$) across any bipartition of the regions.
Note that by assumption \eqref{eq:no2toall}, 2-to-all (as well as all-to-2) boundary scattering is forbidden, so the edges of $\Gamma_{2\to \textrm{all}}$ indeed represent bulk-only scattering processes. 
We also comment that the connectivity of $\Gamma_{2\to \textrm{all}}$ is weaker than demanding the existence of a bulk $n$-to-$n$ scattering region, since non-emptiness of the $n$-to-$n$ scattering region implies  each 2-to-all scattering region is nonempty (by virtue of being contained in the latter). 
So, $n$-to-$n$ scattering does guarantee a connected wedge, as one might expect from the 2-to-2 case.

In the following sections, we examine holographic scattering in the $n$-to-$n$ case and find counterexamples to the CWT converse.

\subsection{\texorpdfstring{3-to-3 in pure AdS$_3$}{3-to-3 in pure AdS3}}
\label{sec:3to3pure}

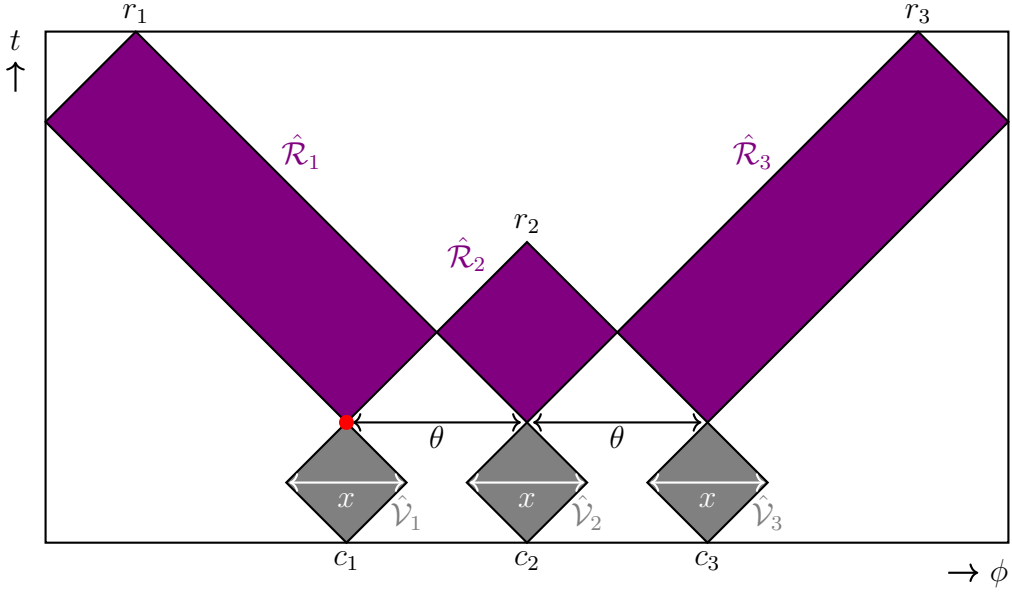
\begin{figure}
    \centering
    \begin{tikzpicture}[scale=0.8]
\draw[thick] (-8, 5) rectangle (8, -3.5);

\draw[->, thick] (-8.5, 4) -- (-8.5, 4.5) node[above] {$t $};
\draw[->, thick] (7,-4) -- (7.5,-4) node[right] {$\phi$};

\draw[fill=gray, thick]
  (-4,-2.5) -- (-3,-1.5) -- (-2, -2.5) -- (-3,-3.5) -- cycle;
  \node[white] at (-3,-2.8) {$x$};
  \draw[<->, thick, white] (-3.95,-2.5) -- (-2.05, -2.5);
  \node[gray] at (-2, -3) {$\mh{V}_1$};
  \node at (-3, -3.8) {$c_1$};
  
\draw[fill=gray, thick]
  (-1,-2.5) -- (0,-1.5) -- (1, -2.5) -- (0,-3.5) -- cycle;
  \node[white] at (0,-2.8) {$x$};
  \draw[<->, thick, white] (-0.95,-2.5) -- (0.95,-2.5);
  \node[gray] at (1, -3) {$\mh{V}_2$};
  \node at (0, -3.8) {$c_2$};

\draw[fill=gray, thick]
  (4,-2.5) -- (3,-1.5) -- (2, -2.5) -- (3,-3.5) -- cycle;
  \node[white] at (3,-2.8) {$x$};
  \draw[<->, thick, white] (3.95,-2.5) -- (2.05, -2.5);
  \node[gray] at (4, -3) {$\mh{V}_3$};
  \node at (3, -3.8) {$c_3$};

\draw[fill=violet, thick]
  (-1.5,0) -- (-3,-1.5) -- (-8, 3.5) -- (-6.5, 5) -- cycle;
\node[violet] at (-3.75, 3) {$\hat{\m{R}}_1$};
\node at (-6.5, 5.3) {$r_1$};

\draw[fill=violet, thick]
  (-1.5,0) -- (0,1.5) -- (1.5,0) -- (0,-1.5) -- cycle;
\node[violet] at (-1, 1.3) {$\mh{R}_2$};
\node at (0, 1.8) {$r_2$};

\draw[<->, black, thick] (-2.9,-1.5) -- (-0.1,-1.5);
\node[black] at (-1.5,-1.75) {$\theta$};
\draw[<->, black, thick] (2.9,-1.5) -- (0.1,-1.5);
\node[black] at (1.5,-1.75) {$\theta$};

\draw[fill=violet, thick]
  (1.5,0) -- (3,-1.5) -- (8, 3.5) -- (6.5, 5) -- cycle;
\node[violet] at (3.75, 3) {$\mh{R}_3$};
\node at (6.5, 5.3) {$r_3$};

\node[circle, fill=red, inner sep=2pt] at  (-3,-1.5) {};

\end{tikzpicture}
\caption{Example configuration of 3-to-3 scattering regions on the cylinder.
The left and right boundaries of this figure are identified.
We take the origin of our coordinate system as the red dot shown.
Note that here $\theta<2\pi/3$, so $t_{r_2}<t_{r_1}=t_{r_3}$.
Note also that unlike the 2-to-2 case of figure \ref{fig:optimal}, in 3-to-3 scattering we have that for each $i$, $\mh{R}_i$ is contained in the future of $\mh{V}_i$.}
\label{fig:setup3to3}
\end{figure}

In this section, we consider the 3-to-3 setup pictured in figure \ref{fig:setup3to3}.
Here, we have three input regions of equal width $x$ on an equal time slice of the cylinder, where $\mh{V}_1$ and $\mh{V}_3$ are arranged symmetrically about $\mh{V}_2$.
As in the 2-to-2 case of figure \ref{fig:optimal}, the parameter $\theta$ controls the separation between the midpoints of $\mh{V}_1$ and $\mh{V}_2$ (and hence also between the midpoints of $\mh{V}_2$ and $\mh{V}_3$), and we are maximizing the $\mh{R}_i$ subject to eq.~\eqref{eq:no2toall} and the particular choice of $\mh{V}_i$.
This again leads to a configuration where the $\mh{R}_i$ partition a Cauchy slice of the cylinder.

We comment that the ${\mh R}_i$ output regions shown in figure \ref{fig:setup3to3} are larger than the ${\mh W}_i$ output regions defined in eq.~\eqref{eq:allforone}.
One can check by the equivalence of causal and entanglement wedges in pure AdS$_3$ that $J^-({\mh W}_i)=J^-(r_i)=J^-({\mh R}_i)$, and hence we are free to perform the replacement ${\mh W}_i\to {\mh R}_i$ without modifying the scattering region of interest.
At a deeper level, we expect that it should be possible to upgrade the $n$-to-$n$ CWT as originally stated to accommodate larger output regions, since \eg upgrading ${\mh W}_i\to {\mh R}_i$ does not affect the focusing or homology arguments in the CWT proof.\footnote{We thank Alex May for discussion on this topic. We emphasize that the 2-to-2 CWT of \cite{May:2021nrl} (our theorem \ref{thm:cwt}), accommodates output regions larger than the $\mh{W}_i$; we used this already in figure \ref{fig:iffsetup}.}

Here we are considering holographic scattering in pure AdS$_3$ in global coordinates:
\begin{equation}
    \rd s^2 = -(r^2 +1) \rd t^2 + \frac{\rd r^2}{r^2+1} + r^2 \rd \phi^2\,.
    \label{eq:pureads0}
\end{equation}
Hence the boundary metric is simply $\rd s^2 = - \rd t^2 + \rd \phi^2$. For completeness, we set the origin of our boundary coordinates at the future tip of the input region $\mh{V}_1$, as indicated by the red dot in figure \ref{fig:setup3to3}, and we tabulate the positions of the input and output points:
\begin{eqnarray}
    c_1\ \ :&\ (t,\phi)&=(-x,\ 0)\,,
    \qquad\qquad
    r_1\ \ :\ (t,\phi)=\left(\pi-\tfrac\theta2,-\pi+\tfrac{3\theta}2\right)\,,
    \nonumber\\[1ex]
    c_2\ \ :& &=(-x,\ \theta)\,,
    \qquad\qquad
    r_2\ \ :\qquad\quad\, =\ (\theta,\ \theta)\,,
    \label{eq:points}\\[1ex]
    c_2\ \ :& &=(-x,2\theta)\,,
    \qquad\quad\ \ \
    r_3\ \ : \qquad\quad\, =\left(\pi-\tfrac\theta2,\pi+\tfrac\theta2\right)\,.
    \nonumber
\end{eqnarray}

Our goal is to determine the range of $x$ and $\theta$ for which $\mh{V}_i$ and $\mh{R}_i$ satisfy the conditions of the 3-to-3 CWT -- see previous subsection.
We then compare these results with the associated entanglement wedge phase diagrams to determine if a generalization of $u<d$ controls holographic scattering in this situation. We begin by making a few additional observations about  our setup in figure \ref{fig:setup3to3}: First note we must demand $\theta > x$ and $x+2\theta < 2\pi$, so that the input regions do not collide. 
These constraints define the upper and lower boundaries in the phase diagrams of figure \ref{fig:ntonphase} below.
They also ensure that 2-to-all scattering is forbidden in the boundary theory, and they imply $\theta\le \pi$.
Next, we comment that the special choice $\theta=\frac{2\pi}{3}$ makes $c_1, c_2$, and $c_3$ all equidistant. 
By design and from eq.~\eqref{eq:points}, we see $c_1$, $c_2$, and $c_3$ always lie on an equal time-slice. 
However, we note that for $\theta<\frac{2\pi}{3}$, $r_2$ lies below $r_1$ and $r_3$ (as in figure \ref{fig:setup3to3}), while for $\theta<\frac{2\pi}{3}$, $r_2$ lies above $r_1$ and $r_3$. Hence $\theta = 2\pi/3$ is also special in that is the only angle for which the output points lie in the same time slice and are equally spaced around the $\phi$ circle.

We now ask, when are the conditions of the 3-to-3 CWT satisfied?
Recall that for any choice of $\theta$, $c_1$ and $c_3$ are always equidistant from $c_2$.
Hence by the symmetry of our setup, if the $(a)$ region in \eqref{eq:abc} is not empty then  $(b)$ must be as well. Now recall that in the present case, a connected causal graph \eqref{eq:causalgraf} requires at least two of the three 2-to-all scattering processes \eqref{eq:abc}. Hence, a necessary and sufficient condition to check for the CWT is nonemptiness of $(a)$, namely 
\begin{equation}
    \m{J}^+(c_1)\cap \m{J}^+(c_2)\cap \m{J}^-(r_1)\cap \m{J}^-(r_2)\cap \m{J}^-(r_3)\,,
\end{equation}
where we have used eq.~\eqref{eq:equivalence} to replace regions with points.

To simplify this problem, we define spacelike geodesics $\gamma_{ij}$ as the intersection of the past light sheets from $r_i$ and $r_j$. Similarly we define $\gamma^{ij}$ as the intersection of the future light sheets from $c_i$ and $c_j$. Then we may write
\begin{equation}
    \m{J}^-(r_{i})\cap \m{J}^-(r_{j})=\m{J}^-(\gamma_{ij})\,,
    \qquad
    \m{J}^+(c_{i})\cap \m{J}^+(c_{j})=\m{J}^+(\gamma^{ij})\,,
    \qquad i \ne j\,.
    \label{eq:ijs}
\end{equation}
We observe that if $r_i$ and $r_j$ lie on the same time-slice, then $\gamma_{ij}$ is a radial spacelike geodesic, or \textit{diametral} geodesic (see eq.~\eqref{eq:diameter} in appendix \eqref{app:basics}); similarly, $\gamma^{ij}$ is diametral for all $i$ and $j$.
From eq.~\eqref{eq:ijs}, the $(a)$ scattering region can be written as
\begin{equation}
    \m{J}^+(\gamma^{12}) \cap \m{J}^-(\gamma_{ij}) \cap \m{J}^-(r_k) \,, \qquad i,j,k\textrm{ all distinct.}
    \label{eq:fancy}
\end{equation}
Now we will find windows of parameter space (namely, $\theta<2\pi/3$ and $\theta>2\pi/3$) where for a particular $k$, $\m{J}^-(r_k)$ is irrelevant for the scattering, \ie nonemptiness of (a) is equivalent to the intersection
\begin{equation}
   \m{J}^+(\gamma^{12})   \cap \m{J}^-(\gamma_{ij})\,,
   \label{eq:fancy2}
\end{equation}
not being empty, because a point in this region will automatically lie in $\m{J}^-(r_k)$.

To summarize, we will diagnose holographic scattering by first showing \eqref{eq:fancy2} is nonempty and then establishing that it contains a point in $\m{J}^-(r_k)$. 
To do this in practice, recall that $\gamma^{12}$ is a diametral geodesic, and from eqs.~\eqref{eq:points},\eqref{eq:diameter}, it lies in the plane $\phi=\theta/2$ (and $\phi=\pi+\theta/2$).
Now, as we will show, in all cases of interest, $\gamma_{ij}$ intersects this plane, and we call the intersection point $p$.
Clearly, \eqref{eq:fancy2} is nonempty if and only if $p$ lies in \eqref{eq:fancy2},
and it is this point that we will show lies within $\m{J}^-(r_k)$.

We begin by considering the regime where $\frac{2\pi}{3} \leq \theta \leq \pi$, and we argue that $r_2$ is irrelevant in determining the scattering threshold there.
Recall from eq.~\eqref{eq:points} that $t_{r_2}$ is larger than $t_{r_1}$ and $t_{r_3}$ in this regime, and so it is reasonable to expect that scattering to $r_1$ and $r_3$ implies scattering to $r_2$. As noted just above, we prove this by showing that $p$, the $\phi=\theta/2$ point on $\gamma_{31}$, automatically lies in $\m{J}^-(r_2)$.

Now, $\gamma_{31}$ is diametral because $r_1$ and $r_3$ lie in the same boundary time-slice.
Using the coordinates in eq.~\eqref{eq:points}, the  boundary interval associated to $\gamma_{31}$ extends between $(t,\phi)=(0,\theta)$ and $(\pi-\theta,\pi+\theta)$. Hence in the notation of appendix \ref{app:basics}, $\gamma_{31}$ is characterized by $\Delta t=\pi-\theta$, $\Delta \phi=\pi$ and midpoint $(t_0, \phi_0)=((\pi - \theta)/2,\pi/2+\theta)$. Since this geodesic is diametral but not contained in the $\phi=\theta/2$ plane, it can only intersect this plane at $r=0\equiv r_p$ which occurs at the midpoint of $\gamma_{31}$ with
\begin{equation}
    t_p = \frac{\pi-\theta}{2}\,.
    \label{eq:hobby}
\end{equation}
Now $p$ automatically lies in $\m{J}^-(r_2)$ when $\theta>2\pi/3$,
because it takes coordinate time $\pi/2$ for a radial light ray to travel from the AdS boundary to $r=0$, and
\begin{equation}
    t_{r_2}-t_p
    = \frac{3\theta}{2}-\frac{\pi}{2} >\frac{\pi}{2}\,,
\end{equation}
for all $\theta>2\pi/3$.
This shows $r_2$ is irrelevant for characterizing the scattering threshold.

To calculate the scattering threshold, we must find the values of $\theta$ and $x$ for which the intersection in eq.~\eqref{eq:fancy2} shrinks to a point, \ie where the two geodesics just intersect. Again this intersection must occur at $r=0$ and so null rays from $c_1$ and $c_2$ must reach $r=0$ before $t_p$ in order for the scattering region to be nonempty. 
Such null rays reach the center at coordinate time $t_{c_1} + \pi/2$, and 
using $t_{c_1}=-x$ from eq.~\eqref{eq:points}, the scattering condition simply reads
\begin{equation}
    \frac{\pi}{2}-x \le \frac{\pi-\theta}{2}
\end{equation}
or equivalently,
\begin{equation}
    \theta\le 2x\,.
    \label{eq:hop314}
\end{equation}
This is equivalent to the condition $t_{r_i}-t_{c_j} > \pi$, so in this regime the holographic scattering threshold simply corresponds to all input systems travelling along radial null geodesics to arrive at $r=0$ (which takes $\Delta t=\pi/2$), and travelling out again along radial null geodesics to reach the output points at $r\to\infty$ (taking another $\Delta t=\pi/2$).

Next, we consider the regime $0\leq\theta \leq \frac{2\pi}{3}$, and we argue that $r_1$ is irrelevant in determining the scattering threshold there.
Again, this amounts to showing that $p$, the $\phi=\theta/2$ point on $\gamma_{23}$, automatically lies in the past of $r_1$.
Using the coordinates given in eq.~\eqref{eq:points}, the geodesic $\gamma_{23}$ is associated with the boundary interval extending between $(t,\phi)=(0,0)$ and $(\theta/2,3\theta/2)$. Hence in the notation of appendix \ref{app:basics}, it is characterized by
$\Delta t=\theta/2$, $\Delta \phi=3\theta/2$, and $(t_0,\phi_0) = (\theta/4,3\theta/4)$.
Then, from eq.~\eqref{eq:RTpure}, $p$ is characterized by
\begin{equation}
    t_p = \frac{\theta}{4} - \tan^{-1}\left(\frac{\tan^2\frac{\theta}{4}}{\tan \frac{3\theta}{4}} \right) \,,\qquad
    r_p = \frac{\cos \theta + 1/2}{\sin \theta} \,,
    \qquad
    \phi_p =\frac{\theta}{2}\,,
    \label{eq:12point}
\end{equation}
where recall that we are in the regime $0\le\theta\le 2\pi/3$ and so, as we expect, $r_p\ge 0$.

To verify that $p$ lies in the past of $r_1$, we introduce the notion of a future lightcone cut.
The future lightcone cut of a bulk point $p$ is defined as the intersection of $p$'s future lightcone (that is, the boundary of the future of $p$) with the conformal boundary of the manifold.
In the context of AdS$_3$, the lightcone cut simply defines a closed curve on the two-dimensional boundary given by eq.~\eqref{eq:cut} in the appendix.
Now, since the point \eqref{eq:12point} lies in $\gamma_{23}$, its future lightcone cut automatically contains both $r_2$ and $r_3$.
As noted in the appendix, the lightcone cut of $p$ has a global maximum at the antipodal angle to $p$, hence in the present context, the lightcone cut has a global maximum at $\phi=\theta/2+\pi$ and hence at $r_3$.
Since $t_{r_3}=t_{r_1}$, the lightcone cut clearly passes to the past of the point $r_1$, so $r_1$ is irrelevant for characterizing the scattering threshold.

Now to find the scattering threshold in the regime of interest, $\theta\le2\pi/3$, we need to determine the values of $\theta$ and $x$ at which $\gamma^{12}$ and $\gamma_{23}$ just intersect and the region in eq.~\eqref{eq:fancy2} reduces to a point. This is a straightforward calculation\footnote{For a quick route to the answer, construct a 2-to-2 scattering configuration like in figure \ref{fig:iffsetup}, with the same $c_1$ and $c_2$ as the 3-to-3 scattering problem and with $\gamma_{\mh{R}_1}=\gamma_{23}$. 
Then, the 3-to-3 scattering inequality of interest reduces to a $u<d$ inequality for the input regions of the 2-to-2 setup. \label{foot:tip}}
yielding the scattering threshold $\sin(\theta -\frac{x}{2}) = 2\sin \frac{x}{2}$.

\begin{figure}[t]
    \centering
    \includegraphics[width=0.3\linewidth]{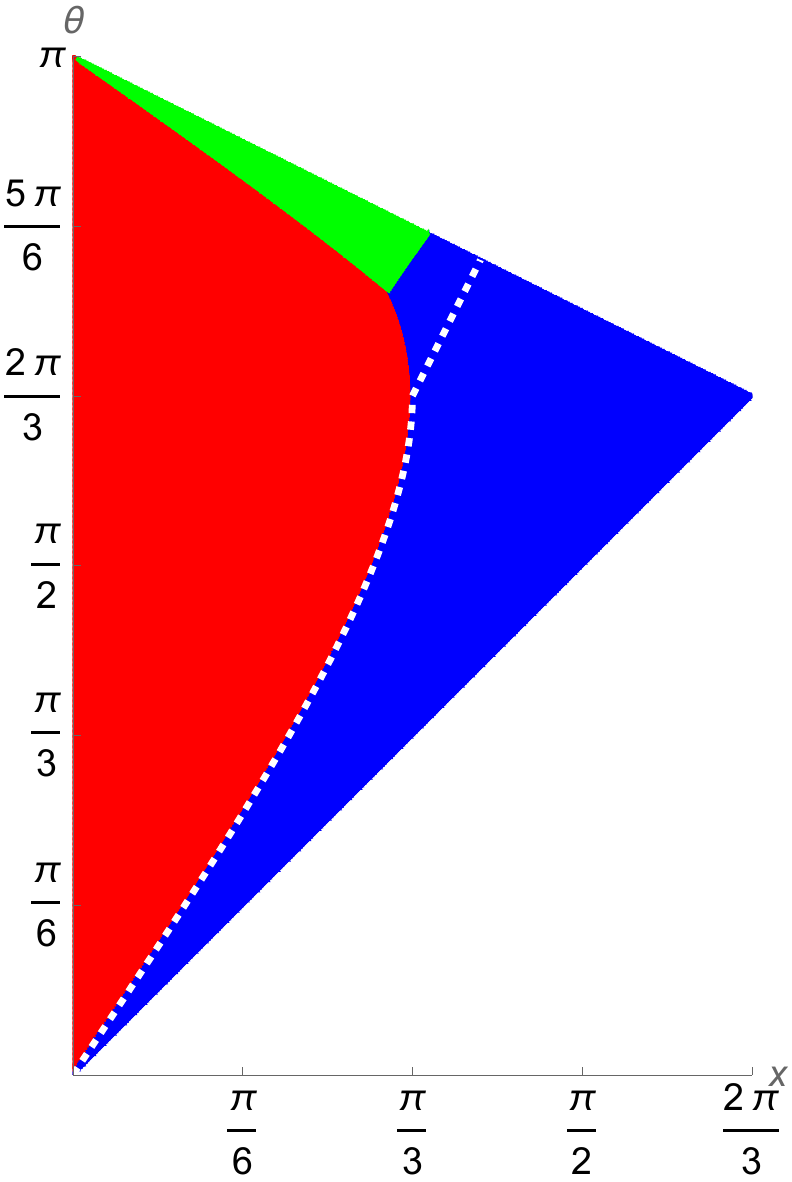}
    \caption{Entanglement wedge phase diagram for $\m{E}_{\mh{V}_1 \cup \mh{V}_2 \cup \mh{V}_3}$, with the 3-to-3 scattering threshold overlaid as a dashed white line. 
    Blue indicates the fully connected phase, red indicates the fully disconnected phase, and green indicates the partially connected phase.
    To the right of the dashed line, scattering is possible, and the 3-to-3 CWT implies $\m{E}_{\mh{V}_1 \cup \mh{V}_2 \cup \mh{V}_3}$ is fully connected.
    To the left of the dashed line, scattering is impossible, and yet there are some setups for which $\m{E}_{\mh{V}_1 \cup \mh{V}_2 \cup \mh{V}_3}$ is still fully connected.
    For $\theta\le\frac{2\pi}{3}$, the scattering threshold agrees with the boundary between fully disconnected and fully connected entanglement wedge phases, while for $\theta\ge\frac{2\pi}{3}$, the scattering threshold does not coincide with any transition among extremal surfaces obeying the appropriate homology constraints. Recall that with our setup, the phase diagram is bounded below by $\theta=x$ and above by $2\theta+x=2\pi$ to ensure that the input regions to not overlap.  }
    \label{fig:phasediagram3to3AdS}
\end{figure}

To summarize, the threshold for 3-to-3 scattering in our setup is
\begin{equation}
    \theta = 
    \begin{cases}
        \frac{x}{2}+\sin^{-1}\left(2\sin \frac{x}{2}\right)\,,\qquad &\theta\le2\pi/3
        \\
        2x \,, \qquad &\theta\ge2\pi/3\,.
    \end{cases}
    \label{eq:thresh3to3}
\end{equation}
We note that these two curves meet at $\theta=2\pi/3$, since they both yield $x=\pi/3$ there. The threshold is shown as the dashed white line in figure \ref{fig:phasediagram3to3AdS}.

How does the scattering inequality compare to the phases of the $\hat{\m{V}}_1 \cup \hat{\m{V}}_2 \cup \hat{\m{V}}_3$ entanglement wedge?
This entanglement wedge can be fully disconnected, partially connected, or fully connected.
We use the length formula $\log \frac{2}{\epsilon^2}(\cos \Delta t - \cos \Delta \phi)$
in pure AdS$_3$, which reduces to the usual $2\log \left[\frac{2}{\epsilon}\sin\frac{\Delta \phi}{2}\right]$ for $\Delta t=0$. Then the length associated with each of the various configurations reads
\begin{align}
    2 \log \left[\frac{8}{\epsilon^3} \sin^3\frac{x}{2}\right] &\quad \text{fully disconnected}\label{eq:fuldisc0}\\
    2 \log \left[\frac{8}{\epsilon^3} \sin(\frac{x}{2})\sin(\frac{ 2\theta - x}{2})\sin(\frac{2\theta + x}{2})\right] &\quad \text{partially connected}\label{eq:sopc}\\
    2 \log \left[\frac{8}{\epsilon^3} \sin^2(\frac{\theta - x}{2})\sin(\frac{2\theta+x}{2})\right] &\quad \text{fully connected.}\label{eq:fulcon0}
\end{align}
Note that the partially connected wedge of eq.~\eqref{eq:sopc} is connected between $\mh{V}_1$ and $\mh{V}_3$, and in principle one could imagine a partially connected wedge which is instead connected between $\mh{V}_1$ and $\mh{V}_2$ (or the symmetrically related choice, $\mh{V}_2$ and $\mh{V}_3$).\footnote{Its length reads $2 \log \left[\frac{8}{\epsilon^3} \sin(\frac{x}{2})\sin(\frac{ \theta - x}{2})\sin(\frac{\theta + x}{2})\right]$.}
However, this candidate is never minimal, so the only partially connected phase is eq.~\eqref{eq:sopc}. 

The phase diagram is shown in figure \ref{fig:phasediagram3to3AdS}. Comparing eqs.~\eqref{eq:fuldisc0} and \eqref{eq:fulcon0},
we find that the boundary between the fully connected and fully disconnected phases is
\begin{equation}
     \sin^2\frac{\theta-x}{2}\sin\frac{x+2\theta}{2}=\sin^3\frac{x}{2}\,. \label{eq:top1}
\end{equation}
With a careful examination, we find that this equality coincides\footnote{To establish that these equations are equivalent on the domain of interest, first note that because $x,\theta \in(0,\pi)$, the second equation is equivalent to
\begin{equation}
    \sin\frac{x}{2}=\frac{\sin\theta}{\sqrt{5+4\cos\theta}}\,,
    \label{eq:intermediate}
\end{equation}
where we have used double-angle identities and the relation $\cos\frac{x}{2}=\sqrt{1-\sin^2\frac{x}{2}}$.
Substituting eq.~\eqref{eq:intermediate} into the the first equation, one finds it is automatically satisfied. 
We emphasize that the choice of domain $x,\theta \in(0,\pi)$ is crucial here.
In particular, at $\theta=0$, eq.~\eqref{eq:top1} is valid for all values of $x$, while eq.~\eqref{eq:top2} requires $x=0$. 
However, the equations agree away from $\theta=0$ (or $2\pi n$) and in our physical setup, we require $0\le x\le\theta\le \pi$.
} with the first line of eq.~\eqref{eq:thresh3to3},
\begin{equation}
   \sin\left( \theta - \frac{x}{2}\right)=2\,\sin \frac{x}{2}\,.
   \label{eq:top2}
\end{equation}
Hence, for $\theta\le2\pi/3$, a fully connected entanglement wedge is necessary and sufficient for holographic scattering.
For $\theta\ge2\pi/3$, however, the scattering threshold is given by the second line of eq.~\eqref{eq:thresh3to3}, and we find that this does not correspond to any transition among extremal surface candidates.
In particular, a fully connected entanglement wedge is necessary but not sufficient for holographic scattering. 
We conclude that the 3-to-3 CWT does not have a converse in pure AdS$_3$.

\subsection{\texorpdfstring{$n$-to-$n$ in pure AdS$_3$}{n-to-n in pure AdS3}}

\begin{figure}[t]
    \centering
    \includegraphics[width=.4975\linewidth]{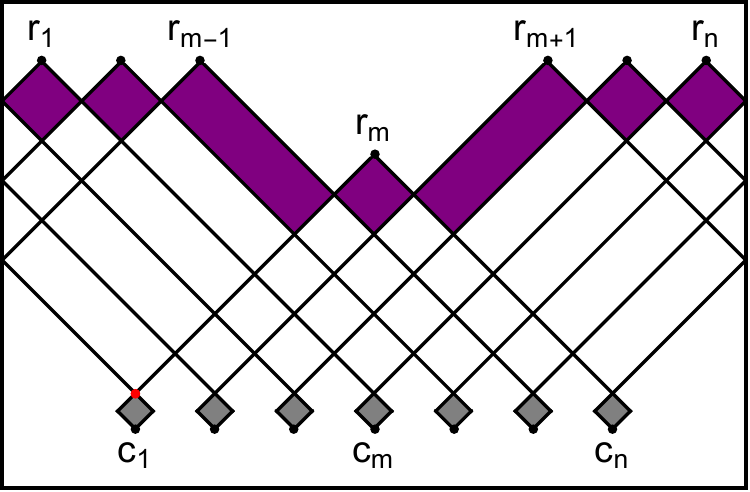}
    \hfill
    \includegraphics[width=.49\linewidth]{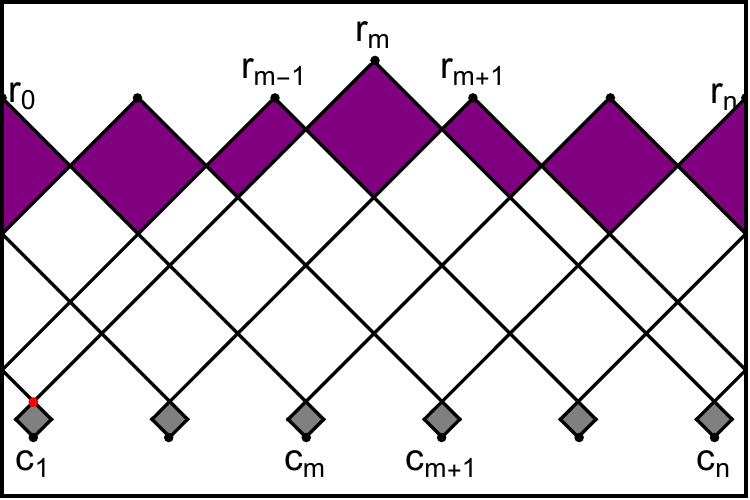}
    \caption{Configuration of $n$-to-$n$ scattering regions on the cylinder.
    The left and right boundaries of each rectangle are identified, and the $(t,\phi)$ coordinate origin is shown in red.
    Left: With $n\equiv 2m-1$ odd, $r_m$ and $c_m$ share the same $\phi$ coordinate.
    Here, $n=7$, $m=4$, and $\theta<2\pi/n$.
    Right: With $n\equiv2m$ even, $r_m$ lies midway between $c_m$ and $c_{m+1}$. Here, $n=6$, $m=3$, and $\theta>2\pi/n$.
    With $n$ even, we identify $r_0=r_n$.}
    \label{fig:oddandeven}
\end{figure}

Here we extend the calculations of section \ref{sec:3to3pure} to the general $n$-to-$n$ case, with $n\ge2$.
We will consider the particular setup shown in figure \ref{fig:oddandeven}. 

\begin{figure}[t]
    \centering
    \includegraphics[width=0.4\linewidth]{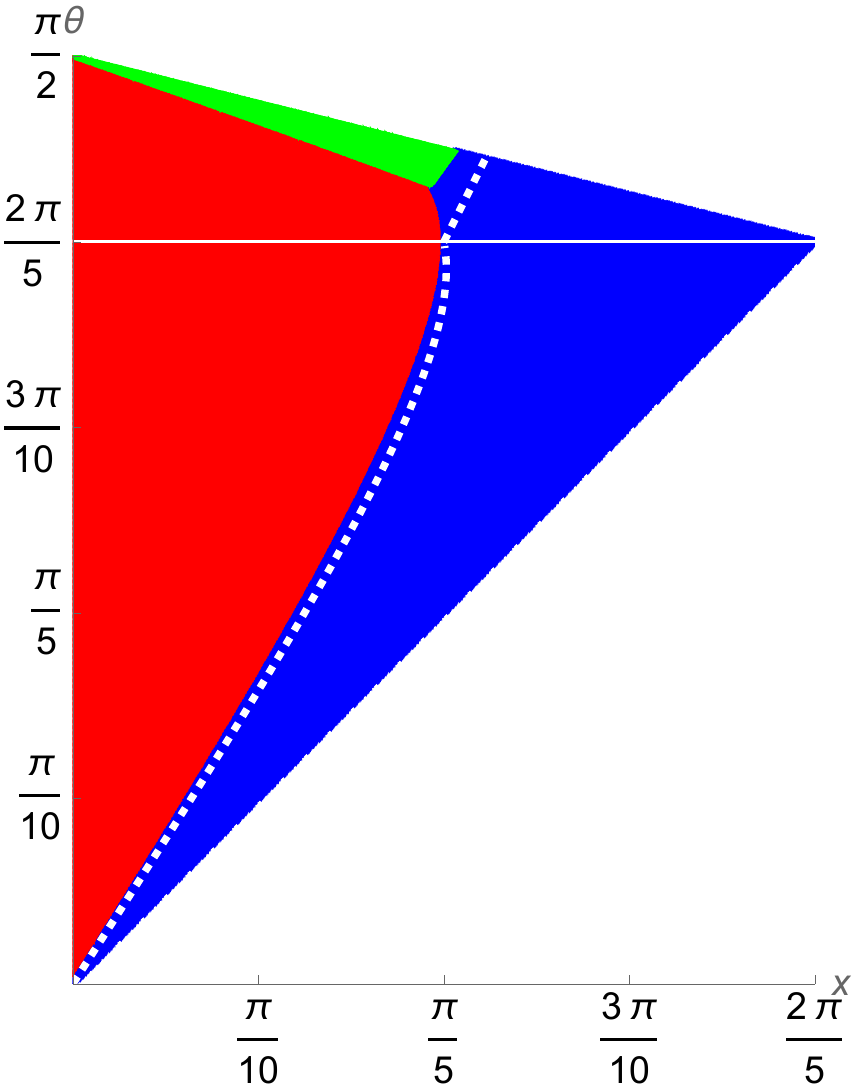}\hspace{1.5cm}
    \includegraphics[width=0.44\linewidth]{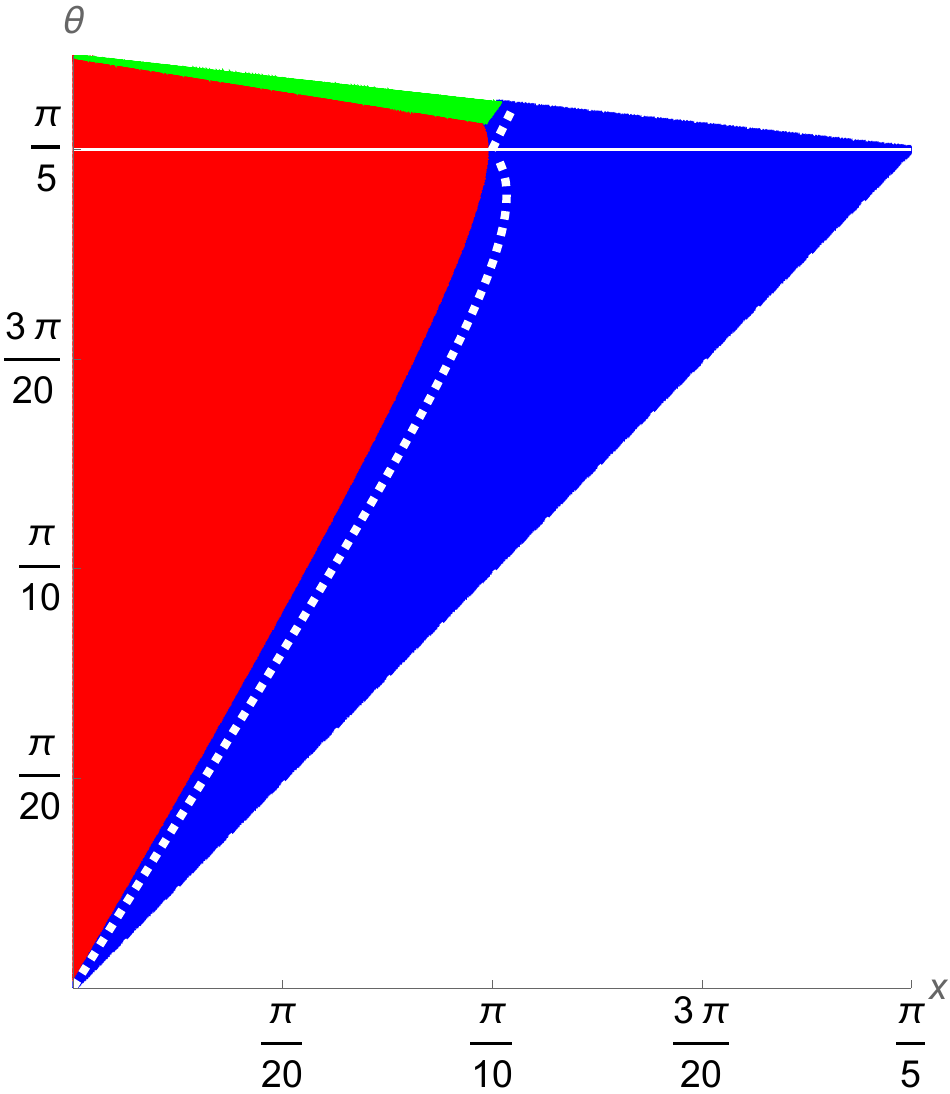}
    \caption{Holographic scattering with $n=5$ (left) and $n=10$ (right).
    The thin white line represents $\theta=2\pi/n$.
    The dashed white line represents the scattering threshold, which is given by eq.~\eqref{eq:truethresh} for $\theta\ge2\pi/n$ and by eq.~\eqref{eq:ugly} for $\theta\le2\pi/n$.
    To the right of this curve, the conditions of the $n$-to-$n$ CWT are satisfied.
    The red, blue, and green regions represent the competition between the fully connected (blue), fully disconnected (red), and partially connected (green) phases of the entanglement wedge discussed in the main text.
    This competition coincides with the holographic scattering threshold only for $n=2$ and for $n=3$ when $\theta\le2\pi/n$ -- see figure \ref{fig:phasediagram3to3AdS}.}
    \label{fig:ntonphase}
\end{figure}

As before, our $n$ input regions $\mh{V}_i$ all lie on an equal time slice, each has width $x$, and the midpoints of neighbouring regions are separated by $\theta$. Hence the separation of the midpoints of $\mh{V}_1$ and $\mh{V}_n$ is $2\pi-(n-1)\theta$. Therefore to avoid $\mh{V}_i$ and $\mh{V}_{i+1}$ from overlapping, we require $\theta\le x$ and to avoid $\mh{V}_1$ and $\mh{V}_n$ from overlapping, we require $2\pi-(n-1)\theta\le x$. These two bounds define the limits of the phase diagrams which we show later in figure \ref{fig:ntonphase}. Of course, these limits are equivalent to demanding that $2$-to-all scattering is forbidden in the boundary theory. As shown in figure \ref{fig:oddandeven}, the (purple) output regions $\mh{R}_i$ are chosen to be maximal subject to the constraint in eq.~\eqref{eq:no2toall} and our choice for the $\mh{V}_i$ (grey), and as in the $3$-to-$3$ section, these regions are larger than the $\mh{W}_i$ regions  defined in eq.~\eqref{eq:allforone}. 
These choices lead to a configuration where the $\mh{R}_i$ partition a Cauchy slice of the cylinder. Note, however, that the output regions do not touch the input regions except in the special cases $n=2$ and $n=3$ shown in figures \ref{fig:optimal} and \ref{fig:setup3to3}, respectively. 
Further, the regions exhibit different behaviours depending on the parity of $n$. 
For example, for odd $n=2m-1$, the midpoints of $\mh{V}_m$ and $\mh{R}_m$ lie at the same angular coordinate. In contrast for even $n=2m$, the $\mh{R}_m$ midpoint is midway between those of $\mh{V}_m$ and $\mh{V}_{m+1}$. 
See figure \ref{fig:oddandeven} for further details.

Next, recall from section \ref{sec:3to3pure} that the $n$-to-$n$ CWT in pure AdS$_3$ is equivalent to a points-based version of the CWT. 
Hence, as above, we label the past tips of the input regions as $c_i$ and the future tips of the output regions as $r_i$. 
We are again considering pure AdS$_3$ with global coordinates, as shown in eq.~\eqref{eq:pureads0}, and hence the boundary metric is simply $\rd s^2 = - \rd t^2 + \rd \phi^2$ where the angle $\phi$ is periodic with period $2\pi$. 
As in the previous subsection, we set the origin of our boundary coordinates at the future tip of the input region $\mh{V}_1$, as indicated by the red dots in figure \ref{fig:oddandeven}. 
For clarity, we tabulate the positions of the various input and output points. First, independent of the parity of $n$, the input points are:
\begin{equation}
    c_i\ \ :\ (t,\phi)=(-x,\ (i-1)\theta)
    \qquad\qquad\qquad\;\;\; {\rm for} \ \ i=1,\cdots,n\,.
    \label{eq:pointsc1}
\end{equation}
For odd $n=2m-1$, the output points are:
    \begin{eqnarray}
    r_i\ \ :&\ (t,\phi) &=\left(\pi-\tfrac{\theta}2,-\pi+\left(i-1+\tfrac{n}2 \right)\theta\right)
    \qquad\quad{\rm for} \ \ 
    i=1,\cdots,m-1\,,
    \nonumber\\[1ex]
    r_m\ \ :& &=\left(\tfrac{n-1}2 \,\theta,\tfrac{n-1}2\,\theta\right)
    \label{eq:pointsrodd}\\[1ex]
    r_i\ \ :& &=\left(\pi-\tfrac{\theta}2,\pi+\left(i-1-\tfrac{n}2 \right)\theta\right)
    \qquad\quad\ \ {\rm for} \ \ 
    i=m+1,\cdots,n\,,
    \nonumber
\end{eqnarray}
while for even $n=2m$, the output points are:
    \begin{eqnarray}
    r_i\ \ :&\ (t,\phi)&= \left(\pi-\tfrac{\theta}2,-\pi+\left(i+\tfrac{n-1}2 \right)\theta\right)
    \qquad\qquad{\rm for} \ \ 
    i=0,\cdots,m-1\,,
    \nonumber\\[1ex]
    r_m\ \ :& &=\left(\tfrac{n-1}2\,\theta,\tfrac{n-1}2\,\theta\right)
    \label{eq:pointsreven}\\[1ex]
    r_i\ \ :& &=\left(\pi-\tfrac{\theta}2,\pi+\left(i-\tfrac{n+1}2 \right)\theta\right)
    \qquad\qquad\ \ {\rm for} \ \ 
    i=m+1,\cdots,n\,.
    \nonumber
\end{eqnarray}
For convenience, we have introduced $r_0=r_n$ in the latter case.

We observe that a $\mathbb{Z}_n$-symmetric configuration is realized when $\theta = 2\pi/n$.
In this case, all of the output points lie on the same time-slice: $t_{r_1}=...=t_{r_n}=\tfrac{n-1}{n}\,\pi$.
Away from the $\mathbb{Z}_n$-symmetric configuration, all of the output points $r_i$ lie on the same time-slice, except for the center point $r_{m}$. Given the coordinates in eqs.~\eqref{eq:pointsrodd} and \eqref{eq:pointsrodd}, we have
\begin{equation}
    \begin{cases}
        t_{r_m}<  t_{r_{i}}\,, &\theta < \tfrac{2\pi}{n}\\
        t_{r_m}>  t_{r_{i}}\,, &\theta > \tfrac{2\pi}{n}\,,
    \end{cases}
\end{equation} 
for $i\ne m$, \ie $t_{r_m}$ is the largest output time when the spacing of the inputs is large, and the smallest output time when the spacing is small, as illustrated in figure \ref{fig:oddandeven}.

To study holographic scattering, we first focus on the case of large spacing, with $2\pi/n\le\theta\le2\pi/(n-1)$.\footnote{Recall the upper bound is required to ensure that $\mh{V}_1$ and $\mh{V}_n$ do not overlap.} Following the discussion of the previous subsection, we define the spacelike geodesics $\gamma_{ij}$ as the intersection of the past light sheets from $r_i$ and $r_j$, and we define $\gamma^{ij}$ as the intersection of the future light sheets from $c_i$ and $c_j$. 
Now let us determine the condition for a general $c_i,c_j\to all$ scattering processes. 
We note that $c_i,c_j\to all$ scattering implies
$c_i,c_j\to r_k,r_\ell$ scattering, and following our strategy from the previous subsection, the threshold for this process comes when $\gamma^{ij}$ and $\gamma_{k\ell}$ intersect. 
However, since $c_i$ and $c_j$ lie in the same time-slice, namely $t=-x$, $\gamma^{ij}$ is a diametral geodesic -- see appendix \ref{app:basics}. Similarly, if we choose $k,\ell\ne m$, $r_k$ and $r_\ell$ both lie in the time slice $t=\pi-\tfrac{\theta}2$ and $\gamma_{k\ell}$ is also a diametral geodesic. Hence the intersection of $\gamma^{ij}$ and $\gamma_{k\ell}$ can only happen at $r=0$. That is, at the threshold, $c_i,c_j\to r_k,r_\ell$ scattering proceeds by sending null rays from $c_i$ and $c_j$ to the scattering point $p$ at $r=0$ and then sending null rays from $p$ back out to $r_k$ and $r_\ell$. Now since the time required for a null ray to venture from the boundary to $r=0$ is $\Delta t=\pi/2$, the scattering threshold is
\begin{equation}
    t_{r_{k}}-t_{c_i}\ge \pi\implies \theta \le 2x\,.
    \label{eq:truethresh}
\end{equation}
Now this reasoning applies for any pair $i,j$ and any $k,\ell\ne m$. Further since $t_{r_m}\ge t_{r_i}$ in the regime $\theta\ge2\pi/n$, it is also true that the scattering point $p$ is automatically in the past of $r_m$.
Therefore eq.~\eqref{eq:truethresh} gives the threshold for $c_i,c_j\to all$ scattering for any pair $i,j$ and hence guarantees a connected causal graph $\Gamma_{2\rightarrow\text{all}}$.

Next, we turn to the case of small spacing with $0\le\theta < 2\pi/n$. 
In this case, for holographic scattering, it is not enough to have $t_{r_k}-t_{c_i}\ge \pi$ for all $k\ne m$, since $r_m$ takes place earlier than the other outputs; accordingly, we expect a stronger condition than $\theta<2x$ (as we found above for $n=3$).
Instead, to obtain the scattering condition, we will show that in this regime, $c_1,c_2\to r_m,r_{m+1}$  scattering is necessary and sufficient for holographic scattering.

As above, the threshold for $c_1,c_2\to r_m,r_{m+1}$ scattering will be determined by the intersection of $\gamma^{12}$ with $\gamma_{m,m+1}$. Now as noted above, $\gamma^{12}$ is a diametral geodesic in the plane $\phi=\theta/2$.\footnote{To see this: from the coordinates in eq.~\eqref{eq:pointsc1}, the boundary endpoints of $\gamma^{12}$ are $(t,\phi)=(\theta/2-x,\theta/2)$ and $(\pi-\theta/2-x,\pi+\theta/2)$. Hence this geodesic is given by eq.~\reef{eq:diameter} with $\Delta t = \pi-\theta$, $t_0=\tfrac\pi2-x$, and $\phi_0 = \frac{\pi+\theta}{2}$.}
Meanwhile, the geodesic $\gamma_{m,m+1}$ has boundary endpoints $(0,0)$ and $(\tfrac{n-2}{2}\theta, \tfrac{n}{2}\theta)$ and so is characterized by $\Delta t = \frac{n-2}{2}\theta$, $\Delta \phi = \frac{n}{2}\theta$, $t_0=\frac{n-2}{4}\theta$, and $\phi_0 = \frac{n}{4}\theta$.
From eq.~\eqref{eq:RTpure}, it reaches the $\phi=\theta/2$ plane at
\begin{align}
    t_p = \frac{n-2}{4}\theta
    -\tan^{-1}\left(\frac{\tan^2{\frac{(n-2)\theta}{4}}}{\tan \frac{n\theta}{4}} \right)\,, \qquad
    r_p = \frac{\sin\frac{n\theta}{2}}{2\sin \frac{\theta}{2}\sin\frac{(n-1)\theta}{2}} \,,
\end{align}
and in particular this is where $\gamma_{m,m+1}$ intersects $\gamma^{12}$ at the scattering threshold.
We establish this threshold in a moment. 

First we consider the future lightcone cut \eqref{eq:cut} of $p$, which passes through $r_m$ and has a maximum at the point $r_{m+1}$, since $\phi_{r_{m+1}}=\theta/2+\pi$ is antipodally related to $p$ with $\phi_p=\theta/2$.
All of the other output points share the same time coordinate as $r_{m+1}$, so the scattering point is automatically in the past of all the output points.
This shows $c_1,c_2\to \gamma_{m,m+1}$ scattering implies $c_1,c_2\to all$ scattering. 

Moreover, having seen that $c_1,c_2\to r_m,r_{m+1}$ scattering implies $c_1$ and $c_2$ can scatter to a point $p$ whose lightcone cut passes through $r_m$ and achieves a maximum at $r_{m+1}$, we now consider rotating this configuration to the right by $\delta\phi=(k-1)\,\theta$ with $k<n$. 
This rotates $\gamma^{12}$ to $\gamma^{k,k+1}$ and the scattering point $p$ to a new point $p'$ on the shifted radial geodesic. The lightcone cut of $p'$ will now have a 
 a maximum at $r_{m+k}$, and this cut either passes to the past of $r_m$ or, in the special case $k=n-1$, intersects $r_m$.
To put it more intuitively, with this rotation, the lightcone cut of $p'$ still lies to the past of all the output points.
So, $c_1,c_2\to \gamma_{m,m+1}$ scattering implies $c_k,c_{k+1}\to all$ scattering, for all $k<n$, and hence $c_1,c_2\to \gamma_{m,m+1}$ scattering implies holographic scattering.

We now argue that this is an if-and-only-if condition, \ie holographic scattering implies $c_1,c_2\to all$ scattering in the $\theta<2\pi/n$ regime. Since any connected causal graph contains a $c_1, c_k\to all$ scattering process for some $k$, the desired result follows if we can show that $c_1, c_k\to all$ scattering implies $c_1,c_2\to all$ scattering for any $k>1$.
Now clearly, $c_1, c_k\to all$ scattering implies $c_1, c_k\to r_m,r_{m+1}$ scattering, and one can check $\gamma_{m,m+1}$ intersects the  $\phi=(k-1)\theta/2$ plane (which contains $\gamma^{1k}$) at a point $p$.\footnote{One might wonder whether $\gamma_{m,m+1}$ instead intersects the $\phi=(k-1)\theta/2+\pi$ plane, which contains the portion of $\gamma^{1k}$ after it passes through the origin. However, we noted previously that the boundary endpoints of $\gamma_{m,m+1}$ lie at $\phi=0$ and $\phi= \frac{n}{2}\theta$. Since the opening angle is $\Delta \phi = \frac{n}{2}\theta < \pi$,
$\gamma_{m,m+1}$ always lies to the side of the origin which is closest to the points $c_1,...,{c_{m}}$, using eq.~\eqref{eq:RTpure}. Hence, $\gamma_{m,m+1}$ the intersects $\phi=(k-1)\theta/2$ plane, and not the $\phi=(k-1)\theta/2+\pi$ plane.}
Now, for $k$ odd, $p$ lies to the side of the origin in which $\gamma^{1k}$ lies to the future of $c_{(k+1)/2}$.\footnote{To see this, note that both $\gamma^{1k}$ and the radial light-ray emanating from $c_{(k+1)/2}$ lie in the plane $\phi=(k-1)\theta/2$.
They intersect at $r=0$ and $\gamma^{1k}$ leaves from the boundary to the future of $c_{(k+1)/2}$, so it must be that $\gamma^{1k}$ is to the future of the radial light-ray from $c_{(k+1)/2}$ until $r=0$ is reached.
}
Hence, if the $c_1,c_k\to all$ scattering region is nonempty, then it contains $p$, and $p$ is contained in the future of $c_{(k+1)/2}$, so $c_1,c_k\to all$ scattering implies $c_1,c_{(k+1)/2}\to all$  scattering. 
For $k$ even, one can similarly show that $p$ lies to the side of the origin where $\gamma^{1k}$ lies in the future of $c_{k/2}$, and hence $c_1,c_k\to all$ scattering implies $c_1,c_{k/2}\to all$  scattering.\footnote{This follows since $c_{k/2}$  lies in the past of the endpoint of $\gamma^{1k}$ with $\phi=\tfrac{k}{2}\theta$, $t=\tfrac{k}{2}\theta-x$.}
Induction then gives the desired result.

Having shown that $c_1,c_2 \to \gamma_{m,m+1}$ scattering is equivalent to holographic scattering, we use this to calculate the scattering threshold.
We write down the light ray which starts from $c_1=(-x,0)$ and reaches $\phi=\theta/2$ and $r=r_p$ in the bulk.
From eq.~\eqref{eq:r1rayapp}, with $x_p\equiv r_p\cos \phi_{c_1}$ and $y_p\equiv r_p\sin \phi_{c_1}$, it is given by
\begin{equation}
\begin{aligned}
    r(\lambda) &= 
        \sqrt{y_p^2
        + 
        \frac{\lambda^2}{1+y_p^2}}\,,\\
    t(\lambda)&=\tan^{-1}\left(\frac{\lambda}{1+y_p^2}\right) + \frac{\pi}{2} + t_{c_1} \,,\\
    \phi(\lambda)&=\tan^{-1}\left(\frac{\lambda }{|y_p|\sqrt{1+y_p^2}}\right) + \frac{\pi}{2} + \phi_{c_1}\,,
\end{aligned}
\end{equation}
and the ray attains $r=r_p$ at $\lambda=-x_p\sqrt{1+y_p^2}\equiv \lambda_*$. From this result, we find
\begin{equation}
\begin{aligned}
    t(\lambda_*)
    &=
    -\tan^{-1}\left(\frac{x_p}{\sqrt{1+y_p^2}}\right) + \frac{\pi}{2} + t_{c_1}\\
    &=
    \frac{\pi}{2}-x
    -\tan^{-1}\left(\frac{\sqrt{2}\sin\frac{n\theta}{2}}{ \tan \frac{\theta}{2}\sqrt{5-4\cos((n-1)\theta)-\cos n\theta}}\right)\,.
\end{aligned}
\end{equation}
Hence, the scattering threshold $t_p=t(\lambda_*)$ becomes
\begin{equation}
    \tan(x+\tfrac{n-2}{4}\theta) 
    =
    \frac{\tan^2\frac{(n-2)\theta}{4} + \sqrt{1+\left(\frac{2\sin\frac{(n-1)\theta}{2}}{\sin \frac{n\theta}{2}}\right)^2}\tan\frac{\theta}{2} \tan\frac{n\theta}{4}}{\tan\frac{n\theta}{4}-\sqrt{1+\left(\frac{2\sin\frac{(n-1)\theta}{2}}{\sin \frac{n\theta}{2}}\right)^2}\tan\frac{\theta}{2}\tan^2 \frac{(n-2)\theta}{4}}\,.
    \label{eq:ugly}
\end{equation}
This curve is shown for $\theta\le2\pi/n$ as the white curve in figure \ref{fig:ntonphase} for $n=5$ and 10.

How does the scattering inequality compare to the phases of the $\hat{\m{V}}_1 \cup ... \cup \hat{\m{V}}_n$ entanglement wedge?
From explicit analyses of $n\le4$, we expect that the only phases which contribute are the fully connected phase (red), the fully disconnected phase (blue), and the partially connected phase (green) with $\mh{V}_1$ connected to $\mh{V}_n$, and all other input regions fully disconnected.
Using the length formula above eq.~\eqref{eq:fuldisc0}, the phase for a given $\theta$ and $x$ is given by the minimal choice of the following: 
\begin{equation}
    \begin{cases}
        \sin^{n-1}\(\frac{\theta-x}{2}\)\sin(\frac{(n-1)\theta+x}{2})\,,\qquad &\textrm{fully connected (blue)}\,,\\
        \sin^n\( \frac{x}{2}\)\,,\qquad 
        &\textrm{fully disconnected (red)}\,,\\
        \sin^{n-2}\(\frac{x}{2}\)\sin(\frac{(n-1)\theta+x}{2})\sin(\frac{(n-1)\theta-x}{2})\,,\qquad &\textrm{partially connected (green)}\,,\\
    \end{cases}
    \label{eq:nton72}
\end{equation}
and we plot the phases in figure \ref{fig:ntonphase}.

The phase boundaries clearly do not match the $\theta=2x$ curve which characterizes holographic scattering in the regime $\theta\ge2\pi/n$, with the holographic scattering imposing a more stringent constraint. Further, in general the phase boundary and scattering threshold do not even match in the regime $\theta\le2\pi/n$,
unlike in the special cases $n=2,3$ (see figure \ref{fig:phasediagram3to3AdS}). 
Figure \ref{fig:ntonphase} shows that for $n\ge4$, the scattering threshold is close to the  boundary between the fully connected and fully disconnected phases:
\begin{equation}
   \sin^{n-1}\(\frac{\theta-x}{2}\)\sin(\frac{(n-1)\theta+x}{2}) =
   \sin^n\( \frac{x}{2}\)\,.
   \label{eq:pboundn}
\end{equation}
However, the scattering threshold imposes a stronger constraint.\footnote{One way to see this discrepancy analytically is to examine the regime $0\le x\le \theta\ll1$. In this regime, the scattering threshold \eqref{eq:ugly} yields $\theta\simeq m_n\,x$ where $m_n=\tfrac{2n}{\sqrt{5n^2 -8n+4}-n+2}$. Further as $n\to\infty$, $m_n\to \tfrac{1+\sqrt{5}}{2}\simeq 1.6180$ where the slope approaches this limit (\ie the golden ratio) from below. Examining the phase boundary \eqref{eq:pboundn} in the same regime yields $\theta\simeq \tlm_n\,x$ where $\tlm_n$ is the unique real root greater than one of the $n$-th order polynomial: $(\tlm_n-1)^{n-1} ((n-1)\tlm_n+1)=1$. For $n=4$, $\tlm_4\simeq1.6187>m_{n\to\infty}$. The slope $\tlm_n$ grows monotonically for larger $n$ and approaches 2 in the limit $n\to\infty$. Of course, both analyses yield identical results for $n=2$ and 3, \ie $\theta\simeq\sqrt{2} x$ for $n=2$ and $\theta\simeq 3x/2$ for $n=3$. }
We conclude that the $n$-to-$n$ CWT does not have a converse for $n\ge 3$.

\subsection{\texorpdfstring{$n$-to-$n$ in conical defect background}{n-to-n in conical defect background}}

Here we briefly comment on the extension of these results to the static conical defect background, \ie eq.~\eqref{eq:spinny} with $J=0$ and $M\in(-1,0)$, considering again the $n$-to-$n$ scattering setup of figures \ref{fig:setup3to3} and \ref{fig:oddandeven}.

The case $n=2$ was considered in \cite{Caminiti:2024ctd}, where a significant difference was found between the points-based and regions-based formulations of the CWT.
However for our $n$-to-$n$ setup with $n\ge 3$, our input and output regions always have a width which is less than $\pi$. This observation implies that the entanglement wedge of each region equals its causal wedge \cite{Caminiti:2024ctd}, and therefore the regions-based CWT is equivalent to the points-based CWT for $n\ge3$.

To state our main result, we find that with 
\begin{equation}
M>-\frac{n^2}{(n+1)^2}\,,
\label{eq:Mcrit}
\end{equation}
holographic scattering is impossible for all choices of $\theta$ and $x$.
Hence, no competition among RT surfaces will suffice to imply holographic scattering.\footnote{For the sake of completeness, we here include the phase transition from the fully disconnected RT candidate to the fully connected RT candidate, with $n=3$:
\begin{align}
    \sin^{2}\(\sqrt{|M|}\frac{\theta-x}{2}\)\sin(\sqrt{|M|}\frac{2\theta+x}{2})<\sin^3\( \sqrt{|M|}\frac{x}{2}\)\,.
    \label{eq:bigear}
\end{align}}
This generalizes the case of 2-to-2 points-based scattering in the defect geometry, where as discussed in \cite{Caminiti:2024ctd}, scattering is impossible with $M>-4/9$.

To explain the value \eqref{eq:Mcrit} of the critical mass, first consider the regime $\theta\ge2\pi/n$.
In this regime, the analog of the scattering point discussed above eq.~\eqref{eq:truethresh} is
\begin{equation}
    r=0\qquad t=t_{r_i}-\frac{\pi}{2\sqrt{|M|}}\,,
\end{equation}
and so the scattering inequality reads
\begin{equation}
    t_{r_i}-t_{c_j}=\pi+x-\frac{\theta}{2}\ge \frac{\pi}{\sqrt{M}}\,.
\end{equation}
If we set $M=-n^2/(n+1)^2$, then this is equivalent to 
\begin{equation}
    \theta \le 2x-\frac{2\pi}{n} = x-(\frac{2\pi}{n}-x) \,,
    \label{eq:bound}
\end{equation}
which is never satisfied since we always have $\theta>x$ and $x<2\pi/n$.
Further, in the $\theta\le2\pi/n$ regime, eq.~\eqref{eq:bound} is necessary for scattering since scattering to a number of output points at $t=\pi-\theta$ and one output point with $t_{r_m}<\pi-\theta$ is more difficult than, and hence implies, scattering to output points all at or above $t=\pi-\theta$.
We conclude that in the mass regime \eqref{eq:Mcrit}, it is impossible to have holographic scattering in our $n$-to-$n$ setup, generalizing the points-based scattering result of \cite{Caminiti:2024ctd}.

\section{Discussion}
\label{sec:discuss}

In this paper, we explored the generality of the relationship between holographic scattering and nonminimal surfaces identified in \cite{Caminiti:2024ctd}.
Starting with the observation that the 2-to-2 connected wedge theorem (CWT) admits a converse in pure AdS$_3$, we established a more general connection between holographic scattering and nonminimal surfaces in pure states for any spacetime where the geometry near the boundary is locally AdS$_3$. This extends the $u<d$ result of \cite{Caminiti:2024ctd}.

We emphasize that our general arguments apply well beyond the static BTZ and conical defect solutions considered in \cite{Caminiti:2024ctd}. Other nontrivial solutions where our results would apply include, \eg spinning defects \cite{Miskovic:2009uz, Compere:2018aar, Edelstein:2011vu, Briceno:2021dpi, Li:2024rma} (see appendix \ref{sec:spinning}), spinning BTZ black holes \cite{Banados:1992wn,Banados:1992gq}, and orbiting or colliding defects and black holes \cite{Balasubramanian:1999zv}. Further, in these spacetimes, our scattering setup can be generalized beyond that considered in \cite{Caminiti:2024ctd} and illustrated in figure \ref{fig:optimal}, in which $\mh{V}_1$ and $\mh{V}_2$ have equal size and lie on the same time-slice. 
Our general arguments allow us to choose the input regions $\mh{V}_1$ and $\mh{V}_2$ to be any spacelike separated causal diamonds on the boundary cylinder, where  the output regions $\mh{R}_1$ and $\mh{R}_2$ are the causal diamonds extending between the future tips of these inputs.
Consequently, $\mh{V}_1$ and $\mh{V}_2$ need not be the same size or lie on a constant time slice, \eg see figure \ref{fig:iffsetup}.

To go beyond vacuum solutions, we examined the particular example of a static AdS$_3$ geometry containing a shell of matter.
We found that the close correspondence between holographic scattering and nonminimal surfaces persists in certain regimes, specifically where the relevant RT surfaces remain outside of the shell. However, the correspondence ultimately fails when the RT surfaces probe the matter shell.
Finally, we demonstrated that the converse of the CWT does not hold for $n$-to-$n$ scattering with $n\ge 3$, even in pure AdS$_3$.

\subsection*{Where the $u<d$ converse falls short}

In section \ref{sec:genulessd}, we showed that for locally AdS$_3$ backgrounds, whenever the $u$-type RT candidate and the RT surface $\gamma_{\mh{R}_1}$ both belong to a locally AdS$_3$  region $\m{N}$, we have that $u<d$ is both necessary and sufficient for holographic scattering, provided that the boundary state is pure.
We now discuss various features and limitations of this result, as well as a potential way to strengthen it.
\vspace{.5ex}

\noindent{\bf Maximal outputs:}
A key feature of the $u<d$ construction is that the output regions $\mh{R}_i$ are chosen to be maximal, given fixed inputs $\mh{V}_1$ and $\mh{V}_2$.
We emphasize that if the outputs are smaller, the $u<d$ relation need not imply scattering. 
In particular, consider sliding the endpoints of $\mh{R}_i$ up along the $\mh{J}^-(r_i)$ lightcones to produce smaller $\mh{W}_i\subset \mh{R}_i$ -- for reference, see figure \ref{fig:iffsetup}. 
It turns out that shrinking $\mh{W}_i$ too far may result in an empty scattering region.
A dramatic illustration of this situation is given by considering $\mh{V}_1$ and $\mh{V}_2$ to be of equal size, on the boundary of a conical defect geometry with $M >-4/9$.
If we set $\mh{W}_1={r_1}$ and $\mh{W}_2={r_2}$, then holographic scattering is impossible for all choices of $\theta$ and $x$, even when we have $u<d$ \cite{Caminiti:2024ctd}.

In the particular setting of section \ref{sec:genulessd}, $\m{E}(\mh{R}_1)$ is contained in the locally AdS$_3$  region $\m{N}$, and so the equivalence of entanglement wedges and causal wedges in pure AdS$_3$ imply $\m{J}^-(\m{E}(\mh{R}_1))=J^-(r_1)=\m{J}^-(\m{E}(\mh{W}_1))$.
Hence, we can actually shrink $\mh{R}_1$ to a point without affecting the bulk scattering region.
On the other hand, generically $\m{E}(\mh{R}_2)$ is not contained in the locally AdS$_3$  region $\m{N}$, and we may find $\m{J}^-(\m{E}(\mh{R}_2))\ne \m{J}^-(\m{E}(\mh{W}_2))$.
Hence the $u<d$ converse of the CWT will only extend to choices of a smaller output region $\mh{W}_2$ such that $\m{J}^-(\m{E}(\mh{R}_2))= \m{J}^-(\m{E}(\mh{W}_2))$.\footnote{For example, in conical defect geometries, this happens if $\mh{R}_2$ and $\mh{W}_2$ have width greater than $\pi$.}
\vspace{.5ex}

\noindent{\bf Closed timelike curves:} 
Another notable feature of the $u<d$ result is that  in the spinning defect backgrounds, the $u$ surface may probe a region of the geometry with closed timelike curves, even if the scattering itself happens outside of the CTC region.\footnote{In the notation of appendix \ref{sec:spinning}, this arises when $\theta+x\in(\Delta \phi_{CTC}, \pi/b_+)$ and $\theta<\Delta \phi_{CTC}$.}
In physical scenarios, one expects that the spinning defect solution describes the spacetime only outside of a localized matter source, such as the spinning string of \cite{Maxfield:2022rry}, and no CTC region actually exists.
Hence, in these physical spacetimes, we expect that the RT surfaces will encounter the matter sources in certain regimes, disrupting the $u<d$ connection to holographic scattering, similarly to the situation in section \ref{sec:starb}. Still it would be interesting to understand, from a bulk reconstruction point of view, what it means for the $u$ entanglement wedge to probe the CTC region, and why this appears to be useful for diagnosing holographic scattering.
\vspace{.5ex}

\noindent{\bf No matter:}
The disruption of the $u$ surface in the physical spinning defect geometries above or in the shell geometry \eqref{eq:metrics} illustrate a major issue with the $u<d$ converse of the CWT, namely its failure to apply in the presence of matter sources.

Specifically, in geometries which are locally AdS$_3$ outside of certain matter sources, the $u$ surface reaches deeper and deeper into the bulk as $\theta+x$ increases, and it may become disrupted by the presence of bulk matter at a critical value of $\theta+x$ which we call $\Delta \phi_s$.
For static defect geometries with $M<-1/4$, \cite{Caminiti:2024ctd} found $\Delta \phi_s = \frac{\pi}{\sqrt{|M|}}$, and for spinning defect geometries with $M<-\frac{1}{4}-J^2$, we generalized this to 
\begin{equation}
\Delta \phi_s = \frac{\pi}{b_+}= \frac{2\pi}{\sqrt{|M| +J}+\sqrt{|M| -J}}    
\end{equation}
in appendix \ref{sec:spinning}.\footnote{Here we assume that the background includes the CTC region. Of course, in the physical solutions mentioned above, which cap off before the CTC region, we should instead write $\Delta \phi_s \gtrsim \Delta \phi_{CTC}$.} 
In all of these cases, the scattering threshold is given by the naive extension of the $u=d$ formula, and yet the $u$ surface ceases to exist for $\theta+x>\Delta \phi_s$ so $u<d$ cannot be said to control scattering.

Indeed, this failure of the $u<d$ converse could have been anticipated from section \ref{sec:genulessd}, where implicitly we showed that whenever the $u<d$ converse holds, the scattering region does not contain matter sources, which deform the local geometry; this was a clue that the presence of matter sources can disrupt the relationship between extremal surfaces and bulk scattering.

Our goal in studying the shell geometry in section \ref{sec:starb} was to further examine the effects of matter on the $u<d$ converse.
In this case, we have $\Delta \phi_s=\frac{2}{\mu}\tan^{-1}\frac{\mu}{R}$, and we found examples where holographic scattering is controlled by a naive extension of the $u=d$ formula and yet $\theta + x >\Delta \phi_s$. Hence in these situations, the $u$ surface does not exist and so cannot be said to control the scattering.
This reinforces the result that bulk matter disrupts the tight relationship between extremal surfaces and bulk scattering introduced in \cite{Caminiti:2024ctd}.

On the other hand, the shell example points to a possible strengthening of the $u<d$ converse.
Recall that the formulation of the $u<d$ converse in section \ref{sec:genulessd} requires that $\m{N}$ contains not only the connected RT candidate $u$ but also the RT surface $\gamma_{\mh{R}_1}$, and one may wonder whether we can relax the assumption about $\gamma_{\mh{R}_1}$, and simply say that whenever $u$ exists and $u<d$, then scattering is possible.
In the shell geometry, we find that the $\gamma_{\mh{R}_1}$ assumption is indeed redundant.
To see this, recall that in the shell geometry, $\gamma_{\mh{R}_1}$ is shell-avoiding for $\theta<\Delta \phi^*$, and $u$ exists for $\theta +x < \Delta \phi_s$.
So, the $u=d$ curve is truncated at the point where it intersects the $\theta +x = \Delta \phi_s$ curve, 
\begin{equation}
    \theta = \frac{2}{\mu}\cos^{-1}\sqrt{\frac{R^2 + 3\mu^2 +  \frac{4\sqrt{2}\,R}{\sqrt{R^2 + \mu^2}}\mu^2 }{R^2 + 9 \mu^2}}\,.
    \label{eq:below}
\end{equation}
Using the numerical and graphical methods of appendix \ref{sec:deetail}, we find that $\Delta \phi^*$ generally exceeds this value of $\theta$, and hence all along the line $u=d$, we find that $\gamma_{\mh{R}_1}$ automatically lies outside the shell, in $\m{N}$.
Hence, we are led to conjecture that the $u<d$ converse can be strengthened to exclude the assumption about $\gamma_{\mh{R}_1}$.

Conversely, one might conjecture that whenever $u<d$ fails to control scattering, the $u$ surface does not exist at the scattering threshold.
This is almost implied by our discussion of eq.~\eqref{eq:below}, except for the possibility that the $\theta +x = \Delta \phi_s$ curve intersects the scattering curve not only at eq.~\eqref{eq:below} but also at some $\theta$ above $\Delta \phi^*$, which is the regime where scattering is no longer controlled by the naive extension of the $u=d$ formula (see figure \ref{fig:scatterme}).
To validate this conjecture in cases where $R$ is very large ($R\gg 1$), or the shell is very light ($\mu \approx 1$), note that  $\Delta \phi_s \approx \Delta \phi^*$ in these limits.\footnote{To be more quantitative: for $R>1$, we find $\Delta \phi_s - \Delta \phi^*< \pi/20\simeq 0.157$ for all choices of $\mu$.
The same bound holds with $\mu>0.9$ and any choice of $R$.}
Hence, for $\theta > \Delta \phi^*$, the $u$ surface only exists ($\theta <\Delta \phi_s -x$) within a tiny island of parameter space near $x=0$, and it is reasonable to expect that this island does not extend to the scattering threshold curve.
Turning now to cases where $R$ is small and the shell is heavy, we observe that in these cases, the (extended) $u=d$ curve provides a lower bound on the scattering threshold, since the shell introduces time delays which make scattering more difficult compared to scattering in the $R=0$ geometry.\footnote{Interestingly, this feature does not hold for all parameters; with the introduction of a sufficiently large shell, it is possible that scattering becomes easier -- see figure \ref{fig:scatterme}. This is because, roughly, light rays travel faster through pure AdS$_3$ than they do in the conical defect geometry, and so a large pure AdS$_3$ region can compensate for the time delays introduced at the shell interface.}
In these cases, the conjecture is validated by eq.~\eqref{eq:below} being less than $\Delta \phi^*$.
To fully validate this conjecture, one must be more careful about intermediate values of $R$, $\mu$, but it appears to hold on a case-by-case basis (\eg $R=0.75$, $\mu=0.4$).

To summarize, we are conjecturing that for the $u<d$ converse in pure states, it suffices to study the $u$ surface, rather than requiring extra data about $\gamma_{\mh{R}_1}$, and this conjecture is supported by our analysis of the shell model.
Towards a physical interpretation of this, recall that the $u$ surface fails to exist as we increase the size $x$ of the $\mh{V}_1$ and $\mh{V}_2$ regions and their separation $\theta$, because the component of the RT surface with opening angle $\theta + x$ eventually reaches so deep into the bulk that it is disrupted by matter or topologically nontrivial features of the geometry.
Since the $u$ surface fails to exist only as the input regions begin to probe sufficiently low energy scales, we propose that the $u$ surface serves as a proxy for a universal feature of the entanglement between $\mh{V}_1$ and $\mh{V}_2$ at high energy scales, in states where the near-boundary geometry of the bulk is that of pure AdS$_3$.
Now, it is precisely the entanglement structure between $\mh{V}_1$ and $\mh{V}_2$ which characterizes the possibility of performing nonlocal quantum computations (which encode bulk scattering) in the boundary theory, so we should not be surprised if $\gamma_{\mh{R}_1}$, which characterizes the entanglement structure of a different Cauchy slice, plays a secondary role to the $u$ surface here.

In the shell geometry of section \ref{sec:starb}, the geometry is locally AdS$_3$ for large radii, \ie $r>R$, and we found the $u<d$ criterion continues to control the scattering process as long as the relevant RT surfaces only probe this region.
So, according to our interpretation above, the matter shell only affects the key entanglement structure of the corresponding boundary state at low energies or at large separations.
Now, conical defects with $M=-1/n^2$ are $\mathbb{Z}_n$ orbifolds, and following \cite{Balasubramanian:2014sra,Balasubramanian:2016xho}, one may ``ungauge'' the discrete $\mathbb{Z}_n$ gauge symmetry to produce a covering theory where the $O(c)$ internal degrees of freedom in the original CFT are now split up into $n$ boundary subregions in covering CFT, with $O(c/n)$ internal degrees of freedom.
This ``entwinement'' trick was used in \cite{Caminiti:2024ctd} to convert $u<d$ to a statement about internal degrees of freedom.
So, we are claiming that for the shell geometry, the internal entanglement structure which \cite{Caminiti:2024ctd} associated with holographic scattering is unchanged in the UV regime, and disrupted only in the IR.
We comment that the disruption of entanglement by bulk matter is reminiscent of the discussion of BTZ black holes in \cite{Caminiti:2024ctd}; in this case, the entanglement structure is disrupted by correlations to the thermofield double degrees of freedom.
We also observe that it would be interesting to understand how the entwinement trick extends to stationary examples such as spinning defects (as discussed in appendix \ref{sec:spinning}) or to dynamic configurations with multiple defects \cite{Balasubramanian:1999zv}.

\subsection*{Matters of focus}

Given the various limitations of the $u<d$ result discussed above, we now discuss possible extensions.

First, we observe that the proof of the connected wedge theorem in \cite{May:2019odp, May:2021nrl} involves a lower bound on $I(\mh{V}_1:\mh{V}_2)$ in terms of the focusing\footnote{More generally, quantum focusing \cite{Bousso:2015mna}.} properties of the lightsheets defining the scattering region.
In pure AdS$_3$, the expansion of these lightsheets vanishes,
leading to a saturation of the bound.
We expect one could leverage this saturation in order to prove the CWT converse in pure AdS$_3$ (as we did in section \ref{subsec:converseAdS}, albeit via direct calculations).
Now, in the presence of matter, the expansion of the lightsheets can be nonvanishing, and we wonder whether a careful analysis can be used to upgrade $u<d$ to a matter-dependent statement; for example, $d-\tilde{u}>m$, where $m>0$ depends on the matter content of the geometry.
As we have seen in the shell geometry of section \ref{sec:starb}, in some cases, the $u$ surface does not exist, so here $\tilde{u}$ represents a possible generalization of the $u$ surface which does always exist.

To propose a possible definition of $\tilde{u}$, let us revisit the key conclusions of section \ref{sec:simpleEx}.
In this section, we found that there is no equivalence between holographic scattering and a competition among nonminimal RT surfaces in the shell geometry.
However, we did find that holographic scattering \textit{implies} certain relations among nonminimal RT candidates when $\theta<\Delta \phi^*$.
(For $\theta>\Delta \phi^*$, analogous results were obtained using the numerical methods of appendix \ref{sec:deetail}, albeit on a case-by-case basis.)
In particular, if the $u$ surface exists, we found holographic scattering implies $u<d$; otherwise, if the $i$ surface exists, it implies $i<d$; and otherwise,  $o<d$.
These statements are nontrivial because in some cases, $o$ is the minimal RT candidate, and so holographic scattering need only imply $o<d$, and yet we find the additional relations $i<d$ and $u<d$.
As mentioned in section \ref{sec:simpleEx}, inspired by the upcoming work \cite{ap,ap2}, we can summarize these statements by writing $u_{simple}<d$, where we define $u_{simple}$ as the unique candidate in $\{o,u,i\}$ which is contained in the simple wedge of the boundary interval with opening angle $\theta+x$.
From this result, a natural proposal is to define $\tilde{u}$ with reference to the $\theta+x$ simple-wedge, such that in the shell geometry, $\tilde{u}=u_{simple}$.
Then, an idea for future work is to understand if focusing implies the existence of some $m$ such that $d-\tilde{u}<m$ matches the scattering condition.

A promising indication in this direction is that focusing arguments can directly explain the $u_{simple}<d$ effect just described, as we have learned from upcoming work \cite{ap,ap2}.
This is a sign that focusing arguments indeed have more to teach us about the relationship between holographic scattering and nonminimal surfaces.
Another interesting topic for future work is to investigate the possibility of using similar focusing arguments to prove the aforementioned conjecture about the redundancy of the $\gamma_{\mh{R}_1}$ assumption in the $u<d$ converse of section \ref{sec:genulessd}.

Stepping back, we note that since a general converse to the CWT is not known, efforts have been made by other authors to introduce modifications of the CWT that admit a converse. 
In this paper, in the spirit of \cite{Caminiti:2024ctd}, we aimed to find extra conditions, in addition to connectedness of the entanglement wedge, which are sufficient to imply scattering. 
However, another possible approach would be to find a new definition of the scattering region (typically, larger than the traditional scattering region) such that connectedness of the entanglement wedge alone is sufficient to imply ``scattering''; for example, see \cite{Leutheusser:2024yvf,ap}.

\subsection*{$n$-to-$n$ scattering}

The converse of the 2-to-2 CWT in pure AdS$_3$, and the partial $u<d$ converse in more general backgrounds, motivated us to study holographic scattering with more inputs in section \ref{sec:3to3}.
There, we found that the naive converse of the CWT does not hold, even in pure AdS$_3$, for $n$-to-$n$ scattering with $n \ge 3$, indicating that one should not expect $u<d$-type inequalities to be relevant for higher-$n$ scattering.
Physically, we can say that the details of the entanglement which enable the $n$-to-$n$ scattering are more complicated than in the 2-to-2 case, perhaps because it shared among $n$ regions.

To elaborate on our analysis, we showed that the only cases where the scattering threshold matches the boundary between connected and disconnected entanglement wedges in pure AdS$_3$ is with $n=2$, $n=3$ with $\theta\le 2\pi/n$, and $n>3$ at the $\mathbb{Z}_n$-symmetric point: $\theta= 2\pi/n$.
It would be interesting to understand what is so special about the entanglement structure at the $\mathbb{Z}_n$-symmetric point, such that having a connected wedge is always enough to imply scattering.
It would also be interesting to compare the scattering curves with phase diagrams for more exotic RT candidates, such as RT candidates which do not obey the standard homology constraint, or alternatively, to the behavior of proposed multipartite entanglement measures in holography such as \cite{Balasubramanian:2014hda,Bao:2018gck,Bao:2019zqc,Akers:2019gcv,Dutta:2019gen,DeWolfe:2020vjp,Gadde:2022cqi,Agon:2022efa,Hayden:2011ag,Ju:2024kuc,Balasubramanian:2024ysu}.
Alternatively, rather than looking for a precise match to the scattering curve, one could first try using focusing arguments to show that scattering \textit{implies} relations among nonminimal surfaces for the $n$-to-$n$ case, as in the $u_{simple}<d$ discussion above.

\acknowledgments

We thank Pablo Bueno, Guglielmo Grimaldi, Veronika Hubeny, Renate Loll, Pedro J. Martínez, Alex May, Monica Kang, Sabrina Pasterski, Simon Ross, Benjamin Sogaard, Alejandro Vilar López and Chris Waddell  for useful discussions.
Research at Perimeter Institute is supported in part by the Government of Canada through the Department of Innovation, Science and Economic Development Canada and by the Province of Ontario through the Ministry of Colleges and Universities. 
JC and CL acknowledge the support from the Natural Sciences and Engineering Research Council of Canada through Vanier Canada Graduate Scholarships [Funding Reference Numbers: CGV--514036 and CGV--192752, respectively].
RCM is also supported in part by a Discovery Grant from the Natural Sciences and Engineering
Research Council of Canada, and by funding from the BMO Financial Group.

\appendix
\section{\texorpdfstring{Geodesics and cuts in pure AdS$_3$}{Geodesics and cuts in pure AdS3}}
\label{app:basics}

In this appendix, we provide some standard technical results which are useful for our calculations in section \ref{sec:3to3} and appendix \ref{sec:spinning}. We begin with a brief review of spacelike and null geodesics in pure AdS$_3$, working in global coordinates:
\begin{equation}
    \rd s^2 = -(r^2 +1) \rd t^2 + \frac{\rd r^2}{r^2+1} + r^2 \rd \phi^2\,.
    \label{eq:pureads}
\end{equation}

A spacelike geodesic (\ie a candidate RT surface) obeys
\begin{align}
    t(s) &=  \tan^{-1}\left(\tan\frac{\Delta t}{2}\tanh s\right) + t_0 \nonumber\\
    r(s) &= \sqrt{\frac{\cosh 2s+\cos \Delta \phi }{\cos \Delta t - \cos \Delta \phi}} \label{eq:RTpure_s}\\
    \phi(s) &= \tan^{-1}\left(\tan\frac{\Delta \phi}{2}\tanh s\right)+\phi_0 \,,
    \nonumber
\end{align}
where $t_0$ and $\phi_0$ parameterize the midpoint of the boundary interval. 
The geodesic reaches its endpoints at $s\to\pm\infty$ where $(t,\phi)=(t_0\pm\Delta t/2,\phi_0\pm\Delta \phi/2)$, corresponding to the endpoints of the boundary interval. 
Further, the geodesic reaches its point of minimum radius at $s=0$ with $(r,t,\phi)=(r_{\min},t_0,\phi_0)$ where $r_{\min}=\sqrt{\frac{1+\cos \Delta \phi }{\cos \Delta t - \cos \Delta \phi}}$.
For $\Delta \phi<\pi$, eq.~\eqref{eq:RTpure_s} can be rewritten as
\begin{align}\label{eq:RTpure}
    t(\phi) &= \tan^{-1}\left(\frac{\tan \frac{\Delta t}{2}}{\tan \frac{\Delta \phi}{2}}\tan(\phi-\phi_0) \right) +t_0 \nonumber\\
    r(\phi) &= \frac{\sin\Delta \phi}{\sqrt{\left(\cos \Delta t - \cos \Delta \phi\right)\left(\cos(2(\phi-\phi_0)) - \cos \Delta \phi\right)}}\,,
\end{align}
where note $t_0 = t(\phi_0)$ and $r_{\min}=r(\phi_0)$.
For $\Delta\phi=\pi$, eq.~\eqref{eq:RTpure_s} should instead be rewritten as
\begin{equation}
    \phi(t)=\phi_0\pm\frac{\pi}2
    \quad {\rm and} \quad
    r(t) = \pm\frac{\sqrt{2}\sin(t-t_0)}{\sqrt{\cos2(t-t_0) - \cos \Delta t}}\,,
    \label{eq:diameter}
\end{equation}
where the plus (minus) refers to the portion of the geodesic with $t>t_0$ ($t<t_0$).
In the main text, we refer to these radial geodesics with $\Delta \phi=\pi$ as \textit{diametral} geodesics since they fall radially inwards, pass directly through the center of the AdS$_3$ and move radially outward again to the boundary. 

A general null geodesic in pure AdS$_3$ obeys
\begin{equation}
\begin{aligned}
    r(\lambda) &= 
        \sqrt{(1-\ell^2) \lambda^2 + \frac{\ell^2}{1-\ell^2} }\,,\\
    t(\lambda)&=\tan^{-1}\left((1-\ell^2)\lambda\right) -\frac{\pi}{2} + t_{\infty} \,,\\
    \phi(\lambda)&=\tan^{-1}\left(\frac{1-\ell^2}{\ell}\lambda\right) -\frac{\pi}{2}\textrm{sgn}(\ell) + \phi_{\infty}\,,
    \label{eq:r1rayapp}
\end{aligned}
\end{equation}
where $\ell$ corresponds to the angular momentum, \ie $\ell=r^2\partial_\lambda\phi$. The geodesic reaches the future boundary endpoint $(t,\phi)=(t_{\infty},\phi_{\infty})$ with $\lambda\to\infty$. One also sees that the initial or past boundary point at $\lambda\to-\infty$ is $(t,\phi)=(t_{\infty}-\pi,\phi_{\infty}-\pi)$, as expected.
If one is interested in labeling the geodesic with this past-most point, $t_{-\infty}$, $\phi_{-\infty}$, one need only flip the sign on the $\pi/2$ terms. The minimum radius point corresponds to $\lambda=0$ where $(r,t,\phi)=(r_{\min},t_{\infty}-\pi/2,\phi_{\infty}-\pi/2\,{\rm sgn}(\ell))$ where $r_{\min}=|\ell|/\sqrt{1-\ell^2}$. Further, we note that with $\ell=0$, the geodesic passes through the center of the space (\ie $r_{\min}=0$), while with $\ell=\pm1$, the geodesic remains on the boundary.\footnote{In this case, it is convenient to reparametrize eq.~\eqref{eq:r1rayapp} in terms of $\tilde{\lambda}=(1-\ell^2)\lambda$.} 

In section \ref{sec:3to3}, we also make use of the notion of future lightcone cuts  \cite{Engelhardt:2016wgb}.
As stated in the main text, the future lightcone cut of a bulk point $p$ is defined as the intersection of $p$'s future lightcone (that is, the boundary of the future of $p$) with the conformal boundary of the manifold. That is, an event at $p$ in the bulk can only influence boundary points on or to the future of this cut. In the context of AdS$_3$ in global coordinates \eqref{eq:pureads}, the lightcone cut simply defines a closed curve on the two-dimensional boundary.
Using eq.~\eqref{eq:r1rayapp}, one straightforwardly obtains the lightcone cut of the bulk point at $({r_p},{t_p}, {\phi_p})$ as \cite{Engelhardt:2016wgb}
\begin{equation}
    \cos (t-{t}_p) = \frac{{r_p}}{\sqrt{1+r_p^2}}\cos (\phi-{\phi_p})\,.
    \label{eq:cut}
\end{equation}
The minimum value of $t$ along the cut is achieved at $\phi=\phi_p$, and its maximum value is achieved at the antipodal angular coordinate, $\phi=\phi_p +\pi$.

\section{Example: spinning defect}
\label{sec:spinning}

Here, we examine how the $u<d$ converse of the CWT discussed in section \ref{sec:genulessd} applies in the spinning conical defect geometry in AdS$_3$.

\subsection*{Spinning defect geometry}

Spinning defects in AdS$_3$ (\eg see  \cite{Miskovic:2009uz, Compere:2018aar, Edelstein:2011vu, Briceno:2021dpi, Li:2024rma}) 
are described by the following stationary but not static geometry:
\begin{align}
     \dd{s}^2= -(r^2 - M) \dd{t}^2 + \frac{\dd{r}^2}{r^2-M+\frac{J^2}{4 r^2}}+r^2 \dd{\phi}^2 - J\dd{t}\dd{\phi}\,,
     \label{eq:spinny}
\end{align}
where $-1<M<0$, $r>0$, $\phi \sim \phi+ 2\pi$, and $t \in \mathbb{R}$.
We focus our attention on $|M|\ge |J|$, which is motivated by the recent KSW criteria \cite{Kontsevich:2021dmb,Witten:2021nzp} for the admissibility of complex metrics in the gravitational path integral \cite{Basile:2023ycy}.
Further, unitarity of the dual CFT implies $|J|\leq 1+M$ \cite{raeymaekers2011, raeymaekers2010},
so we will restrict ourselves to the interval $|J| \le \min(|M|,1-|M|)$. 
We note the same metric yields the spinning BTZ black hole when $M\geq 0$ \cite{Banados:1992wn,Banados:1992gq}, while  we recover pure AdS$_3$ with $M=-1, J=0$. 

To understand the locally AdS$_3$ structure of this spacetime, note that the spinning defect geometry may be produced by removing a wedge of fixed coordinate angle from pure AdS$_3$ and identifying the exposed faces up to a shift in the time coordinate.
To see this, we exchange $M$ and $J$ for the parameters $b_+$ and $b_-$:
\begin{align}
    b_{\pm} = \frac{1}{2}\left(\sqrt{|M|+J}\pm\sqrt{|M|-J}\right)
    \iff
    \begin{cases}
        |M|= b_-^2 + b_+^2\\
        J= 2 b_- b_+\,,
    \end{cases}
    \label{eq:bpm}
\end{align}
The metric then reads
\begin{align}
     \dd{s}^2
     &= -(r^2 + b_+^2+b_-^2) \dd{t}^2 + \frac{r^2 \dd{r}^2}{(r^2+ b_+^2)(r^2 + b_-^2)}+r^2 \dd{\phi}^2 - 2 b_+ b_-\dd{t}\dd{\phi}.
\end{align}
After the following change of coordinates, adapted from \cite{Li:2024rma},\footnote{Introducing null coordinates $u_0 =t-\phi$, $v_0 =t+\phi$, $u=\tau - \varphi$, $v = \tau + \varphi$, the coordinate change on the boundary reads simply $u=(b_+ - b_-) u_0$, $v=(b_+ + b_-) v_0$.\label{foot:null}}
\begin{equation}
    \rho = \sqrt{\frac{r^2+b_-^2}{b_+^2-b_-^2}}\qquad
    \tau = t b_+ + \phi b_- \qquad
    \varphi =t b_-  + \phi b_+ \,,
    \label{eq:jumpy}
\end{equation}
we obtain
\begin{align}
    \dd{s}^2&= -(\rho^2 +1) \dd{\tau}^2 + \frac{\dd{\rho}^2}{\rho^2+1}+\rho^2 \dd{\varphi}^2\,,
    \label{eq:adssimple}
\end{align}
which looks like pure AdS$_3$, up to the atypical identification
\begin{align}
    (\tau, \varphi) &\simeq (\tau + 2\pi b_-, \varphi + 2\pi b_+)\,.
\end{align}
As advertised, this effectively removes a wedge of coordinate angle $2\pi(1- b_+)$ from pure AdS$_3$ and identifies the exposed faces with a relative time shift of $ 2\pi b_-$.

A subtlety here is that this identification leads to a spacetime with closed timelike curves.
To see why, recall that pure AdS$_3$ comes with the coordinate range $0\leq \rho^2 < \infty$.
Under eq.~\eqref{eq:jumpy}, this corresponds to $- b_-^2 \leq r^2 < \infty$, and in particular, $r^2$ can be negative. 
Now consider the closed curve $\phi(s)= 2\pi s$,  $s\in[0,1]$, with $t$ and $r$ held fixed, or in global AdS$_3$ coordinates
\begin{equation}
    \tau(s) = 2\pi b_- s
    \qquad
    \rho(s) = \rho
    \qquad
    \varphi(s) = 2 \pi b_+ s\,.
\end{equation}
The tangent to the curve satisfies
\begin{equation}
   g_{\mu\nu}\, \partial_s x^\mu\, \partial_s x^\nu 
    =
    4\pi^2 r^2
    =
    4\pi^2(b_+^2-b_-^2)\left(\rho^2 - \frac{b_-^2}{b_+^2-b_-^2}\right)\,,
\end{equation}
and so this closed curve is timelike for $\rho^2<{\frac{b_-^2}{b_+^2-b_-^2}}$, \ie $r^2<0$.
On the other hand, the spacetime \eqref{eq:spinny} is well-behaved in the  $r^2>0$ region.
In what follows, we will always be sure to comment on when the geodesics of interest venture into the CTC region.

\subsection*{Entanglement wedge phase diagram}
To explicitly check that $u<d$ controls scattering, we first find spacelike geodesics in the spinning defect geometry \eqref{eq:spinny} in order to evaluate the holographic entanglement entropy. 
These can be obtained by mapping pure AdS$_3$ RT surfaces \eqref{eq:RTpure_s}
to the spinning defect geometry, using eq.~\eqref{eq:jumpy}.
For a spacelike geodesic which extends between two points $(t_i, \phi_i)$ and $(t_f=t_i, \phi_f)$ on the boundary,\footnote{In principle, geodesics in the spinning defect geometry can wind around $r=0$ multiple times \cite{Martinez:2019nor}, but we will not be interested in these cases.}
we obtain
\begin{align}\label{eq:RTspinningcondef}
    t(s) &= \frac{1}{b_+^2-b_-^2} \!\left[b_+ \tan^{-1}\! \left(\!\tan\frac{b_-\Delta\phi}{2} \tanh{s} \right)\!-\!b_- \tan^{-1}\!\left(\!\tan \frac{b_+ \Delta\phi}{2} \tanh{s} \!\right)\!\right] \!+\! t_i \nonumber\\
    \phi(s) &= \frac{1}{b_+^2-b_-^2} \!\left[b_+ \tan^{-1}\! \left(\!\tan\frac{b_+ \Delta\phi}{2}  \tanh{s} \right)\!-\!b_- \tan^{-1}\!\left(\!\tan\frac{b_- \Delta\phi}{2} \tanh{s} \!\right)\!\right] \!+\! \frac{\Delta\phi}{2}\!+\!\phi_i \nonumber\\
    r(s) &= \sqrt{-b_-^2+(b_+^2 - b_-^2)\frac{\cos (b_+ \Delta\phi)+\cosh (2 s)}{\cos(b_- \Delta\phi)-\cos(b_+ \Delta\phi)}}\,,
\end{align}
where $\Delta \phi = \phi_f - \phi_i$ and $s\in (-\infty,\infty)$.
We observe that while $t_i=t_f$, the function $t(s)$ in \eqref{eq:RTspinningcondef} is not constant; starting from $t(-\infty)=t_i$, it increases to a global maximum, then decreases to a global minimum, and finally increases again towards $t(\infty)=t_i$.\footnote{However, let us comment that this behaviour is a coordinate artifact. By transforming to the coordinates $(\rho, \tau, \varphi)$ in eq.~\eqref{eq:jumpy}, we would see that the geodesics lie in a fixed flat Cauchy slice, \eg $\sqrt{r^2+\ell^2}\,\cos (t-t_0) \,\cosh\eta
+ r\cos(\phi-\phi_0) \,\sinh\eta
=
0$.  }
Note that here and throughout, we are assuming, without loss of generality, $J>0$.

We also observe that the geodesic reaches its deepest radius in the bulk, $r_{\min}=r(0)$, when $s=0$.
One can check that $r_{\min}=0$ 
when $\Delta \phi=\Delta \phi_{CTC}$, with $\Delta \phi_{CTC}$ defined by\footnote{A priori there are multiple solutions to this equation, but this ambiguity is fixed by demanding $\Delta \phi_{CTC}<\pi/b_+$ (this constraint is explained momentarily).}
\begin{equation}
    b_+^2 (1+\cos(b_+ \Delta \phi_{CTC})) = b_-^2 (1+\cos(b_- \Delta \phi_{CTC})) \,.
    \label{eq:hit}
\end{equation}
Hence, we must have $\Delta \phi\le \Delta \phi_{CTC}$ for the geodesic in eq.~\eqref{eq:RTspinningcondef} to avoid the CTC region.\footnote{As an aside, using numerical methods, we find that for defects with $b_+ + 0.825\,b_- \lesssim 0.5$, we have $\Delta \phi_{CTC}>2\pi$, meaning there is no risk of geodesics entering the CTC region.
In other words, spacelike geodesics in these geometries never extend deeper than $\rho=b_-/\sqrt{b_+^2 - b_-^2}$.}
If $\Delta \phi> \Delta \phi_{CTC}$, we can follow the geodesic into the CTC region either by changing from the $r$ to the $\rho$ coordinate, or by formally allowing $r(s)$ in eq.~\eqref{eq:RTspinningcondef} to become imaginary, \ie $r\in i(0, b_-)$, after crossing $r=0$.
These CTC-probing solutions exist up to a maximum value of $\Delta \phi$, namely 
\begin{equation}
    \Delta \phi_s = \pi/b_+\,,
\end{equation}
which is the opening angle for which $r_{\min}=ib_-$, \ie $\rho_{\min}=0$, and so the geodesic hits the defect at $\rho_{\min}=0$.
(Of course, if $\pi/b_+$ exceeds $2\pi$, then the maximum opening angle is simply $\Delta \phi = 2\pi$.)
Crucially, even though the geodesics with $\Delta \phi \in (\Delta \phi_{CTC}, \Delta \phi_s)$ cross into the CTC region, the associated entanglement wedges are still equivalent to a portion of pure AdS$_3$.
To see why: for $\Delta \phi< \Delta \phi_s$, we have the radial coordinate $0<\rho_{\min}<\infty$ and the associated entanglement wedge can still be entirely described in pure AdS$_3$ coordinates \eqref{eq:adssimple}.

However, for general $\Delta \phi$, there may be a second RT candidate, obtained from eq.~\eqref{eq:RTspinningcondef} by replacing $\Delta\phi \to 2\pi - \Delta\phi$. The entanglement wedge for this candidate contains the defect and so is \textit{not} equivalent to a portion of pure AdS$_3$.
From the discussion above, this second candidate first appears at a minimum value of $\Delta \phi$, namely $\Delta \phi = 2\pi-\Delta \phi_s$. For $\Delta \phi < 2\pi -\Delta \phi_{CTC}$, the geodesic probes the CTC region, while for $\Delta \phi > 2\pi -\Delta \phi_{CTC}$, it avoids the CTC region.

Using the length formula above eq.~\eqref{eq:fuldisc0}, the length of the first candidate is 
\begin{align}\label{eq:lengthspinning}
    L &= \log \left[\frac{2}{\epsilon^2 (b_+^2-b_-^2)}\left(\cos(b_- \Delta\phi)-\cos(b_+\Delta\phi)\right)\right]\\
    &= \log\left[\frac{2}{\epsilon (b_+ + b_-)}\sin\left(\frac{b_+ + b_-}{2} \Delta\phi\right)\right] + \log\left[\frac{2}{\epsilon (b_+ - b_-)}\sin\left(\frac{b_+ - b_-}{2} \Delta\phi\right)\right],
\end{align}
where $\epsilon$ sets the UV, \ie we choose the cutoff surface close to the boundary at $r=\frac{1}{\epsilon}$, and recall $b_+ \pm b_- = \sqrt{|M| \pm J}$. 
The length of the second candidate is given by eq.~\eqref{eq:lengthspinning} upon substituting 
$\Delta\phi \mapsto 2\pi - \Delta\phi$. 
With $J=0$, we recover the expressions for the static conical defect \cite{Caminiti:2024ctd}.

Now consider the conventional arrangement of $\mh{V}_1$ and $\mh{V}_2$ shown in figure \ref{fig:optimal}.
In particular, we define $\mh{V}_1$ and $\mh{V}_2$ as the causal developments of two intervals of equal size $x$ on a constant time-slice, and 
we define $\theta$ as the separation between the midpoints of the two intervals.
Without loss of generality, we focus on the case $\theta<\pi$.\footnote{To elaborate, consider a scattering setup with $\theta>\pi$ and an equivalent setup but with $\theta' = 2\pi-\theta$. Clearly, rotating the latter setup by $\pi$ yields a setup equivalent to the former.
Note that this argument does not rely on any $\phi\to-\phi$ reflection symmetry (which the spinning defect spacetime does not have).} 
Of course, we also have $x<\theta$, otherwise the two decision regions would overlap and scattering would be possible on the boundary.
Now, as in figure \ref{fig:dcircs}, there are three possible configurations for the entanglement wedge of $\mh{V}_1 \cup \mh{V}_2$: the $d$ (disconnected), and $u$ and $o$ (connected) candidates.
The $u$ wedge is contained in an open portion $\mathcal{N}$ of pure AdS$_3$ (hence obeying the definition of $u$ in section \ref{sec:genulessd}) while the $o$ wedge is not.

Using eq.~\eqref{eq:lengthspinning}, we can construct the entanglement wedge phase diagram in the spinning defect geometry with $M=-0.5, J=0.499$.
The phase boundaries are shown in black in figure \ref{fig:duspinningconicaldefect} below.
Compared to the $J=0$ case, there is a larger portion of the phase diagram for which the entanglement wedge is connected.
However, this effect is very small -- on the order $\Delta x \sim 10^{-3}$, and so the phase diagram closely resembles figure 8 of \cite{Caminiti:2024ctd}.

From eq.~\eqref{eq:lengthspinning}, we have $u \leq d$ when
\begin{align}\label{eq:dutransition}
   \frac{\sin^2\left(\frac{b_+-b_-}{2}x \right)}{\sin^2\left(\frac{b_+-b_-}{2}\theta\right)}+ 
   \frac{\sin^2\left(\frac{b_++b_-}{2}x \right)}{\sin^2\left(\frac{b_++b_-}{2}\theta\right)}
   \geq 1.
\end{align}
We determine the scattering condition below and find a perfect match to \eqref{eq:dutransition}, as expected from section \ref{sec:genulessd} and illustrated in figure \ref{fig:duspinningconicaldefect}. 
This is therefore an example where $u<d$ implies holographic scattering.
Note that in this case, $\Delta \phi_{CTC}\approx 1.1 \pi <2 \pi$, so there are choices of $\theta$ and $x$ for which the $u$ surface crosses into the CTC region (namely, when $\theta + x > \Delta \phi_{CTC}$). As pointed out in \cite{Caminiti:2024ctd}, there exist regimes where $u<d$ controls scattering even when $o$ is the minimal candidate, and hence in these cases scattering implies a relation among non-minimal RT candidates.

\subsection*{Bulk scattering}
\label{app:spinningcondefscattering}

In the previous subsection, we derived the constraint \eqref{eq:dutransition} on $\theta$ and $x$ for obtaining a $u<d$ relation among RT candidates in the spinning defect geometry.
Then, we claimed that eq.~\eqref{eq:dutransition} matches the holographic scattering constraint.
We now confirm this by computing the holographic scattering threshold in terms of $\theta$ and $x$.

Using the notation of subsection \ref{subsec:converseAdS} and figure \ref{fig:iffsetup}, the configuration of scattering regions on the boundary is characterized by 
\begin{align}
c_1 &= \left(\tfrac{\theta}{2} - x, -\tfrac{\theta}{2}\right) &
\qquad
b_1 &= \left(\tfrac{\theta}{2}, -\tfrac{\theta}{2}\right) \\
c_2 &= \left(\tfrac{\theta}{2} - x, \tfrac{\theta}{2}\right) &
\qquad
b_2 &= \left(\tfrac{\theta}{2}, \tfrac{\theta}{2}\right)
\end{align}
in $(t,\phi)$ coordinates, or equivalently, 
\begin{align}
    c_1&=\left(\theta -x, -x\right)
    &\qquad 
    b_1&=\left(\theta, 0\right)\\
    c_2&=\left(-x, \theta - x\right)
    &\qquad
    b_2&=\left(0, \theta\right)\,,
\end{align}
in null coordinates,  $(u_0 = t-\phi, v_0 = t+\phi)$.

Transforming to pure AdS$_3$ coordinates $(u=\tau-\varphi, v=\tau+\varphi)$ using eq.~\eqref{eq:jumpy} and footnote \ref{foot:null}, we obtain
\begin{align}
        c_1&=\left((b_+-b_-)(\theta-x), -(b_+ + b_-)x\right)
        &\qquad
        b_1 &= \left((b_+ - b_-)\theta,0\right)\\
        c_2&=\left(-(b_+ - b_-)x, (b_+ + b_-)(\theta-x)\right)
        &\qquad
        b_2 &= \left(0, (b_+ + b_-) \theta \right)\,.
    \label{eq:c12b12}
\end{align}
Recall that $b_\pm$ were defined in eq.~\eqref{eq:bpm}.
Given these results \reef{eq:c12b12}, we can easily transform to Poincar\'e coordinates, using $U=\tan(u/2)$, $V= \tan(v/2)$; as mentioned in the main text, the transformation between Poincar\'e AdS$_3$ coordinates and global AdS$_3$ coordinates corresponds, on the boundary, to the conformal transformation between Minkowski spacetime and the cylinder.
We obtain
\begin{align}
    c_1&=\left(\tan\tfrac{(b_+-b_-)(\theta-x)}{2}, -\tan\tfrac{(b_+ + b_-)x}{2}\right)
    &\qquad
    b_1 &= \left(\tan\tfrac{(b_+ - b_-)\theta}{2},0\right)\\
    c_2&=\left(-\tan\tfrac{(b_+ - b_-)x}{2}, \tan\tfrac{(b_+ + b_-)(\theta-x)}{2}\right)
    &\qquad
    b_2 &= \left(0, \tan\tfrac{(b_+ + b_-) \theta}{2} \right)\,.
\end{align}

We would now like to diagnose when scattering is possible by applying the Poincar\'e patch scattering results of subsection \ref{subsec:converseAdS}.
As in subsection \ref{subsec:converseAdS}, $\gamma_{\mh{R}_1}$ is contained in the lightcone that emanates from the boundary at Poincaré coordinates $(T,X)=(0,0)$.
Unlike in subsection \ref{subsec:converseAdS}, however, here $b_1$ and $b_2$ do not generally share the same Poincar\'e coordinate time $T$.
Instead, to obtain the RT surface in this case, we can understand it as the boost of an RT surface at constant time, described by
\begin{align}
    &T=R \nonumber\\
    &X^2 + z^2 = R^2\,.
\end{align}
After boosting, one obtains the curve
\begin{align}
    &T\cosh{\beta} + X\sinh{\beta} = R \label{eq:boostedRTtime}\\
    &  (X\cosh{\beta} + T\sinh{\beta})^2 + z^2 = R^2 \label{eq:boostedRTspace}. 
\end{align}
Plugging in $b_1, b_2$ from the equations above, we get:
\begin{align}
    \cosh{\beta}&=\frac{\sin(b_+ \theta)}{\sqrt{\cos^2(b_- \theta) - \cos^2(b_+ \theta)}}\\
    \sinh{\beta}&=-\frac{\sin(b_-\theta)}{\sqrt{\cos^2(b_- \theta) - \cos^2(b_+ \theta)}}\\
    R&=\frac{\cos(b_- \theta)-\cos(b_+ \theta)}{2\sqrt{\cos^2(b_- \theta)-\cos^2(b_+ \theta)}}\,.
\end{align}
Hence, the RT surface extending between $b_1$ and $b_2$ lies in the plane
\begin{equation}
    T \sin(b_+\theta) - X
    \sin(b_-\theta)= \frac{1}{2}\left(\cos(b_-\theta)-\cos(b_+\theta)\right)\,.
     \label{eq:saga} 
\end{equation}

\begin{figure}[htbp]
    \centering
    \includegraphics[width=0.5\linewidth]{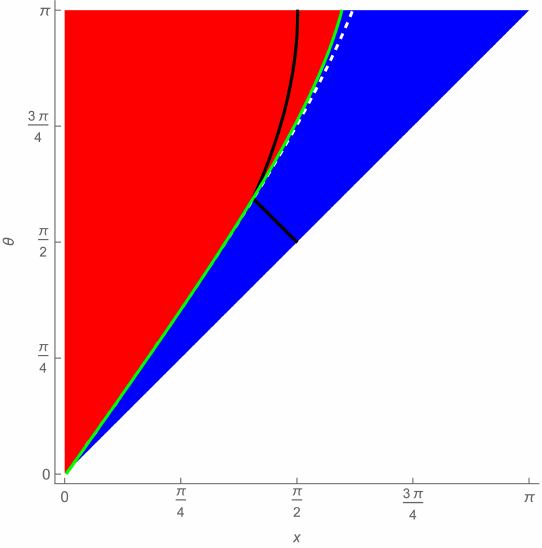}
    \caption{
    Competition between $d$ (red) and $u$ (blue) RT candidates, for $M=-0.5, J=0.499$.
    The $u=d$ boundary, shown in green, coincides with the scattering threshold; to the left of the green curve, scattering is forbidden, while to the right, it is permitted.
    In solid black, we have overlaid the phase boundaries between $o$ (top right), $u$ (bottom right), and $d$ (left) phases for the entanglement wedge.  
    Meanwhile, the dashed white curve is the scattering threshold with $J=0$.
    We see that introducing $J$ increases the portion of the phase diagram in which scattering is possible; this effect is on the order of $\Delta x \sim 0.1$.
    For completeness, we comment that $u$ does not exist in the regime $\theta + x > \pi/b_+$, which, in this case, with $\pi/b_+\approx6.09$, is a small triangle at the top right of the phase diagram  that does not affect the $u=d$ boundary.}
    \label{fig:duspinningconicaldefect}
\end{figure}

Now recall that at the scattering threshold, null rays leaving from $c_1$ and $c_2$ meet precisely at $\gamma_{\mh{R}_1}$.
Therefore, for the holographic scattering to proceed, the intersection point $(U_I, V_I)$ must satisfy
\begin{align} \label{eq:constraintcondef}
    U_I\cot(\tfrac{(b_+-b_-)\theta}{2})+V_I \cot(\tfrac{(b_++b_-)\theta}{2})&\le 1\,,
\end{align}
where we have simply rewritten eq.~\eqref{eq:saga} in the null coordinates \eqref{eq:NULL} and used some trigonometric identities. The inequality follows from allowing the intersection point to arise at or before the plane \reef{eq:saga}.

Using eq.~\eqref{eq:intersectionlightcones}, with $c_1= (U_Q, V_P)$, and $c_2=(U_S, V_R)$, the intersection between the lightcones emanating from $c_1$, $c_2$, and $(0,0,0)$ has
\begin{align}
    U_I&=\frac{\sin ((b_+-b_-)x+b_-\theta)-\sin ((b_+-b_-)x-b_+\theta)-\sin (b_+\theta)-\sin (b_-\theta)}{2 \cos (b_- x) \cos (b_+(\theta-x ))-2 \cos (b_+ x) \cos (b_-(\theta-x))}\,,\\
    V_I&=\frac{\sin ((b_++b_-)x-b_-\theta)-\sin ((b_++b_-)x- b_+\theta)-\sin (b_+\theta )+\sin (b_-\theta)}{2 \cos (b_- x) \cos (b_+(\theta-x))-2 \cos (b_+ x) \cos (b_-(\theta-x))}\,.
\end{align}
Substituting these expressions back into eq.~\eqref{eq:constraintcondef}, we obtain eq.~\eqref{eq:dutransition} which gives $u<d$, as advertised and illustrated in figure \ref{fig:duspinningconicaldefect}.

\section{Details for shell geometry}
\label{sec:deetail}

In this appendix, we provide more details about the shell geometry and the accompanying calculations in section \ref{sec:starb}.
First, we check that as claimed in section \ref{sec:stargeom}, the shell geometry obeys the null energy condition.
Then, we identify spacelike geodesics in the shell geometry, as described by eq.~\eqref{eq:shellsimple} in the main text, and their lengths as given in eqs.~\eqref{eq:interiorLen} and \eqref{eq:exteriorLen}. Finally, we describe the entanglement wedge and holographic scattering phase diagrams.

\subsection*{Null energy condition for shell geometry}

In this section, we demonstrate that the shell geometry \eqref{eq:metrics}, subject to $M\in[-1,0]$ and the condition \eqref{relate}, obeys the null energy condition.\footnote{As mentioned in the main text, $M>0$ would work as well, but we would need to fix $R > \sqrt{M}\ell$ to ensure $R$ is outside the BTZ horizon.}

We use the Israel junction conditions \cite{Israel:1966rt} (\eg see also \cite{Misner:1973prb}) to determine the stress tensor of our shell, such that Einstein's equations are obeyed. First, we evaluate the extrinsic curvature of the $r=R$ surface for the interior and exterior geometries, which is simply given by
\begin{equation}
K_{ab}=\frac12\,n^\mu\partial_\mu g_{ab}
\label{extrinsic1}
\end{equation}
where $n^\mu$ is an outward-pointing unit normal vector (\ie $n^\mu\partial_\mu=\sqrt{f_\mt{i}(R)}\,\partial_r$ for the interior and $n^\mu\partial_\mu=-\sqrt{f_\mt{e}(R)}\,\partial_r$ for the exterior). With some calculation, we find
\begin{equation}
K^{(i)}_{ab}=R\sqrt{f_i(R)}\begin{bmatrix}
-f_e(R)/f_i(R) & \  \\
\  & 1
\end{bmatrix}
\quad{\rm and}\quad
K^{(e)}_{ab}=-R\sqrt{f_e(R)}\begin{bmatrix}
-1 & \  \\
\  & 1
\end{bmatrix}\,,
\label{extrinsic2}
\end{equation}
where both tensors are expressed in ($t,\phi$) coordinates. 
Hence, the discontinuity in the extrinsic curvature is
\begin{equation}
\Delta K_{ab}=R\left(\sqrt{f_i(R)}-\sqrt{f_e(R)}\right)\begin{bmatrix}
\sqrt{f_e(R)}/\sqrt{f_i(R)} & \  \\
\  & 1
\end{bmatrix}\,.
\label{extrinsic3}
\end{equation}
Combining these results then yields the stress tensor
\begin{equation}
S_{ab}=\frac{1}{8\pi G_N}\left(\Delta K_{ab}-h_{ab}\,\Delta K^c{}_c\right)=\frac{\sqrt{f_i(R)}-\sqrt{f_e(R)}}{\pi R}\,
\begin{bmatrix}
f_e(R) & 0 \\
0 & R^4\,\frac{1}{\sqrt{f_i(R)f_e(R)}}
\end{bmatrix}\,,
\label{stress_copy}
\end{equation}
where $h_{ab}$ is the induced metric \eqref{eq:shellm} on the shell, and in the second line we have imposed our convention $8 G_N=1$.

As implicitly noted in eq.~\eqref{ratio}, $\sqrt{f_i(R)}-\sqrt{f_e(R)}>0$. Hence the energy density of the shell is positive, and there is also a positive tension in the angular direction.

Let us test if our shell satisfies the null energy condition. Consider a null vector $v^a$ at the radius of the shell, 
\begin{equation}
h_{ab}\,v^a\,v^b=0\quad\implies\quad v^a\propto \left(
\frac{1}{\sqrt{f_e(R)}},\,\frac{1}{R}
\right)\,.
\label{nullv}
\end{equation}
Hence we find
\begin{equation}
\begin{aligned}
S_{ab}\,v^av^b &\propto 
\frac{\sqrt{f_i(R)}-\sqrt{f_e(R)}}{\pi R}
\left(1 + \frac{R^2}{\sqrt{f_e(R) f_i(R)}} \right)\,.
\label{test}
\end{aligned}
\end{equation}
Examining the prefactor leads to the conclusion that  $S_{ab}\,v^av^b>0$ always. 

Note that here we chose a null vector lying tangential to the sphere. 
If we instead consider a null vector with a radial component, this new component would not contribute to $T_{\mu\nu}\,v^\mu\,v^\nu$ because $T_{\mu r}=0$. Further, the relative contribution of $v^\phi$ (\ie the second term in parentheses in eq.~\eqref{test}) would be reduced. Hence eq.~\eqref{test} produces the most stringent constraint, \ie if this equation is satisfied, then $T_{\mu\nu}\,v^\mu\,v^\nu\ge0$ for all null vectors.

The null energy condition is thus always satisfied (\ie for any $M$ and $R$). 
Note, the assumption $M > -1$ is important here, since otherwise we might have $f_i(R) < f_e(R)$ and a negative null energy.

\subsection*{Spacelike geodesics}

Here, we solve for spacelike geodesics
\begin{equation}
    r=r(s)\qquad \phi=\phi(s)
\end{equation}
in the shell geometry.
Note, it is consistent to demand $\dot{t}(s)\equiv \rd t/\rd s=0$ in this background, \ie that the geodesics stay in a constant-$t$ slice, because the geometry is static (in contrast to the spinning defect of section \ref{sec:spinning}).

A constant-$t$ slice of the shell geometry has metric
\begin{equation}
ds^2=r^2d\phi^2+ dr^2 \left\{\begin{array}{ll}
       1/f_{i}(r)& \qquad\text{for}\ \  0\le r\le R\,,\\
        1/f_{e}(r)& \qquad\text{for}\ \  r\ge R\,.
        \end{array}\right.
\label{eq:metricsappReal}
\end{equation}
The metric components are independent of $\phi$, leading to a conserved quantity:
\begin{equation}
    \ell = r^2 \dot{\phi}\,.
    \label{eq:elleq}
\end{equation}
Because the geodesic is spacelike, we further have
\begin{equation}
1=r^2\dot{\phi}^2+ \dot{r}^2 \left\{\begin{array}{ll}
       1/f_{i}(r)& \qquad\text{for}\ \  0\le r\le R\,,\\
        1/f_{e}(r)& \qquad\text{for}\ \  r\ge R\,,
        \end{array}\right.
\label{eq:metricsappparam}
\end{equation}
where we have chosen the normalization such that $u\cdot u=1$, where $u^{\mu}=\rd x^{\mu}/\rd \lambda$ is the covariant three-velocity.
Using the angular momentum equation \eqref{eq:elleq} and the explicit expressions for the blackening factors in eq.~\eqref{eq:blacken}, we have
\begin{equation}
1=\frac{\ell^2}{r^2} +  \dot{r}^2 \left\{\begin{array}{ll}
       \frac{1}{r^2+1} & \qquad\text{for}\ \  0\le r\le R\,,\\
        \frac{1}{r^2+\mu^2} & \qquad\text{for}\ \  r\ge R\,.
        \end{array}\right.
\label{eq:metricsapp2}
\end{equation}

If we impose $r(0)=r_{\min}$ and $\phi(0)=0$, then
\begin{equation}
\begin{aligned}
    r(s) &=\sqrt{\ell^2\cosh^2 s +\mu^2 \sinh^2 s}\\
    \phi(s) &= \frac{1}{{\mu}}\,\tan^{-1}\left(\frac{{\mu}}{\ell}\tanh s\right)\,,
\end{aligned}
\end{equation}
for the exterior solution, and the same with $\mu\to1$ for the interior solution.
This implies the geodesic formula \eqref{eq:shellsimple} in the main text.

Note that in the preceding discussion, we implicitly assumed $\ell$ is the same on both sides of the shell.
To derive this, observe that the length of the geodesic has two terms:
\begin{equation}
L= \int_{r(s)=r_{\min}}^{r(s)=R} \rd s \sqrt{\frac{\dot{r}}{f_i(r)} +r^2\dot{\phi}^2} +
\int_{r(s)=R}^{r(s)=1/\epsilon} \rd s \sqrt{\frac{\dot{r}}{f_e(r)} +r^2\dot{\phi}^2}\,.
\label{eq:greatleap}
\end{equation}
Varying this action yields a total derivative in both of the bulk terms and hence one arrives at a boundary term at $r=R$. 
Setting the boundary term to zero to ensure the vanishing of the action yields
\begin{equation}
\frac{1}{\sqrt{\frac{\dot{r}^2}{f_{i}(r)} +r^2\dot{\phi}^2}}\left( \frac{\dot{r}}{f_{i}(r)}\,\delta r+r^2\,\dot{\phi}\,\delta\phi\right)\Bigg|_{r=R^-}  
=
 \frac{1}{\sqrt{\frac{\dot{r}^2}{f_{e}(r)} +r^2\dot{\phi}^2}}
 \left( \frac{\dot{r}}{f_{e}(r)}\,\delta r+r^2\,\dot{\phi}\,\delta\phi\right)\Bigg|_{r=R^+}\,, 
 \label{serious2}
\end{equation}
where $R^\pm$ indicates we are approaching $r=R$ from above or below, respectively. 
Now in this expression, the first factor on both sides is simply 1 given our choice of parameterization in eq.~\eqref{eq:metricsappparam}.
Further, in both of the remaining terms, we can set $\delta r=0$ because the matching is made on a constant $r$ surface, \ie $r=R$. So we have
\begin{equation}
r^2\,\dot{\phi}\,\delta\phi \Big|_{r=R^-} =
r^2\,\dot{\phi}\,\delta\phi\Big|_{r=R^+}\,,
\end{equation}
which means that the angular momentum is the same in both regions, \ie
\begin{equation}
r^2\,\frac{d\phi}{d s} = \ell_i= 
r^2\,\frac{d\phi}{ds} = \ell_e\,.
\end{equation}

Another important matching condition at the shell is continuity.
This imposes a constraint on the relative rotation angle $\phi_0$ appearing in eq.~\eqref{eq:shellsimple}.
Specifically, the interior portion of the geodesic hits the shell at $\phi=\pm \phi_{shell,i}$, where
\begin{equation}
    \phi_{shell,i} \defeq \tan^{-1}\left(\frac{1}{\ell}\sqrt{\frac{R^2-\ell^2}{R^2+1}}\right)\,,
    \label{eq:hardnose}
\end{equation}
while the exterior portion hits the shell at $\phi=\phi_{shell,e}$, $\phi= 2\phi_0-\phi_{shell,e}$ where
\begin{equation}
    \phi_{shell,e} \defeq \phi_0 + \frac{1}{{\mu}}\tan^{-1}\left(\frac{{\mu}}{\ell}\sqrt{\frac{R^2-\ell^2}{R^2+\mu^2}}\right)\,.
\end{equation}
We take $\ell>0$, and in this case continuity implies $\phi_{shell,e}=\phi_{shell,i}$, \ie
\begin{equation}
    \phi_0 = \tan^{-1}\left(\frac{1}{\ell}\sqrt{\frac{R^2-\ell^2}{R^2+1}}\right)
    - \frac{1}{{\mu}}\tan^{-1}\left(\frac{{\mu}}{\ell}\sqrt{\frac{R^2-\ell^2}{R^2+\mu^2}}\right)\,.
    \label{eq:horseplay}
\end{equation}

We may now solve for the angular extent of the boundary interval, $\Delta \phi$, as a function of $\ell$.
To do so, it is convenient to 
expand the exterior solution, $r(\phi)$, via
\begin{equation}
\begin{aligned}
    \frac{2 \ell^2 R^2 \mu^2}{r(\phi)^2}&=
    R^2(\mu^2 -\ell^2)
    +
    (\ell^2(R^2+\mu^2)-\mu^2(R^2-\ell^2))\cos (2{\mu}(\phi-\phi_{shell,i}))
    \\&\qquad\qquad\qquad\quad- 2\mu\ell\sqrt{(R^2-\ell^2)(R^2+\mu^2)}\sin (2{\mu}(\phi-\phi_{shell,i}))\,,
\end{aligned}
\end{equation}

\noindent
since we may now easily take the $r \to \infty$ limit and find $\phi(r)\to \Delta \phi/2$, where $\Delta \phi$ is defined by
\begin{equation}
    \frac{\Delta \phi}{2}-\phi_{shell,i}=\frac{1}{{\mu}}\,\tan^{-1}\left({\mu}\,\ell\,\frac{1-\sqrt{\frac{R^2-\ell^2}{R^2 + \mu^2}}}{\ell^2 + \mu^2 \sqrt{\frac{R^2-\ell^2}{R^2 + \mu^2}}}\right)\,.
    \label{eq:DeltaPhi}
\end{equation}
Substituting eq.~\eqref{eq:hardnose} and using $r_{\min}=\ell$, we recover eq.~\eqref{eq:cantinvert} in the main text.

\subsection*{Geodesic lengths}

In this section, we compute the lengths of various RT candidates in the shell geometry.

In general, the line element reads
\begin{equation}
\begin{aligned}
    ds = \sqrt{g_{\mu \nu}\frac{\rd x^{\mu}}{\rd r}\frac{\rd x^{\nu}}{\rd r}}\,\rd r
    =\sqrt{r^2 (\phi')^2 + \frac{1}{f_{i/e}(r)}}\,\rd r \,,
    \label{eq:line_elt}
\end{aligned}
\end{equation}
and we must be careful to multiply by 2 if we only integrate from $r=r_{\min}$ to $r=1/\epsilon\to\infty$.

For a geodesic which does not cross the shell, we may consider the first line of eq.~\eqref{eq:shellsimple} with $\phi_0=0$.
Using eq.~\eqref{eq:line_elt}, we obtain 
\begin{equation}
    \textrm{Length}_{ext} =2 \int_{\ell}^{1/\epsilon} \frac{r}{\sqrt{(r^2+\mu^2)(r^2-\ell^2)}}\,dr
    =
    2\log \frac{2}{\epsilon\sqrt{\ell^2+\mu^2}}\,,
    \label{eq:lout}
\end{equation}
which is eq.~\eqref{eq:exteriorLen} in the main text.

For a geodesic which does cross the shell, the first step is to calculate the length of the portion of the geodesic inside the shell:
\begin{equation}
\begin{aligned}
    L_1 &=2 \int_{\ell}^R \sqrt{r^2 (\phi')^2 + \frac{1}{f_i(r)}}\,dr\\
    &=2 \int_{\ell}^R \frac{r}{\sqrt{(r^2+1)(r^2-\ell^2)}}\,dr\\
    &=2\, \mathrm{tanh}^{-1}\sqrt{\frac{R^2-\ell^2}{R^2+1}}\,.
\end{aligned}
\end{equation}
The second step is to calculate the length of the portion of the geodesic outside the shell:
\begin{equation}
\begin{aligned}
    L_2 &=2 \int_{R}^{1/\epsilon} \sqrt{r^2 (\phi')^2 + \frac{1}{f_e(r)}}\,dr\\
    &=2 \int_{R}^{1/\epsilon} \frac{r}{\sqrt{(r^2+\mu^2)(r^2-\ell^2)}}\,dr\\
    &=2 \log \frac{2}{\epsilon\sqrt{\ell^2+\mu^2}}- 2 \tanh^{-1}\sqrt{\frac{R^2-\ell^2}{R^2+\mu^2}}\,.
\end{aligned}
\end{equation}
All together, we have 
\begin{equation}
\begin{aligned}
    \textrm{Length}_{in} &= L_1+L_2
    = 2\log\left(\frac{2}{\epsilon\sqrt{\ell^2+1}}\,\frac{\sqrt{R^2+1}+\sqrt{R^2-\ell^2}}{\sqrt{R^2+\mu^2}+\sqrt{R^2-\ell^2}}\right)\,,
\end{aligned}
\label{eq:lin}
\end{equation}
which is eq.~\eqref{eq:interiorLen} in the main text, after substituting $r_{\min}=|\ell|$. 

\begin{figure}
 \centering
 \includegraphics[width=0.4\linewidth]{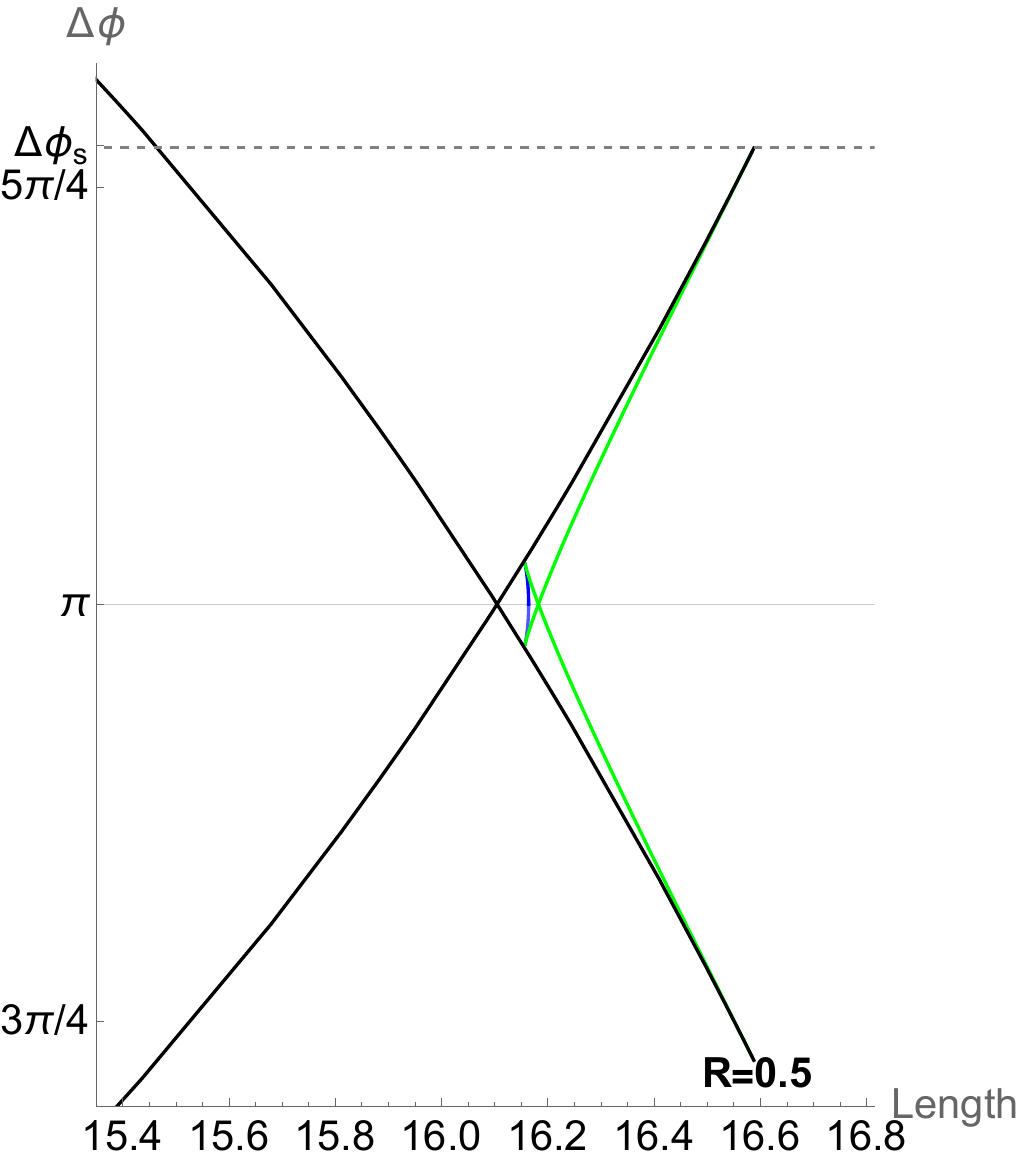}
 \qquad\qquad
 \includegraphics[width=0.36\linewidth]{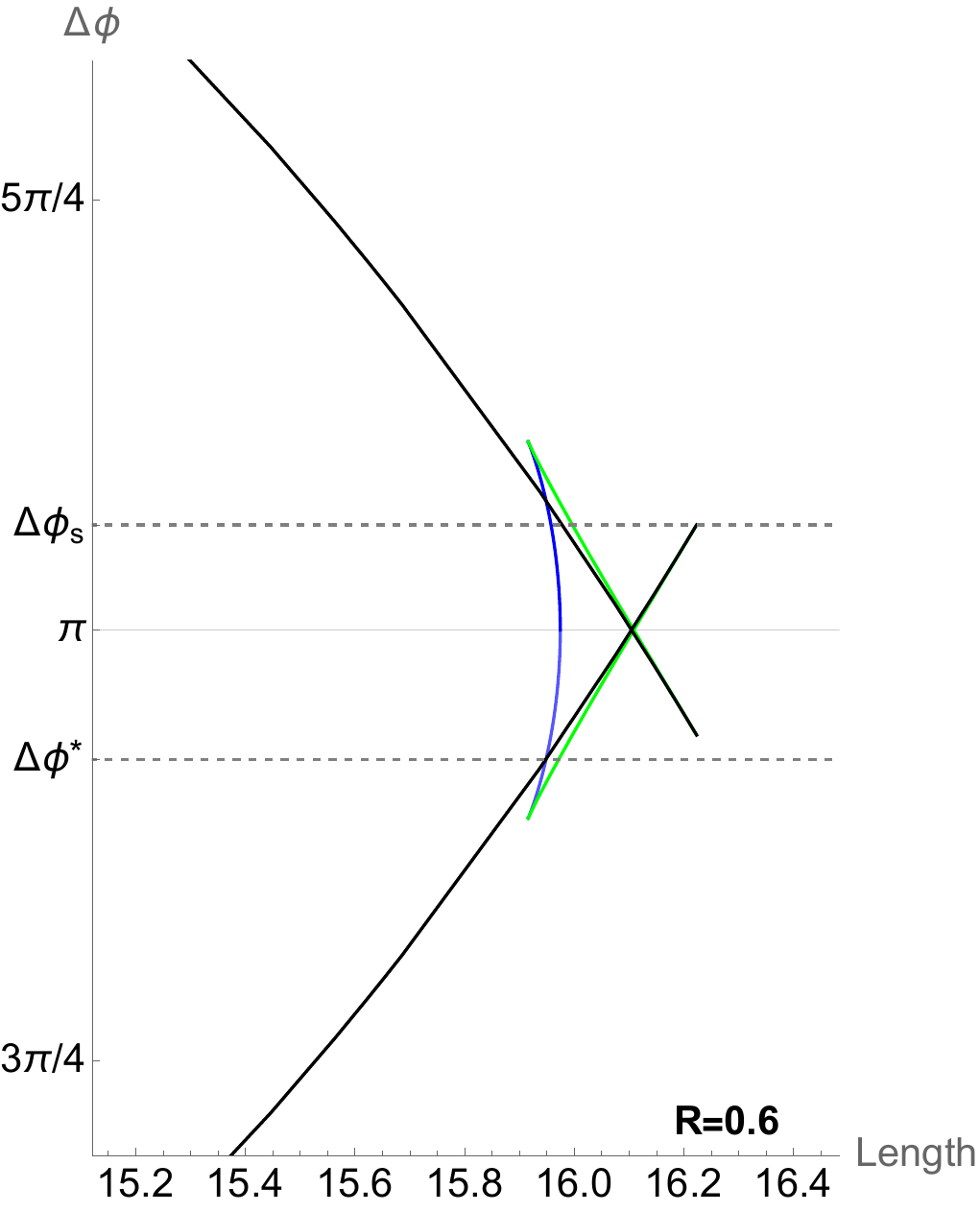}
 \caption{Opening angle versus length, for two choices of $R$ and with $\mu\approx 0$.
 In evaluating the length equations \eqref{eq:lout} and \eqref{eq:lin}, we have set $\epsilon=10^{-3}$.
 The black curves represent geodesics which do not cross the shell.
 For the same boundary angle, there can exist one or two black geodesics, as explained in section \ref{sec:rtstar}; this is related to the $\Delta \phi\mapsto 2\pi-\Delta \phi$ symmetry of the graphs shown.
 The green solution is the shell-crossing solution which is continuously related to the exterior solution at $\Delta \phi=\Delta \phi_s$, meaning it exists for every $r_{\min}= R-\delta$ with $\delta>0$ sufficiently small.
 We see the green solution is never the minimal candidate.
 The blue solution, when it exists, is the shell-crossing solution which is continuously related to the straight line through $r_{\min}=0$ with $\Delta \phi=\pi$. As $R$ increases, $\Delta \phi^*$ decreases, meaning it is more likely for the blue solution to be the minimal candidate.}
 \label{fig:flight}
\end{figure}

Figure \ref{fig:flight} shows Length$_{ext}$ in black and Length$_{in}$ in blue and green, for increasing choices of $R$, with $\mu$ fixed at $\mu\approx 0$.
These are parametric plots in the parameter $\ell=r_{\min}>0$.\footnote{Note, the parameter range $\ell=r_{\min}>0$ yields only half of the curves shown in figure \ref{fig:flight}. The rest of the curves may be obtained using $\Delta \phi \to 2\pi-\Delta \phi$ symmetry.} 
As discussed around figures \ref{fig:cool} and \ref{fig:nike} in the main text, we distinguish the blue and green solutions by whether $\Delta \phi(r_{\min})$ is locally decreasing (blue solution) or locally increasing (green solution).
Labeling $r_{gb}$ as the critical point where $\Delta \phi'(r_{gb})=0$, we see that the blue curve corresponds to the parameter range $r_{\min}\in(0,r_{gb})$, while the green curve corresponds to the parameter range $r_{\min}\in(r_{gb},R)$.
The specific expression for $r_{gb}$ where the green and blue solutions degenerate to a single solution is complicated and we omit it.
Figure \ref{fig:flight}, and others like it, teach us that the minimal length geodesic is always either the exterior solution (black) or the blue interior solution, but never the green solution.

\subsection*{Holographic scattering}

Next, we examine the CWT in detail for the shell geometry, using our usual boundary configuration shown in figure \ref{fig:optimal}. We begin by listing the possibilities for the entanglement wedge phase diagrams for $\mh{V}_1\cup \mh{V}_2$.

The possible entanglement wedge phases include the $o, i, u,$ and $d$ phases discussed in section \ref{sec:simpleEx}, as well as two more possible phases, which become relevant at large $R$:
\begin{itemize}
     \item $i_2$: connected phase such that both geodesics are shell-crossing,
     \item $d_2$: disconnected phase such that both geodesics are shell-crossing.
\end{itemize}
The $2$ subscript indicates the number of geodesics which are shell-crossing.
Note that a hypothetical $d_1$ phase could never describe the entanglement wedge; the symmetry between $\mh{V}_1$ and $\mh{V}_2$ implies that either both $\gamma_{\mh{V}_1}$ and $\gamma_{\mh{V}_2}$ cross the shell, or neither do.

\begin{figure}[ht]
    \centering
    \includegraphics[width=0.32\linewidth]{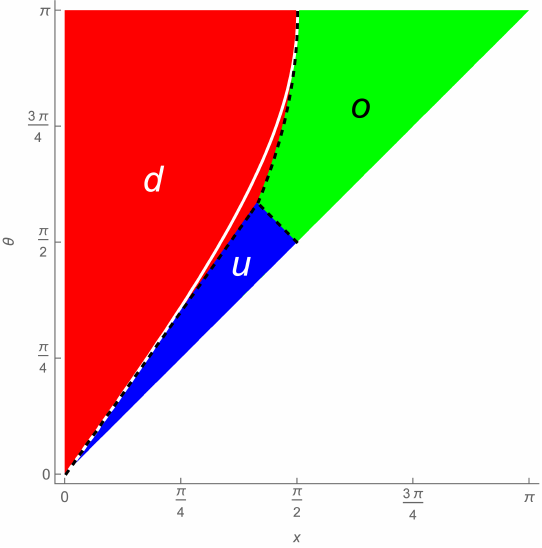}
    \hfill
    \includegraphics[width=0.32\linewidth]{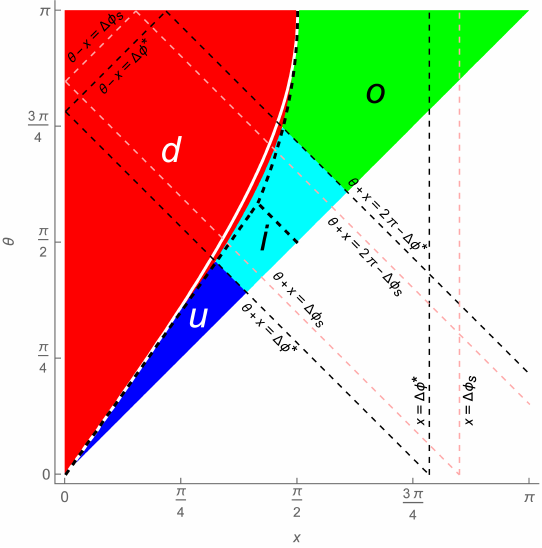}
    \hfill
    \includegraphics[width=0.32\linewidth]{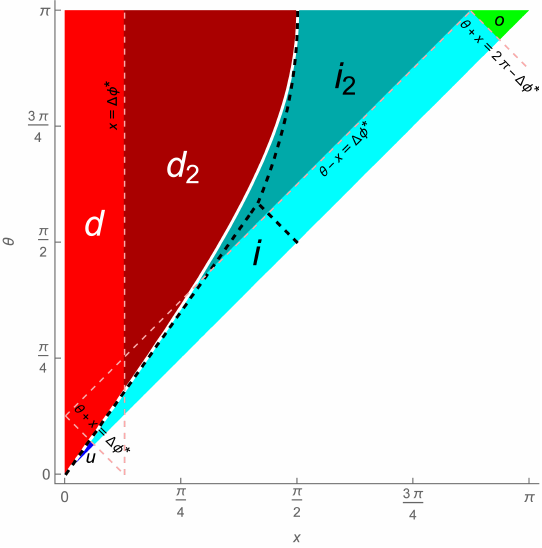}
    \caption{
    Entanglement wedge phase diagrams for shell geometries with $\mu^2=0.001$.
    We have overlaid a white curve corresponding to the boundary between connected and disconnected entanglement wedges in pure AdS, and thick dashed black curves representing the boundary between $d$, $u$, and $o$ phases in the conical defect geometry with $\mu^2=0.001$. 
    The thin dashed lines are useful for keeping track of the existence of and competition between pairs of candidates: $d$, $u$, $o$, $i$, $i_2$, and $d_2$ in the shell geometry. 
    \textbf{Left:} With $R=0.5$, $\Delta \phi^*>\pi$, so all RT surfaces avoid the shell, and the phase diagram is the same as the one for the conical defect geometry.
    \textbf{Middle:} With $R=0.75$, $\Delta \phi^* \approx 2.46 \in (\pi/2,\pi)$, so the $i$ phase is realizable, but the $i_2$ and $d_2$ phases are not. 
    We also have $\Delta \phi_s \approx 2.67 <\pi$, so there is a band of phase space for which the neither the $u$ nor the $o$ candidate exists.
    Compared to the left figure, which has a smaller value of $R$, the boundary between the connected and disconnected phases is closer to that of pure AdS (white).
    \textbf{Right:} With $R=5$, $\Delta \phi^* \approx 0.4 <\frac{\pi}{2}$, so the $i_2$ and $d_2$ phases are realizable. 
    In this case, $\Delta \phi_s \approx \Delta \phi^*$, so the transition between $u$ and $i$ involves a smaller ``jump'' (see below eq.~\eqref{eq:LinLout})
    than in the previous case.
    The boundary between the connected and disconnected phases nearly coincides with that of pure AdS, due to the large value of $R$.
    Further, the $d$, $u$, and $o$ phases are pushed away to the corners of the phase diagram; most RT surfaces are going through the shell.
    In the limit $R\to\infty$, only $i_2$ and $d_2$ matter.}
    \label{fig:phasediagshellmu001R05}
\end{figure}

In figure \ref{fig:phasediagshellmu001R05}, we show entanglement wedge phase diagrams for a fixed, heavy defect ($\mu^2=0.001$) and increasing shell radius $R$.
We see that when $\theta + x \notin (\Delta\phi^*,2\pi-\Delta\phi^*)$, the only possible dominant phases are $d$, $u$, and $o$.
This is always the case when $\Delta\phi^*\ge \pi$, so we need $\Delta\phi^*<\pi$ for the $i$ phase to be realized.
For the $i_2$ and $d_2$ phases to be realized, we must further have $\Delta\phi^*<\pi/2$; this is related to the conditions $\theta-x>\Delta \phi^*$ (for $i_2<i$) and $x>\Delta \phi^*$ (for $d_2<d$).

We now discuss holographic scattering in the shell geometry.
As in the previous examples, our goal is to determine the scattering threshold, \ie when the scattering region \eqref{eq:scatregion} transitions from being empty to nonempty.
Since the shell state is pure, we always have
\begin{align}
    J^-(\m{R}_1)\cap J^-(\m{R}_2)
    &= J^-(\gamma_{\mh{R}_1}).
\end{align}
Moreover, when $x<\Delta \phi^*$, the $\mh{V}_i$ entanglement wedges are equivalent to the $\mh{V}_i$ causal wedges.
As a consequence,
\begin{align}
    J(\m{V}_1,\m{V}_2\to \m{R}_1,\m{R}_2)
    =J^+(c_1) \cap J^+(c_2) \cap
    J^-(\gamma_{\mh{R}_1})\,,
\end{align}
and the scattering threshold takes place when light rays leaving symmetrically from $c_1$ and $c_2$ can meet precisely at the center of $\gamma_{\mh{R}_1}$.

If $\theta < \Delta \phi^*$, then $\gamma_{\mh{R}_1}$ is shell-avoiding, and the scattering problem reduces to the corresponding problem in the conical defect geometry, as anticipated in section \ref{sec:simpleEx}.
If $\theta > \Delta \phi^*$, then $\gamma_{\mh{R}_1}$ is shell-crossing, and the scattering problem becomes nontrivial.

Accordingly, let us study a shell-crossing null geodesic emanating from a point on the conformal boundary. To simplify our analysis, we define
\begin{equation}
    \sigma^2 := \frac{f_i(R)}{f_e(R)}=\frac{R^2+1}{R^2+\mu^2}\,.
\end{equation}
This allows us to make the coordinate transformation $\tilde{t}=t/\sigma$ for the interior metric, following the matching condition in eq.~\eqref{relate}. In what follows then, we use the same time coordinate $t$ in both the interior and exterior regions of the shell geometry \eqref{eq:metrics}.

Now, the null geodesics of interest are characterized by two conserved quantities:\footnote{Extending the discussion following eq.~\eqref{eq:greatleap}, we can show that both $\ell$ and $e$ remain conserved when the null rays cross the shell. 
For $e$, it is essential to use the same $t$ coordinate on both sides.} an angular momentum, $\ell$, and an energy, $e$, which we can freely set to $e=1$ by reparameterizing the curve.
Using eq.~\eqref{eq:metricsappReal} and calling the curve parameter $\lambda$,
\begin{equation}
    \ell = g_{\phi\phi}\frac{\rd \phi}{\rd \lambda}
    = r^2 \dot\phi
\qquad
    1:= e = -g_{tt}\dot{t}
    = \dot{t}\begin{cases}
        \frac{1}{\sigma}f_i(r), &r<R\\
        f_e(r), &r>R\,.
    \end{cases}
    \label{eq:el}
\end{equation}
The relation $ds/d\lambda=0$ implies
\begin{equation}
    0 = \begin{cases}
    -\frac{1}{\sigma}\,f_{i}(r)\,\dot{t}^2+\frac{\dot{r}^2}{f_{i}(r)}+r^2\dot{\phi}^2 \qquad &r<R\\
    -f_{e}(r)\,\dot{t}^2+\frac{\dot{r}^2}{f_{e}(r)}+r^2\dot{\phi}^2 \qquad  &r>R\,,
    \end{cases}
\end{equation}
and using eq.~\eqref{eq:el} to exchange $\dot{\phi}$ and $\dot{t}$ for $e$ and $\ell$, we obtain a purely radial equation.
The solution reads 
\begin{equation}
    r(\lambda_{i}) = 
        \sqrt{(\sigma^2-\ell^2) \lambda_i^2  +\frac{\ell^2}{\sigma^2-\ell^2}}\,,\qquad \lambda_i >\lambda_i(R)\,,
        \label{eq:r_int}
\end{equation}
in the interior, where $\lambda_{i}(r):=-\frac{1}{\sigma^2-\ell^2}\sqrt{(\sigma^2 - \ell^2) r^2- \ell^2}$, and by
\begin{equation}
    r(\lambda_{e}) = 
        \sqrt{(1-\ell^2)\lambda_e^2  + \frac{\mu^2\ell^2}{1-\ell^2}},\qquad
        \lambda_e < \lambda_e(R)\,,
        \label{eq:r_ext}
\end{equation}
in the exterior, where $\lambda_e(r):=-\frac{1}{1 -\ell^2}\sqrt{(1 - \ell^2) r^2 - \ell^2\mu^2}$.
From this, one finds that the null ray crosses into the shell if and only if $\ell^2<\frac{R^2}{R^2+\mu^2}$.
At the shell, null geodesics refract, much like the spacelike geodesics in figure \ref{fig:cool}.

We now compute the time $\Delta t$ that it takes for the null ray to go from $r=\infty$ to some $r=r_p<R$ in the bulk.
We have
\begin{equation}
    \Delta t = \left[t(\lambda_{r_p})-t(\lambda_R)\right]_{i}
    + \left[t(\lambda_{R})-t(-\infty)\right]_{e}\,,
    \label{eq:dsev}
\end{equation}
where the subscripts indicate we are to use the interior solution to calculate $t(\lambda_{r_p})-t(\lambda_R)$ and exterior solution to calculate $t(\lambda_{R})-t(-\infty)$.

In the exterior, eqs.~\eqref{eq:el} and \eqref{eq:r_ext} give
\begin{equation}
    t(\lambda_e)=\frac{1}{{\mu}}\tan^{-1}\frac{(1-\ell^2)\lambda_e}{{\mu}}+t_0\,,
\end{equation}
and hence\footnote{Note that eq.~\eqref{eq:delayss} with $R\mapsto r_p$ describes the total time delay to reach a point with $r_p>R$.}
\begin{equation}
    \left[t(\lambda_{R})-t(-\infty)\right]_e
    = \frac{1}{\mu}\left(\frac{\pi}{2}
    - \tan^{-1}\frac{\sqrt{R^2-\ell^2(R^2+\mu^2)}}{\mu}\right)\,.
    \label{eq:delayss}
\end{equation}
In the interior, eqs.~\eqref{eq:el} and \eqref{eq:r_int} give
\begin{equation}
    t(\lambda_i)={\sigma} \tan^{-1} \frac{(\sigma^2-\ell^2)\lambda_i}{{\sigma}}+t_0'\,,
\end{equation}
and hence
\begin{equation}
    \left[t(\lambda_{r_p})-t(\lambda_R)\right]_i
    =
   {\sigma} \left(\tan^{-1}\frac{\sqrt{R^2 \sigma^2 - \ell^2(R^2+1)}}{\sigma}
    -
    \tan^{-1} \frac{\sqrt{r_p^2 \sigma^2 - \ell^2(r_p^2+1)}}{\sigma}
    \right)\,.
\end{equation}

Similar calculations yield 
\begin{equation}
    \frac{\Delta \phi}{2} = \left[\phi(\lambda_{r_p})-\phi(\lambda_R)\right]_i
    + \left[\phi(\lambda_{R})-\phi(-\infty)\right]_e\,,
\end{equation}
with\footnote{There should be a $\mathrm{sgn}(\ell)$ multiplying the $\pi/2$ term, so we are implicitly assuming $\ell>0$. 
We also comment that eq.~\eqref{eq:rangess} with $R\mapsto r_p$ describes the full angular extent of the trajectory when $r_p>R$.}
\begin{equation}
    \left[\phi(\lambda_R)-\phi(-\infty)\right]_e
    =
    \frac{1}{\mu}\left(\frac{\pi}{2}
    -\tan^{-1}\frac{\sqrt{R^2  - \ell^2 (R^2+\mu^2)}}{\mu\,\ell}\right)\,,
    \label{eq:rangess}
\end{equation}
and
\begin{equation}
\left[\phi(\lambda_{r_p})-\phi(\lambda_R)\right]_i
=
\tan^{-1}\frac{\sqrt{R^2 \sigma^2 - \ell^2(R^2+1)}}{\ell}
-\tan^{-1}\frac{\sqrt{r_p^2\, \sigma^2 - \ell^2(r_p^2+1)}}{\ell}\,.
\end{equation}
From these relations, we can solve numerically for $\ell(r_p, \Delta \phi)$, and hence eliminate the $\ell$-dependence in eq.~\eqref{eq:dsev} to obtain $\Delta t=\Delta t(\Delta \phi,r_p)$.

\begin{figure}[htbp]
    \centering
    \includegraphics[width=0.52\linewidth]{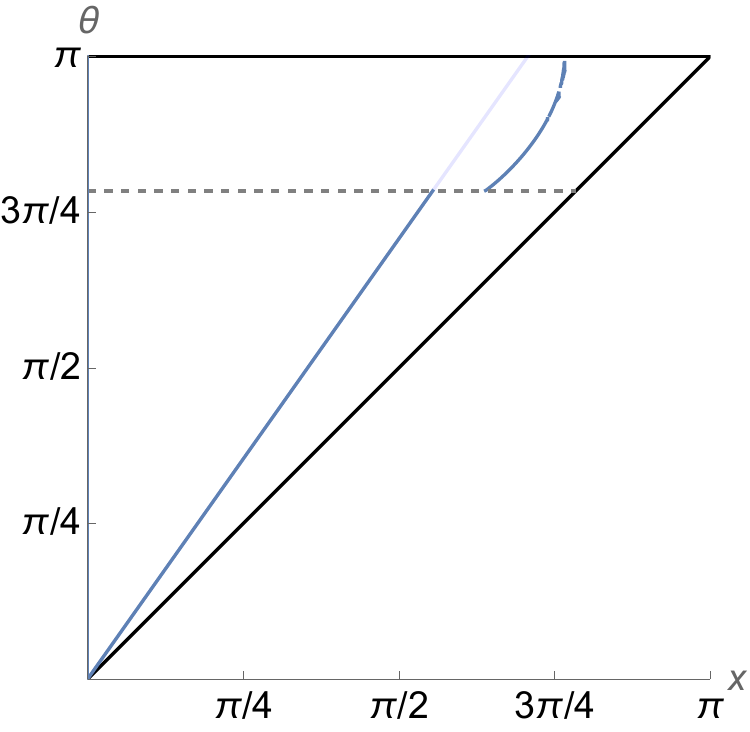}
    \hfill
    \includegraphics[width=0.375\linewidth]{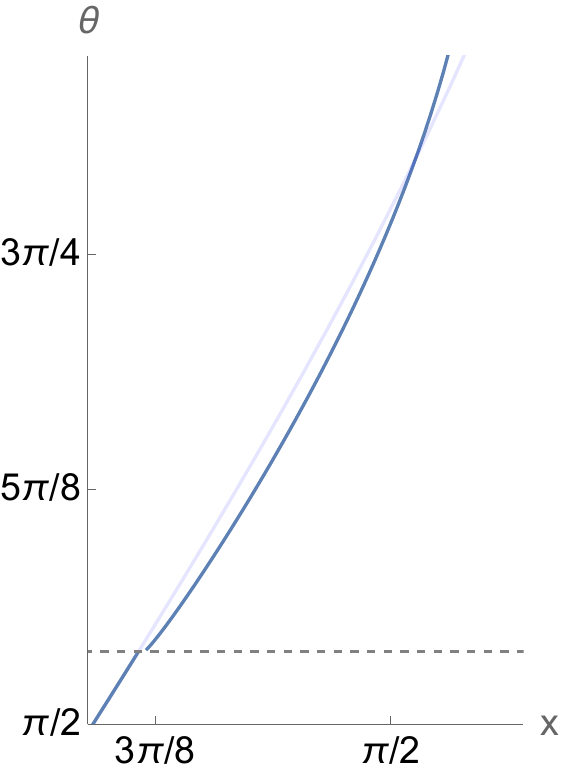}
    \caption{Left: Scattering threshold for a heavy shell ($\mu^2 = 0.001$) with $R=0.75$.
    To the left of the scattering threshold, scattering is impossible; to the right, scattering is possible, and this implies the entanglement wedge is connected; compare figure \ref{fig:phasediagshellmu001R05}, where the middle diagram also has $\mu^2 = 0.001$ and $R=0.75$.
    The horizontal dashed line represents where $\theta$ crosses $\Delta \phi^*$ and so $\gamma_{\mh{R}_1}$ jumps inside the shell, while the light blue line represents the naive extension of the $u=d$ formula, \ie the scattering threshold with $R=0$.
    We see that scattering is more difficult in this shell geometry, compared to the defect geometry of equal mass.
    Right: Scattering threshold for $\mu^2=0.6$, $R=1$.
    We have zoomed into a small window of the phase diagram to clarify the comparison with the $u=d$ formula. 
    We see scattering is sometimes easier in the shell geometry, compared to the defect geometry of equal mass.}
    \label{fig:scatterme}
\end{figure}

In practice, to find the scattering threshold, we set $r_p$ to be the minimum radius of $\gamma_{\mh{R}_1}$ and $\Delta \phi$ to be the width of $\mh{R}_1$; schematically, $\Delta \phi = \theta(r_p)$.
Then, we solve for the choice of $r_p$ such that $\Delta t(\Delta \phi,r_p)=x$, defined as $r_p(x)$.
Then, the curve $\theta(x) = \theta(r_p(x))$ is the scattering threshold.
For an example scattering phase diagram, see figure \ref{fig:scatterme}.

Consistently with the discussion in section \ref{sec:simpleEx}, the phase diagrams show that scattering is controlled by the $u<d$ inequality (or its naive extension, in the cases where $u$ doesn't exist) when $\theta < \Delta \phi^*$. 
When $\theta$ crosses $\Delta \phi^*$, then the $\gamma_{\mh{R}_1}$ surface jumps inside the shell; hence, we see a discontinuity in the scattering curve at $\theta = \Delta \phi^*$.
Numerical analyses show that $i<d$ does not control scattering after the jump is made.
However, in all cases checked, we have that holographic scattering implies $i<d$.

\newpage

\bibliographystyle{JHEP}
\bibliography{biblio.bib}

\end{document}